%% file: qcdnew.tex
\documentclass[12pt,twoside,a4paper]{article}

  \usepackage{graphicx}
  \usepackage{epstopdf}
  \usepackage{calc}
  \usepackage{amssymb}
  \usepackage{amsmath}
  \usepackage{makeidx}

  \input sections/layout.tex
  \numberwithin{equation}{section}
  \input sections/newcommands.tex

  \input sections/title.tex

\makeindex

\begin{document}

\input sections/seeindex.tex   

 \maketitle                        
 
 \input sections/abstract.tex
 \newpage
 \input sections/summary.tex

 \newpage
 \tableofcontents
 \newpage
 \input sections/introduction.tex

\input sections/theoryintro.tex 
 \input sections/theory.tex

\input sections/numanintro.tex
 \input sections/numan.tex

\input sections/program.tex

\input sections/callsintro.tex

\input sections/calls.tex

\input sections/cengine.tex

\input sections/ack.tex

 \appendix
 \input sections/appendix.tex

 \input sections/zmstf.tex

\input sections/hqstf.tex

 \addcontentsline{toc}{section}{References \hfill}
 \input sections/references.tex

 \addcontentsline{toc}{section}{Index \hfill}
 \printindex

\end{document}

%% file: sections/layout.tex
  \newlength{\dinwidth} \newlength{\dinmargin}

%% file: sections/newcommands.tex
\newcommand{\beq}[1]{\begin{equation}\label{#1}}
\newcommand{\eeq}{\end{equation}}
\newcommand{\bea}[1]{\begin{eqnarray}\label{#1}}
\newcommand{\eea}{\end{eqnarray}}

\newcommand{\Eq}[1]{Eq.~(\ref{#1})}

\newcommand{\eq}[1]{(\ref{#1})}
\newcommand{\Fi}[1]{Figure~\ref{#1}}
\newcommand{\Ta}[1]{Table~\ref{#1}}
\newcommand{\Se}[1]{Section~\ref{#1}}
\newcommand{\Ap}[1]{Appendix~\ref{#1}}

  \def\Journal#1#2#3#4{{#1}~{\bf #2}, #3 (#4)}

  \def\NP{Nucl.\ Phys.}
  \def\PL{Phys.\ Lett.}
  \def\PRL{Phys.\ Rev.\ Lett.}
  \def\PR{Phys.\ Rev.}
  \def\ZP{Z.\ Phys.}

  \def\JP{J.\ Phys.}
  \def\EPJ{Eur.\ Phys.\ J.}
  
  \def\SJNP{Sov.\ J.\ Nucl. Phys.}
  \def\JETP{Sov.\ Phys.\ JETP}
  \def\CPC{Comput.\ Phys.\ Commun.}
  \def\RMP{Rev.\ Mod.\ Phys.}
  \def\IJMP{Int.\ J.\ Mod.\ Phys.}

\newcommand{\qcdnum}{\mbox{\sc qcdnum}}
\newcommand{\Qcdnum}{\mbox{\sc Qcdnum}}
\newcommand{\qcdnumold}{\mbox{\sc qcdnum16}}

\newcommand{\qcdnumnew}{\mbox{\sc qcdnum17}}
\newcommand{\Qcdnumnew}{\mbox{\sc Qcdnum17}}

\newcommand{\mbutil}{\mbox{\sc mbutil}}
\newcommand{\netlib}{\mbox{\sc netlib}}
\newcommand{\cernlib}{\mbox{\sc cernlib}}

\newcommand{\fortran}{\mbox{\sc fortran}}

\newcommand{\pegasus}{\mbox{\sc pegasus}}
\newcommand{\hoppet}{\mbox{\sc hoppet}}

\newcommand{\ffns}{\mbox{\sc ffns}}
\newcommand{\vfns}{\mbox{\sc vfns}}

\newcommand{\gmvfns}{\mbox{\sc gm-vfns}}
\newcommand{\zmstf}{\mbox{\sc zmstf}}
\newcommand{\hqstf}{\mbox{\sc hqstf}}
\newcommand{\mbie}{\textit{i.e.}}
\newcommand{\mbeg}{\textit{e.g.}}
\newcommand{\mbetc}{\textit{etc.}}

\newcommand{\pa}{\partial}
\newcommand{\msbar}{\mbox{$\overline{\rm{MS}}$}} 
\newcommand{\qsq}{\mbox{${Q^2}$}}
\newcommand{\ms}{\ensuremath{\mu^2}}
\newcommand{\msz}{\ensuremath{\mu^2_0}}

\newcommand{\cbt}{\ensuremath{\rm c,b,t}}

\newcommand{\masqh}{\ensuremath{m^2_{h}}}

\newcommand{\msh}{\ensuremath{\mu^2_{h}}}
\newcommand{\msc}{\ensuremath{\mu^2_{\rm c}}}

\newcommand{\Rs}{\ensuremath{\mu^2_{\rm R}}}
\newcommand{\Fs}{\ensuremath{\mu^2_{\rm F}}}

\newcommand{\elR}{\ensuremath{L_{\rm R}}}
\newcommand{\elF}{\ensuremath{L_{\rm F}}}

\newcommand{\gevs}{\mbox{GeV$^2$}}
\newcommand{\enef}{\ensuremath{n_{f}}}
\newcommand{\rmv}{\ensuremath{{\rm v}}}
\newcommand{\rmg}{\ensuremath{{\rm g}}}
\newcommand{\rmq}{\ensuremath{{\rm q}}}
\newcommand{\rms}{\ensuremath{{\rm s}}}
\newcommand{\rmqb}{\ensuremath{\bar{\rm q}}}

\newcommand{\rmL}{\ensuremath{{\rm L}}}
\newcommand{\cF}{\ensuremath{\mathcal{F}}}
\newcommand{\Fell}{\ensuremath{F_{\rm L}}}
\newcommand{\as}{\ensuremath{\alpha_{\rm s}}}

\newcommand{\asubs}{\ensuremath{a_{\rm s}}}
\newcommand{\asmz}{\ensuremath{\alpha_{\rm s}(m_{\rm Z}^2)}}
\newcommand{\qbar}{\ensuremath{\bar{q}}}
\newcommand{\dbar}{\ensuremath{\bar{d}}}
\newcommand{\ubar}{\ensuremath{\bar{u}}}
\newcommand{\sbar}{\ensuremath{\bar{s}}}
\newcommand{\cbar}{\ensuremath{\bar{c}}}
\newcommand{\bbar}{\ensuremath{\bar{b}}}
\newcommand{\tbar}{\ensuremath{\bar{t}}}
\newcommand{\Abar}{\ensuremath{\bar{A}}}
\newcommand{\Bbar}{\ensuremath{\bar{B}}}
\newcommand{\Rbar}{\ensuremath{\bar{R}}}
\newcommand{\Sbar}{\ensuremath{\bar{S}}}

\newcommand{\order}{\ensuremath{{\rm O}}}
\newcommand{\epm}{\ensuremath{e^{\pm}}}
\newcommand{\qpm}{\ensuremath{q^{\pm}}}

\newcommand{\qsi}{\ensuremath{q_{\rm s}}}
\newcommand{\qva}{\ensuremath{q_{\rm v}}}
\newcommand{\qnsp}{\ensuremath{q_{\rm ns}^+}}
\newcommand{\qnsm}{\ensuremath{q_{\rm ns}^-}}

\newcommand{\ceeka}[1]{\ensuremath{C^{(#1)}}}

\newcommand{\al}{\alpha}
\newcommand{\be}{\beta}

\newcommand{\de}{\delta}
\newcommand{\De}{\Delta}
\newcommand{\la}{\lambda}
\newcommand{\La}{\Lambda}
\newcommand{\eps}{\epsilon}

\newcommand{\half}{\mbox{$\frac{1}{2}$}}     
\newcommand{\mbfrac}[2]{\mbox{$\frac{#1}{#2}$}}  

\newcommand{\ket}[1]{\ensuremath{| #1 \rangle}}

\newcommand{\der}{\ensuremath{{\rm d}}}
\newcommand{\ve}[1]{\ensuremath{\boldsymbol{#1}}}

\newcommand{\sspace}{^{\;}}
\newcommand{\mbsspace}{\hspace{2pt}}

\newcommand{\xtt}{\texttt}

\newcommand{\rar}{\rightarrow}
\newcommand{\transp}[1]{#1^{\rm T}}
\newcommand{\range}[5]{\xtt{#1}~$#2$~\xtt{#3}~$#4$~\xtt{#5}}
\newcommand{\compa}[3]{\xtt{#1}~$#2$~\xtt{#3}}

\newcommand{\siindex}[1]{\index{#1}}
\newcommand{\snindex}[2]{\index{#1|see{#2}}}
\newcommand{\dbindex}[2]{\index{#1!#2}}

\newcommand{\srindex}[1]
   {\index{qcdnumroutines@\textsc{qcdnum}~routines!#1@\texttt{#1}}}
\newcommand{\scindex}[1]
   {\index{qcdnumroutines@\textsc{qcdnum}~convolution
       engine!#1@\texttt{#1}}}
\newcommand{\sfindex}[1]
   {\index{qcdnumroutines@\textsc{qcdnum}~fast
       convolutions!#1@\texttt{#1}}}

\newcommand{\subr}[1]{\texttt{#1}\srindex{#1}}
\newcommand{\subc}[1]{\texttt{#1}\scindex{#1}}

\newcommand{\iepm}{epm@$e^{\pm}$}

\newcommand{\iffns}{ffns@\textsc{ffns}}
\newcommand{\ivfns}{vfns@\textsc{vfns}}

\newcommand{\iqcdnum}{qcdnum@\textsc{qcdnum}~program}
\newcommand{\izmstf}{zmstf@\textsc{zmstf}~package}
\newcommand{\ihqstf}{hqstf@\textsc{hqstf}~package}
\newcommand{\ipegasus}{pegasus@\textsc{pegasus}}
\newcommand{\imbutil}{mbutil@\textsc{mbutil}}
\newcommand{\inull}{null@\texttt{null}}

\newcommand{\ifellp}{fell@$F_{\rm L}'$}

\newcommand{\iyvariable}{yvariable@$y$-variable}
\newcommand{\itvariable}{tvariable@$t$-variable}

\newcommand{\imsbar}{msbar@$\overline{\rm{MS}}$}

\newcommand{\mbstrut}[1]{\rule{0mm}{#1}}

\newcommand{\subrbox}[2]{\begin{trivlist}\item%
\framebox[\linewidth][c]{\texttt{#1}}\end{trivlist}%
\srindex{#2}}
\newcommand{\subcbox}[2]{\begin{trivlist}\item%
\framebox[\linewidth][c]{\texttt{#1}}\end{trivlist}%
\scindex{#2}}
\newcommand{\subfbox}[2]{\begin{trivlist}\item%
\framebox[\linewidth][c]{\texttt{#1}}\end{trivlist}%
\sfindex{#2}}

\newcommand{\subnbox}[1]{\begin{trivlist}\item%
\framebox[\linewidth][c]{\texttt{#1}}\end{trivlist}}

\newenvironment{tdeflist}[2][\quad]%
 {\begin{list}{}{
  \settowidth{\labelwidth}{#1}%
  \setlength{\leftmargin}{\labelwidth+\labelsep}%
  \setlength{\itemsep}{#2}}}
 {\end{list}}

%% file: sections/title.tex
\title{
  \begin{flushright}
  \tt\normalsize{Nikhef-10-002}\\ 
  \tt\normalsize{arXiv:1005.1481} 
  \end{flushright}
  \vspace{1cm}
  \Large \bf QCDNUM: Fast QCD Evolution and Convolution\\
  \vspace{0.5cm}
  \normalsize \bf QCDNUM Version 17.00
}

\author{
  M. Botje\thanks{ 
  Nikhef, Science Park 105, 1098XG Amsterdam, the Netherlands;
  email m.botje@nikhef.nl}\\
  Nikhef, Science Park, Amsterdam, the Netherlands
}

\date{May 8, 2010\\\vspace{3mm}(Revised October 6, 2010)}

%% file: sections/seeindex.tex

\snindex{weights}{convolution weights}
\snindex{mesh point}{interpolation mesh}
\snindex{fragmentation functions}{time-like evolution}

%% file: sections/abstract.tex
\begin{abstract}
\noindent
The \qcdnum\ program numerically solves the evolution equations for
parton densities and fragmentation functions in perturbative QCD. 
Un-polarised parton
densities can be evolved up to next-to-next-to-leading order in powers
of the strong coupling constant, while polarised
densities or fragmentation functions can be evolved up to
next-to-leading order.  Other types of evolution can be accessed by
feeding alternative sets of evolution kernels into the program.  A
versatile convolution engine provides tools to compute parton
luminosities, cross-sections in hadron-hadron scattering, and deep
inelastic structure functions in the zero-mass scheme or in
generalised mass schemes. Input to these calculations are either the
\qcdnum\ evolved densities, or those read in from an external parton
density repository.  Included in the software distribution are packages
to calculate zero-mass structure functions in un-polarised deep
inelastic scattering, and heavy flavour contributions to these structure
functions in the fixed flavour number scheme. 
\end{abstract}

%% file: sections/summary.tex
{\bf PROGRAM SUMMARY}

\begin{small}
\noindent

{\em Program Title:} \textsc{qcdnum}

{\em Version:} 17.00

{\em Author:} M. Botje

{\em E-mail:} \xtt{m.botje@nikhef.nl}

{\em Program obtainable from:} \xtt{http://www.nikhef.nl/user/h24/qcdnum}

{\em Distribution format:} gzipped tar file

{\em Journal Reference:}

{\em Catalogue identifier:}

{\em Licensing provisions:} GNU Public License

{\em Programming language:} \textsc{fortran}-77

{\em Computer:} all

{\em Operating system:} all

{\em RAM:} Typically 3 Mbytes


{\em Classification:} 
     11.5 Quantum Chromodynamics, Lattice Gauge Theory

{\em External routines/libraries:} none, except the 
     \textsc{mbutil}, \textsc{zmstf} and \textsc{hqstf}
     packages that are part of the \qcdnum\ software
     distribution.

{\em Nature of problem:} Evolution of the strong coupling
     constant and parton densities, up to next-to-next-to-leading
     order in perturbative QCD. Computation of observable
     quantities by Mellin convolution of the evolved densities
     with partonic cross-sections.
     
{\em Solution method:} Parametrisation of the parton densities
     as linear or quadratic splines on a discrete grid, and
     evolution of the spline coefficients by
     solving (coupled) triangular matrix equations
     with a forward substitution algorithm. Fast computation of
     convolution integrals as weighted sums of spline coefficients,
     with weights derived from user-given convolution kernels.
     
{\em Restrictions:} Accuracy and speed are determined by the
     density of the evolution grid.
     
{\em Running time:} Less than 10~ms on a 2 GHz Intel Core 2 Duo
     processor to evolve the gluon density and 12 quark densities
     at next-to-next-to-leading order over a large kinematic range.                                   
     
\end{small}

%% file: sections/introduction.tex
                          
\section{Introduction} \label{Introduction}

In perturbative quantum chromodynamics (pQCD), a hard hadron-hadron
scattering cross section is calculated as the convolution of a
partonic cross section with the momentum distributions of the partons
inside the colliding hadrons. These parton distributions depend
on the Bjorken-$x$ variable (fractional momentum of the partons inside
the hadron) and on a scale \ms\ characteristic of the hard scattering
process. Whereas the $x$-dependence of the parton densities is
non-perturbative, the \ms\ dependence can be described in pQCD by the
DGLAP evolution equations~\cite{ref:dglap}. The perturbative expansion
of the splitting functions in these equations has recently been
calculated up to next-to-next-to-leading order (NNLO) in powers of the
strong coupling constant \as~\cite{ref:nnloa,ref:nnlob}.

\Qcdnum\ is a \fortran\ program that numerically solves the DGLAP
evolution equations on a discrete grid in $x$ and \ms.  Input to the
evolution are the $x$-dependence of the parton densities at some
input mass factorisation scale, and an input value of \as\ at some
input renormalisation scale. To study the scale uncertainties, the
renormalisation scale can be varied with respect to the mass
factorisation scale. All calculations in \qcdnum\ are performed in
the \msbar\ scheme.
\siindex{\imsbar\ scheme}%

\dbindex{\iqcdnum}{history}%
The program was originally developed in 1988 by members of the BCDMS
collaboration~\cite{ref:ouraou} for a next-to-leading order (NLO) pQCD
analysis of the SLAC and BCDMS structure function
data~\cite{ref:marcalain}. This code was adapted by the~NMC for use at
low $x$~\cite{ref:nmcqcd}. A complete revision led to the version
16.12 which was used in the QCD fits by ZEUS~\cite{ref:zeusqcd}, and
in a global QCD analysis of deep inelastic scattering data by the
present author~\cite{ref:mbfit}.

\Qcdnumnew\ is the NNLO upgrade of \qcdnumold.  A new
evolution algorithm, based on quadratic spline interpolation, yields
large gains in accuracy and speed; on a 2~GHz processor it takes less
than 10~ms to evolve  over a large kinematic range the full set
of parton densities at NNLO in the variable flavour number scheme.
\Qcdnumnew\ can evolve
un-polarised parton densities up to NNLO, and polarised densities or
fragmentation functions up to NLO.  Alternative sets of evolution
kernels can be fed into the program to perform other types of
evolution. It is also possible to read a parton density set
from an external library, instead of evolving these from the
input scale. 

A versatile set of convolution routines is provided that can be used
to calculate hadron-hadron scattering cross-sections,
parton luminosities, or deep inelastic
structure functions in either the zero-mass or in
generalised mass schemes.  Included in the software distribution are
the \zmstf\ and \hqstf\ add-on packages to compute un-polarised zero-mass
structure functions and, in the fixed flavour number scheme, the
contribution from heavy quarks to these structure functions.

This write-up is organised as follows. In \Se{se:theory} we summarise
the formalism underlying the DGLAP evolution of parton densities.  The
\qcdnum\ numerical method is described in \Se{se:numan}. Details about
the program itself and the description of an example job can be found
in \Se{se:program}. A subroutine-by-subroutine manual is given in
\Se{se:subroutines}.  The \qcdnum\ convolution engine is presented in
\Se{se:stfuser}.  The \zmstf\ and \hqstf\ packages are described in
the Appendices~\ref{se:zeromasstf} and~\ref{se:riemersma},
respectively.

%% file: sections/theoryintro.tex
\section{QCD Evolution}\label{se:theory}

In pQCD, the strong coupling constant \as\ evolves on the
renormalisation scale \Rs. The starting value is specified
at some input scale, which usually is taken to be $m_{\rm Z}^2$.
 
The parton density functions~(pdf)
\siindex{parton density function, pdf}%
evolve on the factorisation scale \Fs.
The starting point of a pdf evolution is given
by the $x$ dependence of the pdf at some initial scale~\msz.  The
coupled evolution equations that are obeyed by the gluon and the quark
densities can, to a large extent, be decoupled by writing them in
terms of the \emph{singlet} quark density (sum of all active quarks
and anti-quarks) and \emph{non-singlet} densities (orthogonal to the
singlet in flavour space). A~nice feature of \qcdnum\ is that it
automatically takes care of the singlet/non-singlet decomposition of a
set of pdfs.

Another input to the QCD evolution is the number of active flavours
\enef\ which specifies how many quark species (d, u, s, $\ldots$) are
participating in the QCD dynamics.  In the \emph{fixed flavour number
scheme} (\ffns), \enef\ is kept fixed throughout the evolution. Input
to an \ffns\ evolution are then the gluon density and $2\enef$
(anti-)quark densities at the input scale \msz. In the \emph{variable
flavour number scheme} (\vfns), the flavour thresholds $\ms_{\cbt}$ are
introduced and 3~light quark densities (d, u, s) are, together with
the corresponding anti-quark densities, specified below the charm
threshold~\msc. The heavy quarks and anti-quarks (c, b, t) are
dynamically generated by the QCD evolution equations at and above
their thresholds. Both the \ffns\ and the \vfns\ are supported by
\qcdnum. 

The QCD evolution formalism is relatively simple when the
renormalisation and factorisation scales are equal, but it becomes
more complicated when $\Rs \neq \Fs$. \Qcdnum\ supports a linear
relationship between the two scales.

In the following sections we describe the evolution of \as\ and the
pdfs, the renormalisation scale dependence, the singlet/non-singlet
decomposition, and the flavour schemes.

%% file: sections/theory.tex

\subsection{Evolution of the Strong Coupling Constant}
\label{se:alfas}

The evolution of the strong coupling constant reads, up to NNLO,
\siindex{evolution of \as|(}%
\beq{eq:asevolution}
  \frac{\der \asubs(\ms)}{\der \ln \ms} = - \sum_{i=0}^2 \be_i\;
  \asubs^{i+2}(\ms). 
\eeq
Here $\ms = \Rs$ is the renormalisation scale
\siindex{renormalisation scale \Rs}%
and $\asubs = \as/2\pi$.  The $\be$-functions
\siindex{beta-functions}%
in \eq{eq:asevolution} depend on the number \enef\ of active quarks
with pole mass $m < \mu$.
\siindex{number of active flavours \enef}%
In the \msbar\ scheme they are given 
by~\cite{ref:fpzphys,ref:betafunctions}
\bea{eq:betafunctions}
  \be_0 & = & \frac{11}{2} - \frac{1}{3}\; \enef \nonumber \\
  \be_1 & = & \frac{51}{2} - \frac{19}{6}\; \enef \nonumber \\
  \be_2 & = & \frac{2857}{16} - \frac{5033}{144}\; \enef +
  \frac{325}{432}\; \enef^2.
\eea
The leading order (LO) analytical solution of \eq{eq:asevolution} can
be written as
\beq{eq:aslo}
  \frac{1}{\asubs(\ms)} = \frac{1}{\asubs(\msz)} + \be_0\; \ln\left(
    \frac{\ms}{\msz} \right) \equiv \be_0\; \ln \left(
    \frac{\ms}{\La^2} \right). 
\eeq
In \eq{eq:aslo}, the parameter $\La$ is defined as the scale where the
first term on the right-hand side vanishes, that is, the scale where
\as\ becomes infinite.
\siindex{scale parameter $\La$}%
Beyond LO, the definition of a scale parameter is ambiguous so that it
is more convenient to take $\asmz$ as a reference. The value of \as\ 
at any other scale is then obtained from a numerical integration
of~\eq{eq:asevolution},\footnote{I thank A.~Vogt for providing his
  $4^{\rm th}$ order Runge-Kutta routine to integrate
  \eq{eq:asevolution} up to NNLO.} instead of from approximate
analytical solutions parametrised in terms of $\La$.

In the evolution of \as, the number of active flavours is set to $\enef
= 3$ below the charm threshold $\Rs = \msc$ and is changed from
$\enef$ to $\enef + 1$ at the flavour thresholds $\Rs = \ms_{\cbt}$.
\dbindex{flavour thresholds}{on the renormalisation scale}%
At NNLO, and sometimes also at NLO, there are small discontinuities in
the \as\ evolution at the flavour thresholds~\cite{ref:asmatch}; see
\Se{se:vfns} for details.
\siindex{discontinuities in \as\ evolution}%

In \Fi{fig:alphas},
%
\begin{figure}[tbh]
\begin{center}
\includegraphics[width=0.75\linewidth]{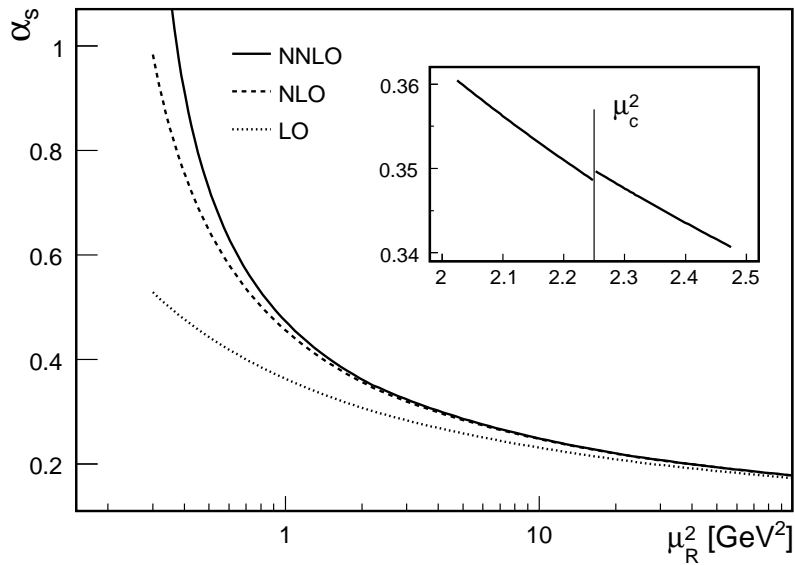}
\end{center}
\caption{\footnotesize
  The strong coupling constant $\as(\Rs)$ evolved downward from $\asmz
  = 0.118$ in LO (dotted curve), NLO (dashed curve) and NNLO (full curve).
  The inset shows an enlarged view of the NNLO discontinuity in \as\ 
  at the charm threshold \msc.}
\label{fig:alphas} 
\end{figure}
%
we plot the evolution of \as\ calculated at LO, NLO and
NNLO.\footnote{With the settings $\asmz = 0.118$ and $\mu_{\cbt} =
(1.5,5,188)$~GeV.} Because pQCD breaks down when \as\ becomes large,
\qcdnum\ will issue a fatal error when~$\as(\ms)$ exceeds a pre-set
limit. For a given value of \asmz, it is clear from the figure
that such a limit will correspond to larger values of \ms\ at larger
perturbative order. 
\siindex{evolution of \as|)}%


\subsection{The DGLAP Evolution Equations} \label{se:dglap}

\siindex{DGLAP evolution equations}%
The DGLAP evolution equations can be written as
\beq{eq:dglap1}
  \frac{ \pa f_i(x,\ms)}{\pa \ln \ms} =
  \sum_{j = q,\qbar,g} \ \int_x^1 \frac{\der
  z}{z} P_{ij} \left( \frac{x}{z},\ms \right) f_j(z,\ms)
\eeq
where $f_i$ denotes an un-polarised parton number density,
\dbindex{parton density function, pdf}{number/momentum density}%
\siindex{un-polarised evolution}%
$P_{ij}$ are the QCD splitting functions,
$x$ is the Bjorken scaling variable
\siindex{Bjorken-$x$ variable}%
and $\ms = \Fs$ is the mass factorisation scale,
\siindex{factorisation scale \Fs}%
which we assume here to be equal to the renormalisation scale $\Rs$.
The indices $i$ and $j$ in \eq{eq:dglap1} run over the parton species
\mbie, the gluon and \enef\ active flavours of quarks and anti-quarks.
In the quark parton model, and also in LO pQCD, the parton densities
are defined such that $f(x,\ms)\der x$ is, at a given $\ms$, the
number of partons which carry a fraction of the nucleon momentum
between $x$ and $x + \der x$.  The distribution $xf(x,\ms)$ is then
the parton momentum density.\footnote{In this section we use the
number densities $f(x,\ms)$.  In \qcdnum\ itself, however, we use
$xf(x,\ms)$.}
\dbindex{parton density function, pdf}{number/momentum density}%
Beyond LO there is no such intuitive interpretation.  The definition
of $f$ then depends on the renormalisation and factorisation scheme in
which the calculations are carried out (\msbar\ in
\qcdnum).\footnote{In the DIS scheme $f$ is defined such that the LO
  (quark-parton model) expression for the $F_2$ structure function is
  preserved at NLO.  But this is true only for $F_2$ and not for
  \Fell\ and $xF_3$.}

Introducing a short-hand notation for the Mellin convolution,
\siindex{Mellin convolution}%
\beq{eq:mellinconv}
  [f \otimes g](x) =  \int_x^1 \frac{\der z}{z} f \left( \frac{x}{z}
  \right) g(z) = \int_x^1 \frac{\der z}{z} f(z)\; g \left( \frac{x}{z}  
  \right),
\eeq
we can write \eq{eq:dglap1} in compact form as (we drop the arguments
$x$ and \ms\ in the following) 
\beq{eq:dglap2}
  \frac{\pa f_i}{\pa \ln \ms} = \sum_{j = q,\qbar,g} P_{ij}
  \otimes f_j.  
\eeq
If the $x$ dependencies of the parton densities are known at some
scale
\msz, they can be evolved to other values of \ms\ by solving this set
of $2\enef+1$ coupled integro-differential equations. Fortunately,
\eq{eq:dglap2} can be considerably simplified by taking the symmetries
in the splitting functions
\dbindex{splitting functions}{symmetries in}
into account~\cite{ref:fpzphys}:
\bea{eq:pqqfun} 
  P_{\rmg \rmq_i} & = & P_{\rmg \rmqb_i} =\ P_{\rmg \rmq}
  \nonumber \\
  P_{\rmq_i \rmg} & = &
  P_{\rmqb_i \rmg} =\  \frac{1}{ 2 \enef} P_{\rmq \rmg} \nonumber \\
  P_{\rmq_i \rmq_k} & = & P_{\rmqb_i \rmqb_k} =\ \de_{ik}P_{\rmq
  \rmq}^{\rm v} + P_{\rmq \rmq}^{\rm s} \nonumber \\
  P_{\rmq_i \rmqb_k} & = & P_{\rmqb_i \rmq_k} =\ \de_{ik}P_{\rmq
  \rmqb}^{\rm v} + P_{\rmq \rmqb}^{\rm s}.
\eea
Inserting \eq{eq:pqqfun} in \eq{eq:dglap2}, we find after some algebra
that the singlet quark density
\siindex{singlet quark density}%
\beq{eq:defsinglet}
  q_{\rm s} = \sum_{i=1}^{\enef} (q_i + \qbar_i)
\eeq
obeys an evolution equation coupled to the gluon density
\siindex{singlet-gluon evolution}%
\beq{eq:dglapsg}
  \frac{\pa}{\pa \ln \ms} 
  \begin{pmatrix} 
  q_{\rm s} \\ g
  \end{pmatrix} 
  = 
  \begin{pmatrix}
  P_{\rmq \rmq} & P_{\rmq \rmg} \\
  P_{\rmg \rmq} & P_{\rmg \rmg}
  \end{pmatrix} 
  \otimes 
  \begin{pmatrix}
  q_{\rm s} \\ g
  \end{pmatrix},
\eeq
with $P_{\rm qq}$ given by
\beq{eq:pqqdef}
  P_{\rm qq} = P_{\rm qq}^{\rm v} +  P_{\rmq \rmqb}^{\rm v} + \enef
  (P_{\rm qq}^{\rm s} +  P_{\rmq \rmqb}^{\rm s}).
\eeq
Likewise, we find that the non-singlet combinations
\siindex{non-singlet quark density}%
\beq{eq:defnonsinglet}
  q^{\pm}_{ij}  =  (q_i \pm \qbar_i) - (q_j \pm \qbar_j)
  \qquad \mbox{and} \qquad 
  q_{\rm v} =  \sum_{i=1}^{\enef} (q_i - \qbar_i)
\eeq
evolve independently from the gluon and from each other according to
\siindex{non-singlet evolution}%
\beq{eq:dglapns}
  \frac{\pa\mbsspace q^{\pm}_{ij}}{\pa \ln \ms} =  
  P_{\pm} \otimes q_{ij}^{\pm}
  \qquad \mbox{and} \qquad 
  \frac{\pa\mbsspace q_{\rm v}}{\pa \ln \ms}  =  
  P_{\rm v} \otimes q_{\rm v},
\eeq
with splitting functions defined by
\beq{eq:pnsdef}
  P_{\pm}  =  P_{\rm qq}^{\rm v} \pm P_{\rmq \rmqb}^{\rm v}
  \qquad \mbox{and} \qquad 
  P_{\rm v} =  P_{\rm qq}^{\rm v} -  P_{\rmq \rmqb}^{\rm v}
  + \enef (P_{\rm qq}^{\rm s} -  P_{\rmq \rmqb}^{\rm s}).
\eeq
The evolution of the $q^{\pm}_{ij}$ is linear in the densities, so that
any linear combination of the $q^{+}_{ij}$ or $q^{-}_{ij}$ also
evolves according to \eq{eq:dglapns}.

The splitting functions can be expanded in a perturbative series in
\as\ which presently is known up to NNLO.
\dbindex{splitting functions}{perturbative expansion of|(}%
For the four splitting functions $P_{ij}$ in \eq{eq:dglapsg} we may
write
\beq{eq:pseries}
  P_{ij}(x,\ms) =
    \asubs(\ms)\; P_{ij}^{(0)}(x) + \asubs^2(\ms)\; P_{ij}^{(1)}(x) +
    \asubs^3(\ms)\; P_{ij}^{(2)}(x) + \order(\asubs^4)
\eeq
where we have set, as in the previous section, $\asubs = \as/2\pi$.
Note the separation in the variables $x$ and \ms\ on the right-hand
side of \eq{eq:pseries}.
We drop again the arguments $x$ and~\ms\ and write the expansion of
the non-singlet splitting functions as
\bea{eq:nsexpansion}
  P_{\pm} & = &
    \asubs\; P_{\rm qq}^{(0)} + \asubs^2\; P_{\pm}^{(1)} 
    + \asubs^3\; P_{\pm}^{(2)} + \order(\asubs^4)  \nonumber \\
  P_{\rm v} & = &
    \asubs\; P_{\rm qq}^{(0)} + \asubs^2\; P_{-}^{(1)} +  
    \asubs^3\; P_{\rm v}^{(2)} + \order(\asubs^4).
\eea
Truncating the right-hand side to the appropriate order in \asubs, it
is seen that at LO the three types of non-singlet obey the same
evolution equations. At NLO, $q_{ij}^-$ and $\qva$ evolve in the same
way but different from $q_{ij}^+$. At NNLO, all three non-singlets
evolve differently.
\dbindex{splitting functions}{perturbative expansion of|)}%

It is evident from \eq{eq:pqqfun}, \eq{eq:pqqdef} and \eq{eq:pnsdef}
that several splitting functions depend on the number of active
flavours \enef.  This number is set to 3 below $\Fs = \msc$ and changed
to~$\enef = (4,5,6)$ at and above the thresholds $\Fs = \ms_{\cbt}$.
\dbindex{number of active flavours \enef}{value at threshold}%
\dbindex{flavour thresholds}{on the factorisation scale}%
\dbindex{flavour thresholds}{on the renormalisation scale}%
In case $\Fs \neq \Rs$, \qcdnum\ adjusts the \Rs\ thresholds such that
\enef\ changes in both the splitting and the beta functions when
crossing a threshold; see also \Se{se:vfns}.

The LO splitting functions are given in \Ap{app:singular}. Those at
NLO can be found in~\cite{ref:fpns} (non-singlet) and~\cite{ref:fpsi}
(singlet).\footnote{Two well-known misprints
in~\cite{ref:fpsi} are: (i) the lower integration limit in the definition
of~$S_2(x)$ must read $x/(1+x)$;
(ii) in the expression for $\hat{P}_{\rm FF}^{(1,{\rm T})}$ the term 
$(10-18x-\mbfrac{16}{3}x^2)$ must read
$(-10-18x-\mbfrac{16}{3}x^2)$.}
The NNLO splitting functions and their parametrisations are
given in~\cite{ref:nnloa} (non-singlet) and~\cite{ref:nnlob}
(singlet). The DGLAP equations also apply to polarised parton
\siindex{polarised evolution}%
densities and to fragmentation functions (time-like evolution),
\siindex{time-like evolution}%
each with their own set of evolution kernels. For the
polarised splitting functions up to NLO 
we refer to~\cite{ref:vogelsang}, and
references therein. The time-like evolution of fragmentation
functions at~LO is described in~\cite{ref:nasonwebber}, see 
also~\Ap{app:singular}.
The NLO time-like splitting functions 
can be found in~\cite{ref:fpns} and~\cite{ref:fpsi}. 


\subsection{Renormalisation Scale Dependence}\label{se:rscale}

\dbindex{renormalisation scale dependence}{of pdfs|(}%
In the previous section, we have assumed that the factorisation and
renormalisation scales are equal. For $\Fs \neq \Rs$ we expand \asubs\
in a Taylor series on a logarithmic scale around \Rs
\beq{eq:astaylor}
  \asubs(\Fs) = \asubs(\Rs) + \asubs'(\Rs) \elR +
  \frac{1}{2}\; \asubs''(\Rs) \elR^2 + \ldots
\eeq
with $\elR = \ln(\Fs/\Rs)$. Using \eq{eq:asevolution} to calculate the
derivatives in \eq{eq:astaylor}, we obtain
\bea{eq:asexpansion}
\asubs(\Fs)   & = & \asubs(\Rs) - \be_0 \elR\;
                    \asubs^2(\Rs) - ( \be_1 \elR - \be_0^2 \elR^2 )\;
                    \asubs^3(\Rs) + \order(\asubs^4) \nonumber \\ 
\asubs^2(\Fs) & = & \asubs^2(\Rs) - 2 \be_0 \elR
                          \;\asubs^3(\Rs) + \order(\asubs^4) \nonumber \\
\asubs^3(\Fs) & = & \asubs^3(\Rs) + \order(\asubs^4).
\eea
To calculate the renormalisation scale dependence of the evolved
parton densities, the powers of $\asubs$ in the splitting function
expansions \eq{eq:pseries} and \eq{eq:nsexpansion} are replaced by the
expressions on the right-hand side of \eq{eq:asexpansion}, with the
understanding that these are truncated
\siindex{truncation prescription}%
to order \asubs\ when we evolve at LO, to order $\asubs^2$ when we
evolve at NLO, and to order $\asubs^3$ when we evolve at NNLO.
\dbindex{renormalisation scale dependence}{of pdfs|)}%


\subsection{Decomposition into Singlet and Non-singlets}
\label{se:decompose} 

In this section we describe the transformations between a flavour
basis and a singlet/non-singlet basis, as is implemented in \qcdnum. For
this purpose we write an arbitrary linear combination of quark and
anti-quark densities as
\beq{eq:linearsum}
  \ket{p} = \sum_{i=1}^{\enef} (\al_i \ket{q_i} + \be_i
    \ket{\qbar_i}),
\eeq
where the index $i$ runs over the number of active flavours.
To make a clear distinction between a coefficient and a pdf, we
introduce here the ket notation $\ket{f}$ for $f(x,\ms)$.
\siindex{ket notation}%

Because a linear combination of non-singlets is again a non-singlet, it
follows directly from the definition~\eq{eq:defnonsinglet} that the
coefficients of any non-singlet satisfy the constraint
\beq{eq:nsconstraint}
  \sum_{i=1}^{\enef} ( \al_i + \be_i ) = 0,
\eeq
that is, a non-singlet is---by definition---orthogonal to the singlet
in flavour space.
\siindex{non-singlet quark density}%

It is convenient to define $\ket{\qpm_i} = \ket{q_i} \pm
\ket{\qbar_i}$ and write the linear combination \eq{eq:linearsum} as
\beq{eq:qplusminsum}
  \ket{p} = \sum_{i=1}^{\enef} (b^+_i \ket{q^+_i} + b^-_i
  \ket{q^-_i} ).
\eeq
The coefficients $b^{\pm}_i$, $\al_i$ and $\be_i$ are related by
\beq{eq:getbcoef}
  b^{\pm}_i = \frac{\al_i \pm \be_i}{2}, \qquad
 \al_i = b^+_i + b^-_i, \qquad \be_i = b^+_i - b^-_i.
\eeq

We define a basis of singlet, valence, and $2(\enef -1)$ additional
non-singlets by
\siindex{singlet/non-singlet basis $\epm$}%
\siindex{\iepm\ basis pdfs}
\beq{eq:evecdef}
  \ket{e^+_1} = \ket{\qsi},\qquad \ket{e^-_1} = \ket{\qva}, \qquad
  \ket{\epm_i} = \sum_{j=1}^{i-1} \ket{\qpm_j} - (i-1)\sspace
  \ket{\qpm_i} \mbox{\ \ for\ \ } 2 \leq i \leq \enef.
\eeq
In matrix notation, this transformation can be written as
\beq{eq:basis}
  \ket{\ve{e}^{\pm}} = \ve{U} \ket{\ve{q}^{\pm}},
\eeq
where $\ve{U}$ is the $\enef \times \enef$ sub-matrix of the $6
\times 6$ transformation matrix
\beq{eq:defm}
  \cal{U} = \left( \begin{array}{rrrrrr}
                   1  &  1  &  1  &  1  &  1  &  1 \\
                   1  & -1  &  0  &  0  &  0  &  0 \\
                   1  &  1  & -2  &  0  &  0  &  0 \\
                   1  &  1  &  1  & -3  &  0  &  0 \\
                   1  &  1  &  1  &  1  & -4  &  0 \\
                   1  &  1  &  1  &  1  &  1  & -5
             \end{array} \right).
\eeq

It is seen that the second to sixth row of \eq{eq:defm} are
orthogonal to the first row (singlet), so that they indeed represent
non-singlets as defined by~\eq{eq:nsconstraint}. In fact, all rows of
$\ve{U}$ are orthogonal to each other, so that scaling by the row-wise
norm yields a rotation matrix, which has the transpose as its inverse.
By scaling back this inverse we obtain
\beq{eq:inversem}
  \ve{U}^{-1} = \ve{U}^{\rm T} \ve{S}^2,
\eeq
where $\ve{U}^{\rm T}$ is the transpose of $\ve{U}$ and $\ve{S}^2$ is
the square of the diagonal scaling matrix:
\beq{eq:defscale}
  S_{ij}^2 = 
  \de_{ij} \left(\; \sum_{k=1}^{\enef}
  U_{ik}^2 \right)^{\!\!-1} = 
  \left\{ \begin{array}{ll}
    \de_{ij}/\enef & \mbox{\ for\ } i = 1 \\
    \de_{ij}/i(i-1)& \mbox{\ for\ } i > 1.
  \end{array} \right.
\eeq
Using \eq{eq:inversem} and \eq{eq:defscale} to invert any $\enef
\times \enef$ sub-matrix of \eq{eq:defm}, it is straight forward to
show by explicit calculation that
\beq{eq:minverse}
  U_{ij}^{-1} = \left\{
  \begin{array}{ll}
   \ \ 1/\enef   & \mbox{\ for\ } j = 1 \\
       -1/j      & \mbox{\ for\ } j = i \neq 1 \\
   \ \ 1/j(j-1)  & \mbox{\ for\ } j > i \\
   \ \  0        & \mbox{\ otherwise}.
  \end{array}
  \right.
\eeq
The inverse of the transformation \eq{eq:evecdef} is thus given by
\bea{eq:backtoq}
  \ket{\qpm_1} & = & \frac{\ket{\epm_1}}{\enef} + \sum_{j=2}^{\enef}\;
  \frac{\ket{\epm_j}}{j(j-1)} \nonumber \\
  \ket{\qpm_i} & = & \frac{\ket{\epm_1}}{\enef} - \frac{\ket{\epm_i}}{i} +
  \sum_{j=i+1}^{\enef}\; \frac{\ket{\epm_j}}{j(j-1)}\qquad i > 1.
\eea
We can now write the linear combination $\ket{p}$ on the 
$\ket{\epm}$ basis as 
\beq{eq:eplusminsum}
  \ket{p}  = \sum_{i=1}^{\enef} ( d^+_i \ket{e^+_i} + d^-_i
  \ket{e^-_i} ), 
\eeq
where the coefficients $d^{\pm}_i$ are related to the $b^{\pm}_i$ of
\eq{eq:qplusminsum} by
\beq{eq:dfromc}
  d^{\pm}_i = \sum_{j=1}^{\enef} b^{\pm}_j U^{-1}_{ji}, \qquad 
  b^{\pm}_i = \sum_{j=1}^{\enef} d^{\pm}_j U_{ji}.
\eeq

Let the starting values of the DGLAP evolutions be given by the gluon
density and $2\enef$ arbitrary quark densities, that is, by $2\enef+1$
functions of $x$ at some input scale \msz.  We can arrange the input
quark densities in a \mbox{$2\enef$-dimensional} vector
$\ket{\ve{p}}$. Likewise, we store the densities $\ket{\qpm_i}$ in a
vector $\ket{\ve{q}}$, the $\ket{\epm_i}$ in a vector $\ket{\ve{e}}$
and the $b^{\pm}$ coefficients of each input density in the rows of a
$2\enef \times 2\enef$ matrix $\ve{B}$. The flavour decomposition of
the input densities can then be written as $\ket{\ve{p}} = \ve{B}
\ket{\ve{q}}$ and the singlet/non-singlet decomposition as
\beq{eq:pisdq}
  \ket{\ve{p}} = \ve{B} \ve{T}^{-1}
  \ket{\ve{e}} \mbox{\ \ with \ \ }
   \ve{T} \equiv  \begin{pmatrix}
                  \ve{U} & \ve{0}       \\
                  \ve{0} & \ve{U}
                  \end{pmatrix}. 
\eeq
Provided that $\ve{B}^{-1}$ exists (\mbie\ the input densities are
linearly independent), the starting values of the singlet and
non-singlet densities are calculated from the inverse relation
\beq{eq:efromp}
  \ket{\ve{e}} = \ve{T} \ve{B}^{-1} \ket{\ve{p}}.
\eeq
%


\subsection{Flavour Number Schemes}
\label{se:vfns}

\Qcdnum\ supports two evolution schemes, known as the fixed flavour
number scheme (\ffns) and the variable flavour number scheme (\vfns).
 
\siindex{\iffns\ fixed flavour number scheme}%
In the \ffns\ we assume that \enef\ quark flavours have zero mass, while
those of the remaining flavours are taken to be infinitely large.  In
this way, only \enef\ flavours participate in the QCD dynamics so that
in the \ffns\ the value of \enef\ is simply kept constant for all~\ms,
with $3 \leq \enef \leq 6$.  In the \ffns, the input scale \msz\
can be chosen anywhere within the boundaries of the evolution grid,
although one should be careful with backward evolution in \qcdnum;
see \Se{se:dglaplin}.

\siindex{\ivfns\ variable flavour number scheme}%
In the \vfns, the number of flavours changes from \enef\ to $\enef +1$
when the factorisation scale is equal to the pole mass of the heavy quarks
$\msh = m^2_h$, $h = (\cbt)$. A heavy quark~$h$ is thus considered to
be infinitely massive below $\msh$ and
mass-less above $\msh$. As a consequence, the heavy flavour
distributions are zero below their respective thresholds and are
dynamically generated by the QCD evolution equations at and above
$\msh$.  Such an abrupt turn-on at a fixed scale is of course
unphysical but this poses no problem since the parton densities
themselves are not observables. The \vfns\ or \ffns\ parton densities
evolved with \qcdnum\ are, in fact, valid input to structure function
and cross section calculations that include mass terms and obey the
kinematics of heavy quark
production~\cite{ref:riemersma,ref:acotchi,ref:trvfns}. Such
calculations are not part of \qcdnum\ itself, but can be coded in
add-on packages; see \Se{se:stfuser}.

An important feature of \vfns\ evolution is that the input scale \msz\
cannot be above the lowest heavy flavour threshold \msc.  This is
because otherwise heavy flavour contributions must be included in the
input parton densities which clearly is in conflict with the dynamic
generation of heavy flavour by the QCD evolution equations.

Another feature of the \vfns\ is the existence of discontinuities at
the flavour thresholds in~\as\ and the parton densities; we will now
turn to the calculation of these discontinuities.
\siindex{discontinuities in \as\ evolution|(}%
Because the beta functions \eq{eq:betafunctions} depend on \enef, it
follows that the slope of the \as\ evolution is discontinuous when
crossing a threshold in the \vfns. Beyond LO there are not only
discontinuities in the slope but also in \as\
itself. In N$^{\ell}$LO, the value of
$\as^{(\enef+1)}$ is, at a flavour threshold, related to
$\as^{(\enef)}$ by~\cite{ref:asmatch,ref:pegasus}
\beq{eq:asjump}
  \asubs^{(\enef+1)}(\kappa \msh) =  \asubs^{(\enef)}(\kappa \msh) +
  \sum_{n=1}^{\ell} \Bigl\{ \left[\asubs^{(\enef)}(\kappa \msh)
  \right]^{n+1} \sum_{j=0}^n C_{n,j} \ln^j \kappa \Bigr\} \qquad
  \ell = 1,2.
\eeq 
Here $\msh$ is the threshold defined on the factorisation scale and
$\kappa$ is the ratio $\Rs/\Fs$ at~$\msh$.  For $\asubs = \as/4\pi$,
the coefficients $C$ in \eq{eq:asjump} read
\[
  C_{1,0} = 0,\qquad C_{1,1} = \mbfrac{2}{3},\qquad 
  C_{2,0} = \mbfrac{14}{3},\qquad C_{2,1} = \mbfrac{38}{3}, \qquad
  C_{2,2} = \mbfrac{4}{9}.
\]
Note that there is always a discontinuity in \as\ at NNLO. At NLO, a
discontinuity only occurs when $\kappa \neq 1$, that is, when the
renormalisation and factorisation scales are different.  In case of
upward evolution, $\as^{(\enef+1)}$ is computed directly from
\eq{eq:asjump} while for downward evolution,
$\as^{(\enef-1)}$ is evaluated by numerically solving the equation
\[
  \asubs^{(\enef)} - \asubs^{(\enef-1)} -
  \De \asubs\left(\asubs^{(\enef-1)}\right) = 0,
\]
where the function $\De \asubs(\asubs)$ is given by the second term on
the right-hand side of~\eq{eq:asjump}.
\siindex{discontinuities in \as\ evolution|)}%

\siindex{discontinuities in pdf evolution|(}%
In the \vfns\ at NNLO, not only \as\ but also the parton densities have
discontinuities at the flavour thresholds~\cite{ref:nnlomatch}:
\bea{eq:pdfjumps}
  g(x,\msh,\enef+1) & = & g(x,\msh,\enef) +
  \De g(x,\msh,\enef) \nonumber \\
  q^{\pm}_i(x,\msh,\enef+1) & = & q^{\pm}_i(x,\msh,\enef) +
  \De q^{\pm}_i(x,\msh,\enef)\qquad i = 1,\ldots,\enef \nonumber \\
  h^{+}(x,\msh,\enef+1) & = & \De h^{+}(x,\msh,\enef)\nonumber \\
  h^{-}(x,\msh,\enef+1) & = & \De h^{-}(x,\msh,\enef) = 0,
\eea
where $h = (\cbt)$ for $\enef = (3,4,5)$. Note that a heavy quark $h$
becomes a light quark $q_i$ above the threshold $\ms_h$.

\dbindex{flavour thresholds}{on the renormalisation scale}%
In \qcdnum, the flavour thresholds on the renormalisation scale are
adjusted such that \enef\ changes by one unit in both the beta
functions and the splitting functions when crossing a threshold. With
this choice, the parton densities are continuous at LO and NLO while at
NNLO the calculation of the discontinuities is considerably simplified
(all terms proportional to powers of $\ln(m^2/\mu^2)$ in
ref.~\cite{ref:nnlomatch} vanish). So we may write
\bea{eq:jumpconv}
  \De g(x,\msh,\enef) & = & \asubs^2 \left\{ [A_{\rm gq} \otimes
  \qsi](x,\msh,\enef) +  [A_{\rm gg} \otimes g](x,\msh,\enef)\right\}
  \nonumber \\ 
  \De q^{\pm}_i(x,\msh,\enef) & = & \asubs^2\  [A_{\rm qq} \otimes
  q^{\pm}_i](x,\msh,\enef) \nonumber \\
  \De h^+(x,\msh,\enef) & = & \asubs^2 \left\{ [A_{\rm hq} \otimes
  \qsi](x,\msh,\enef) +  [A_{\rm hg} \otimes g](x,\msh,\enef)\right\}.
\eea
Here $\asubs$ stands for $\asubs^{(\enef+1)}(\kappa \msh)$ as defined
by \eq{eq:asjump}.  The convolution kernels $A_{ij}$ can be found in
Appendix~B of~\cite{ref:nnlomatch}.\footnote{In the notation
of~\cite{ref:nnlomatch},
  $A_{\rm gq} = A_{\rm gq,H}^{{\rm S},(2)}$ (eq. B.5),
  $A_{\rm gg} = A_{\rm gg,H}^{{\rm S},(2)}$ (B.7), 
  $A_{\rm qq} = A_{\rm qq,H}^{{\rm NS},(2)}$ (B.4), 
  $A_{\rm hq} = \tilde{A}_{\rm Hq}^{{\rm PS},(2)}$ (B.1) and
  $A_{\rm hg} = \tilde{A}_{\rm Hg}^{{\rm }S,(2)}$ (B.3).
  For the latter we use a parametrisation provided by A.~Vogt.}

The discontinuities in the basis vectors $\ket{e^{\pm}_i}$ are
calculated from
\beq{eq:epmjump}
  e^{\pm}_i(x,\msh,\enef+1)  =  e^{\pm}_i(x,\msh,\enef) +
  \De e^{\pm}_i(x,\msh,\enef) + \la_i(\enef) \De
  h^{\pm}(x,\msh,\enef),
\eeq
where the light component $\De e^{\pm}_i$ is given by
\eq{eq:jumpconv}, with $q^{\pm}_i$ replaced by $e^{\pm}_i$. With the
definition \eq{eq:evecdef} of the basis functions, the values of the
coefficients $\la_i(\enef)$ are
\beq{eq:lambdaenef}
  \begin{array}{c|rrrrrr}
  \enef & \la_1 & \la_2 & \la_3 & \la_4 & \la_5 & \la_ 6 \\
  \hline 
    3   & \ \;1   & \ \;0   & \ \;0   &    -3   &         &     \\
    4   & \ \;1   & \ \;0   & \ \;0   & \ \;0   &    -4   &     \\
    5   & \ \;1   & \ \;0   & \ \;0   & \ \;0   & \ \;0   &    -5
  \end{array}
\eeq
When the densities are evolved upward in \ms, it is straight forward
to calculate with~\eq{eq:pdfjumps} and~\eq{eq:jumpconv} the parton
densities at $\enef+1$ from those at $\enef$.  However, \qcdnum\ is
capable to invert the relation between $\enef$ and $\enef+1$ so that
it can also calculate the discontinuities in case of downward
evolution. For this it is convenient to write the calculation of the
singlet and gluon discontinuities in matrix form, similar to
\eq{eq:dglapsg}
\beq{eq:jumpsg}
  \begin{pmatrix} 
  q_{\rm s} \\ g
  \end{pmatrix}^{(\enef+1)}
  =
  \begin{pmatrix} 
  q_{\rm s} \\ g
  \end{pmatrix}^{(\enef)}
  +\;
  \asubs^2
  \begin{pmatrix}
  A_{\rm qq} + A_{\rm hq} & A_{\rm hg} \\
  A_{\rm gq} & A_{\rm gg}
  \end{pmatrix} 
  \otimes
  \begin{pmatrix}
  q_{\rm s} \\ g
  \end{pmatrix}^{(\enef)}.
\eeq
In \Se{se:dglaplin} we will show how \eq{eq:jumpsg} is turned into an
invertible matrix equation.
\siindex{discontinuities in pdf evolution|)}%

\siindex{evolution of heavy quark pdfs|(}%
Note that the heavy quark non-singlets do not obey the DGLAP evolution
equations over the full range in \ms, because the heavy flavours are
simply set to zero below their thresholds, instead of being evolved.
The evolution of the set $\ket{e^{\pm}_i}$ thus proceeds in the \vfns\
as follows: The singlet/valence densities $\ket{e^{\pm}_{1}}$ and the
light non-singlets $\ket{e^{\pm}_{2,3}}$ are evolved both upward and
downward starting from some scale $\msz < \msc$.  The heavy non-singlets
$\ket{e^{\pm}_{4,5,6}}$ are dynamically generated from the DGLAP
equations by upward evolution from the thresholds $\mu^2_{\cbt}$.  At
and below the thresholds, $\ket{e^+_{4,5,6}}$ is set equal to the
singlet and $\ket{e^-_{4,5,6}}$ to the valence. This is equivalent to
setting the heavy quark and anti-quark distributions to zero, except
that at NNLO the heavy flavours do not evolve from zero but from the
non-zero discontinuity given in \eq{eq:pdfjumps}. This is illustrated
in \Fi{fig:discvsx}
%
\begin{figure}[tbh]
\begin{center}
\includegraphics[width=0.65\linewidth]{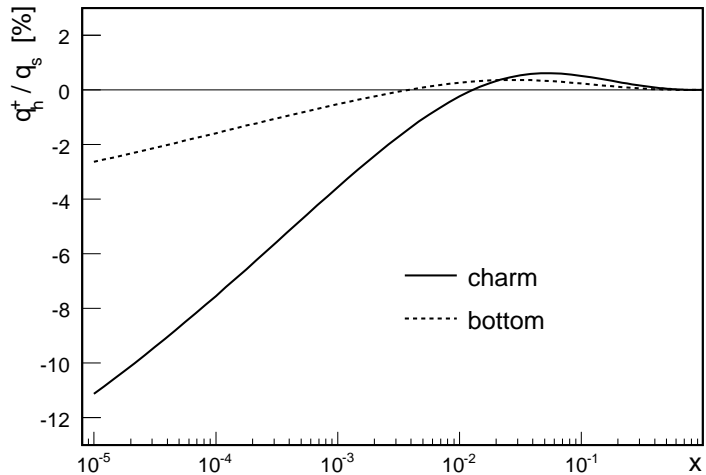}
\end{center}
\caption{\footnotesize
  The NNLO starting densities $q^+_h(x,\ms_h)$, normalised to the
  singlet density $q_{\rm s}(x,\ms_h)$, for charm (full curve) and
  bottom (dotted curve).}
\label{fig:discvsx} 
\end{figure}
%
where we plot the charm and bottom starting distributions, normalised
to the singlet distribution. It is seen that the bottom discontinuity
is less than~3\% of the singlet over the whole range in $x$, while for
charm it is much larger, exceeding 10\% at low $x$. Note that the
starting distributions are negative below $x \approx 10^{-2}$.
\siindex{evolution of heavy quark pdfs|)}%


%% file: sections/numanintro.tex
\section{Numerical Method}\label{se:numan}

The DGLAP evolution equations are in \qcdnum\ numerically solved on a
discrete $n \times m$ grid in $x$ and \ms. In such an approach the
convolution integrals can be evaluated as weighted sums with weights
calculated once and for all at program initialisation. Because of the
convolutions, the total number of operations to solve a DGLAP equation
is quadratic in $n$ and linear in $m$. The accuracy of the solution
depends, for a given grid, on the interpolation scheme chosen (linear
or quadratic).

The advantage of this `$x$-space' approach, compared to
\siindex{\ipegasus\ program}%
`$N$-space'~\cite{ref:pegasus}, is its conceptual simplicity and the
fact that one is completely free to chose the functional form of the
input distribution since it is fed into the evolution as a discrete
vector of input values. A disadvantage is that accuracy and speed
depend on the choice of grid and that each evolution will yield no
less than $n \times m$ parton density values (typically $10^4$)
whether you want them or not.

The numerical method used in \qcdnum\ is based on polynomial spline
interpolation of the parton densities on an equidistant logarithmic
grid in $x$ and a (not necessarily equidistant) logarithmic grid in
$\mu^2$. The order of the $x$-interpolation can in be set to $k = 2$
(linear) or 3 (quadratic).  The interpolation in $\mu^2$ is always
quadratic. With such an interpolation scheme, the DGLAP evolution
equations transform into a triangular set of linear equations in the
interpolation coefficients. This leads to a very fast evolution of
these coefficients from some input scale $\mu^2_0$ to any other scale
$\mu^2_i$ on the grid.   
In the following sections we will describe the
spline interpolation, the calculation of convolution integrals and the
QCD evolution algorithm.
Note that several features of the  \qcdnumnew\
numerical method have been previously proposed in,
for example,~\cite{ref:hoppet,ref:moreprograms}.

%% file: sections/numan.tex

\subsection{Polynomial Spline Interpolation}\label{se:numsplines}

\siindex{spline interpolation|(}%
To interpolate a function $h(y)$,\footnote{In \qcdnum, $h(y)$
  represents a parton \emph{momentum} density in the scaling variable
  $y = - \ln x$. However, for this section the identification of $h$
  with a parton density is not so relevant.} we sample this function on
an ($n+1$)-point grid
\[
  y_0 < y_1 < \ldots < y_{n-1} < y_{n}
\]
and parametrise it in each interval by a piecewise polynomial of
order $k$. Such a piecewise polynomial is turned into a spline by
imposing one or more continuity relations at each of the grid points.
Usually---but not always---continuity is imposed at the internal grid
points on the function itself and on all but the highest derivative,
which is allowed to be discontinuous.  Without further constraints at
the end points, the spline has $k+n-1$ free parameters. Increasing the
order $k$ of the interpolation thus costs only \emph{one} and not~$n$
extra parameters as is the case for unconstrained piecewise
polynomials.

\siindex{B-splines|(}%
It is convenient to write a spline function as a linear combination of
so-called B-splines
\beq{eq:bkasum}
   h(y) = \sum_i A_i Y_i(y).
\eeq
The basis $Y_i$ of B-splines depends on the order $k$, on the
distribution of the grid points along the $y$ axis (equidistant in
\qcdnum) and on the number of continuity relations we wish to impose
at the internal grid points and at the two end points. For how to
construct a B-spline basis and for more details on splines in general
we refer to~\cite{ref:splines}.

In \Fi{fig:bsplines}
%
\begin{figure}[tbh]
\begin{center}
\includegraphics[width=0.70\linewidth]{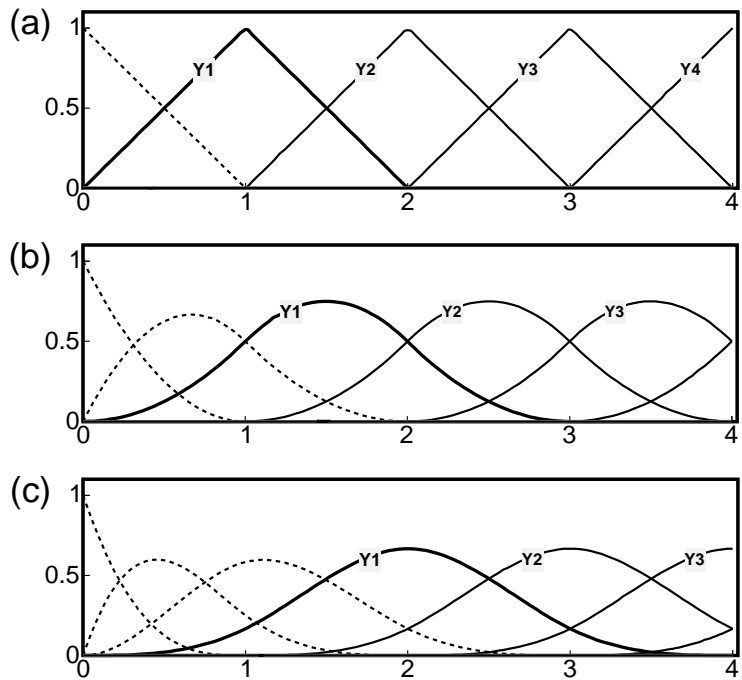}
\end{center}
\caption{\footnotesize
  B-spline bases generated on an equidistant grid. (a) Linear
  B-splines ($k = 2$).  Removing the dashed spline enforces the
  boundary condition $h(y_0) = 0$; (b) Quadratic B-splines ($k =
  3$).  Removing the first two dashed splines enforces the boundary
  condition $h(y_0) = h'(y_0) = 0$; (c) Cubic B-splines ($k =
  4$).  Removing the first three dashed splines enforces the boundary
  condition $h(y_0) = h'(y_0) = h''(0) = 0$. Spline interpolation on
  such a basis is numerically unstable.}
\label{fig:bsplines} 
\end{figure}
%
are shown the B-splines for linear ($k = 2$) quadratic ($k = 3$) and
cubic ($k = 4$) interpolation on an equidistant grid. In case $h(y_0)
= h(0) = 0$---which is always true for parton densities---we may
remove the first B-spline in the plots of \Fi{fig:bsplines}. Removing
the second B-spline in \Fi{fig:bsplines}b gives quadratic
interpolation with an additional boundary condition $h'(y_0) =
0$.\footnote{\siindex{spline oscillation}%
  A parton density parametrisation should thus behave like
  $h(y \rar 0) \propto y^{\la}$ with $\la > 1$ because otherwise the
  condition $h'(0) = 0$ is violated and the spline might oscillate.
  All known pdf parametrisations fulfil this requirement but when
  the parameters are under control of a fitting program
  one should take precautions that $\la$ will always stay above
  unity.} With these boundary
conditions---and because the grid is equidistant---the remaining
B-splines possess translation invariance, that is, the basis can be
generated by successively shifting the first spline one interval to
the right (full curves in \Fi{fig:bsplines}a,b).  Translation
invariance greatly simplifies the evolution algorithm, as we will see
later.  
\siindex{B-splines|)}%

It is therefore tempting to extend the scheme to cubic interpolation
by removing the first three B-splines in \Fi{fig:bsplines}c. This
would yield a translation invariant basis with the boundary conditions
$h(y_0) = h'(y_0) = h''(y_0) = 0$. However, it turns out that such a
cubic spline interpolation tends to be numerically unstable. The cure
is to drop the constraint $h''(y_0) = 0$ and impose a constraint on
$h'(y_n)$ at the other end of the grid. But this does not fit in the
evolution algorithm as it now stands so that we have abandoned cubic
and higher order splines in \qcdnum.

If we number the B-splines $1,2,\ldots,n$ from left to right as
indicated in \Fi{fig:bsplines} it is seen that for both $k = 2$ and 3
the following relation holds (translation invariance):
\beq{eq:btranslate}
  Y_i(y) = Y_1(y-y_{i-1}).
\eeq
Furthermore, for linear interpolation ($k = 2$) we have $Y_i(y_i) = 1$
so that
\bea{eq:bsum2}
  h(y_0) & = & 0 \nonumber \\
  h(y_i) & = & A_i\sspace Y_i(y_i) = A_i \qquad 1 \leq i \leq n.
\eea
Likewise, for quadratic interpolation ($k = 3$) we have $Y_{i-1}(y_i)
= Y_i(y_i) = 1/2$ so that
\bea{eq:bsum3}
  h(y_0) & = & 0 \nonumber \\
  h(y_1) & = & A_1\sspace Y_1(y_1) =  A_1/2 \nonumber \\
  h(y_i) & = & A_{i-1}\sspace Y_{i-1}(y_i) + A_i\sspace Y_i(y_i) =
  ( A_{i-1} + A_i )/2 \qquad 2 \leq i \leq n.
\eea
We denote $h(y_i)$ by $h_i$, the column vector of function values
by $\ve{h} = \transp{(h_1,\ldots,h_n)}$, the corresponding vector
of spline coefficients by \ve{a} and write \eq{eq:bsum2} and
\eq{eq:bsum3} as
\beq{eq:htoa}
  \ve{h} = \ve{S}\;\ve{a}
\eeq
where $\ve{S}$ is the identity matrix in case of linear interpolation
and a lower diagonal band matrix for the quadratic spline.  On a
5-point equidistant grid $y_0,\ldots,y_4$, for instance, we have in case
of quadratic interpolation the vector $\ve{h} = \transp{(h_1,\ldots,h_4)}$
and the matrix
\beq{eq:sband}
  \ve{S} = \frac{1}{2} \left(
   \begin{array}{cccc}
     1 &   &   &   \\
     1 & 1 &   &   \\
       & 1 & 1 &   \\
       &   & 1 & 1
  \end{array}
  \right) \qquad  \mbox{with inverse} \qquad  \ve{S}^{-1} = 2 \left(
  \begin{array}{rrrr}
      1  &      &      &       \\
     -1  &   1  &      &       \\
      1  &  -1  &   1  &       \\
     -1  &   1  &  -1  & \ \; 1
  \end{array}
  \right).
\eeq
Note that $\ve{S}$ is sparse but $\ve{S}^{-1}$ is not.  Thus, when a
parton distribution $\ve{h_0}$ is given at some input scale $\mu^2_0$,
the corresponding vector $\ve{a_0}$ of spline coefficients is found by
solving~\eq{eq:htoa}.\footnote{Obtaining $\ve{a}$ from
  solving \eq{eq:htoa} by forward substitution (\Ap{app:solveit})
  costs $\order(2n)$ operations. This is cheaper than the alternative
  of calculating $\ve{a} = \ve{S}^{-1}\ve{h}$ which costs
  $\order(n^2/2)$ operations.} This
vector is then evolved to other values of $\mu^2$ using the DGLAP
evolution equations as is described in the next two sections.
\siindex{spline interpolation|)}%


\subsection{Convolution Integrals}\label{se:numconvol}

\siindex{convolution integrals in \textsc{qcdnum}|(}%
The Mellin convolution \eq{eq:mellinconv} calculated in \qcdnum\
is not that of a number density~$f$ and some kernel~$g$
but, instead, that of a momentum density $p = xf$ and a kernel
$q = xg$. These convolutions differ by a factor $x$:
\beq{eq:xmellin}
  [p \otimes q](x) = x [f \otimes g](x).
\eeq      
This also true for multiple convolution: for $p = xf$, 
$q = xg$ and $r = xh$ we have
\siindex{multiple convolution}
\beq{eq:xmultmellin}
  [p \otimes q \otimes r](x) = x [f \otimes g \otimes h](x).
\eeq
A change of variable $y = -\ln x$ turns a Mellin convolution into
a Fourier convolution:
\siindex{Fourier convolution}
\beq{eq:fourier}
  [f \otimes g](x) =
  [u \otimes v](y) = \int_0^y \der z\; u(z)\;v(y-z)
                   = \int_0^y \der z\; u(y-z)\; v(z),
\eeq                      
where the functions $u$ and $v$ are defined by
$u(y) = f(e^{-y})$ and $v(y) = g(e^{-y})$.
\siindex{convolution integrals in \textsc{qcdnum}|)}%

In the following we will
denote by $h(y,t)$ a parton \emph{momentum} density in the
logarithmic scaling variables $y = -\ln x$ and $t = \ln \mu^2$.  In
terms of $h$, the DGLAP non-singlet evolution equation \eq{eq:dglapns}
is written as
\siindex{\iyvariable}%
\siindex{\itvariable}%
\beq{eq:dglap0}
  \frac{\pa h(y,t)}{\pa t} = \int_0^y \der z\;
  Q(z,t)\; h(y-z,t) = \int_0^y \der z\; Q(y-z,t)\;
  h(z,t)
\eeq
with a kernel $Q(y,t) = e^{-y}P(e^{-y},t)$. Here $P(x,t)$ is a
non-singlet splitting function, as given in \Se{se:dglap}.
To solve \eq{eq:dglap0} we first have to evaluate the Fourier
convolution
\beq{eq:defI}
  I(y,t) \equiv \int_0^y \der z\; Q(y-z,t)\; h(z,t).
\eeq
Inserting \eq{eq:bkasum} in \eq{eq:defI} we find for the integrals at
the grid points $y_i$ (for clarity, we drop the argument $t$ in the
following)
\beq{eq:defIn}
  I(y_i) = \sum_{j=1}^i A_{j}\int_0^{y_i} \der z\; Q(y_i-z)\;
  Y_{j}(z) \equiv \sum_{j=1}^i W_{ij}\small A_{j}
  \qquad (1 \leq i \leq n).
\eeq

The summation is over the first $i$ terms only,
because B-splines with an index $j > i$ are
zero in the integration domain $z \leq y_i$, see \Fi{fig:bsplines}.

\siindex{convolution weights}%
\Eq{eq:defIn} defines the weights $W_{ij}$ which are calculated as
follows.  Because $Y_{j}(y) = 0$ for $y < y_{j-1}$ the weights can be
written as
\beq{eq:wialf}
   W_{ij} = \int_{y_{j-1}}^{y_i} \der z\; Q(y_i-z) Y_{j}(z)
   = \int_0^{y_i-y_{j-1}} \der z\; Q(y_i-y_{j-1}-z) Y_1(z)
\eeq
where we have used \eq{eq:btranslate} in the second identity.  From
the property of equidistant grids
\[ y_i + y_j = y_{i+j} \]
it follows that $W_{ij}$ depends only on the difference $i-j$
(Toeplitz matrix):
\siindex{Toeplitz matrix}%
\beq{eq:wjdef}
  W_{ij} = w_{i-j+1} \qquad \mbox{with} \qquad
  w_{\ell} \equiv \int_0^{y_{\ell}} \der z\; Q(y_{\ell}-z) Y_1(z)
  \qquad (1 \leq \ell \leq n).
\eeq
The integrand only contributes in the region $k\De$ where $Y_1$ is
non-zero so that in practical calculations the upper integration limit
$y_{\ell}$ is replaced by $\min(y_{\ell},k\De)$, with $\De$ the grid
spacing.  We remark that the calculation of the weights $w_{\ell}$ is
a bit more complicated than suggested by \eq{eq:wjdef} because
singularities in the splitting functions have to be taken into
account; for the relevant formula's we refer to \Ap{app:singular}.

The weights can thus be arranged in a lower-triangular Toeplitz
matrix, as is illustrated by the $4 \times 4$ example below:
\beq{eq:wmat}
  W_{ij} = 
  \left( \begin{array}{cccc}
      w_1  &     &     &     \\
      w_2  & w_1 &     &     \\
      w_3  & w_2 & w_1 &     \\
      w_4  & w_3 & w_2 & w_1 
  \end{array} \right).
\eeq
This matrix is fully specified by the first column, taking $n$ instead
of $n(n+1)/2$ words of storage. This is not only advantageous in terms
of memory usage but also in terms of computing speed since frequent
calculations like summing the perturbative expansion
\beq{eq:dglapev3}
   \ve{W}(t) = \asubs (t)\{\; \ve{W}^{(0)} +
               \asubs (t)\ve{W}^{(1)} + \cdots\; \}
\eeq
takes only $\order(n)$ operations instead of $\order(n^2/2)$. We
write the vector of convolution integrals as $\ve{I}$ and express
\eq{eq:defIn} in vector notation as
\beq{eq:Ivec}
  \ve{I} = \ve{W} \ve{a}.
\eeq

Also multiple convolutions can be calculated as weighted sums.
Let $f(x)$ be a number density and $K_{a,b}(x)$ be two convolution
kernels. The vector of Mellin \mbox{convolutions}
\siindex{multiple convolution}%
\[
  I_i = x_i [f \otimes K_a \otimes K_b](x_i)
\]
can be calculated from \eq{eq:Ivec}, using the weight table
\beq{eq:wawb}
  \ve{W} = \ve{W}_{\! a} \ve{S}^{-1} \ve{W}_{\! b}.
\eeq
Here $\ve{W}_{\! a}$ and $\ve{W}_{\! b}$ are the weight tables of
$K_a$ and $K_b$, respectively, and $\ve{S}$ is the transformation
matrix defined by \eq{eq:htoa}.

Another interesting convolution is that of two number densities $f_a$
and $f_b$
\[
  I_i = x_i [f_a \otimes f_b](x_i).
\]
\siindex{parton luminosity}%
This `parton luminosity'~\cite{ref:eichten} (times $x$) is
calculated from the Fourier convolution
\beq{eq:hacrosshb}
  I(y_i) = \int_0^{y_i} \der z\; h_a(z)\, h_b(y_i-z).
\eeq 
Inserting the spline representation \eq{eq:bkasum} gives an expression
for the convolution integral as a weighted sum over the set of spline
coefficients $\ve{a}$ of $h_a$ and $\ve{b}$ of $h_b$, 
\[
  I(y_i) = \sum_{j=1}^i \sum_{k=1}^i A_j B_k\;W_{ijk}
  \qquad \mbox{with} \qquad
  W_{ijk} \equiv \int_0^{y_i} \der z\; Y_j(z) Y_k (y_i - z).
\]
To reduce the dimension of $W_{ijk}$, we use the translation invariance
\eq{eq:btranslate} and write
\[
  W_{ijk} 
  = \int_0^{y_{i-j+1}} \!\!\! \der z\; Y_1(z)\; Y_k (y_{i-j+1} - z).
\]
Because B-splines with index $k > i-j+1$ do not have their support
inside the integration domain, we obtain an upper limit
$k \leq i-j+1$. Again using translation invariance
yields
\[
  W_{ijk} = 
  \int_0^{y_{i-j-k+2}} \!\!\!\! \der z\; Y_1(z)\; Y_1 (y_{i-j-k+2} - z).
\]
We now have a compact expression for the
convolution integral~\eq{eq:hacrosshb}:
\beq{eq:sumhacrosshb}
  I(y_i) = \sum_{j=1}^i \sum_{k=1}^{i-j+1} A_j B_k\; w_{i-j-k+2}
  \qquad \mbox{with} \qquad
  w_{\ell} = \int_0^{y_{\ell}} \der z\; Y_1(z)\; Y_1(y_{\ell}-z).
\eeq
Because $Y_1$ has a limited support, it turns out that only the
first~3~(5) terms of $w_{\ell}$ are non-zero in case of linear
(quadratic) interpolation. The operation
count to calculate a convolution of parton densities is thus 
not more than $\order(5n)$, for quadratic splines.

%
%

\subsection{DGLAP Evolution}\label{se:dglaplin}

We denote by the vector $\ve{h_0}$ a \emph{non-singlet} quark density
at the input scale $t_0 = \ln{\mu^2_0}$.  The derivative of $\ve{h_0}$
with respect to the scaling variable $t$ is given by the DGLAP
evolution equation \eq{eq:dglap0} which can be written in vector
notation as, from \eq{eq:htoa} and~\eq{eq:Ivec}
\beq{eq:dglapev1} 
   \frac{\der \ve{h_0}}{\der t} = \frac{\der \ve{S a_0}}{\der t} = 
   \ve{W_0}\; \ve{a_0} \qquad \mbox{or} \qquad
   \frac{\der \ve{a_0}}{\der t} \equiv \ve{a'_0} =
   \ve{S}^{-1}\ve{W_0}\;\ve{a_0}.   
\eeq
Likewise we have at $t_1$
\beq{eq:dglapev2}
   \ve{a'_1} = \ve{S}^{-1}\ve{W_1}\;\ve{a_1}. 
\eeq
We have indexed the weight matrices above by a subscript because
they depend on $t$ through multiplication by powers of \asubs, see
\eq{eq:dglapev3}.

Assuming that $\ve{a}(t)$ is quadratic in $t$, we can
relate \ve{a_0}, \ve{a_1}, $\ve{a'_0}$ and $\ve{a'_1}$ by
\beq{eq:dglapev4}
  \ve{a_1} = \ve{a_0} + ( \ve{a'_0} + \ve{a'_1} ) \De_1
\eeq
with $\De_1 = (t_1-t_0)/2$.  If $t_1 > t_0$, $\De_1$ is positive and
we perform forward evolution.  If $t_1 < t_0$, $\De_1$ is negative and
we perform backward evolution.

Inserting \eq{eq:dglapev1} and \eq{eq:dglapev2} in \eq{eq:dglapev4} we
obtain a relation between the known spline coefficients $\ve{a_0}$ and
the unknown coefficients $\ve{a_1}$
\beq{eq:dglapev5}
  (\ve{1}-\ve{S}^{-1}\ve{W_1}\De_1)\; \ve{a_1} = 
  (\ve{1}+\ve{S}^{-1}\ve{W_0}\De_1)\; \ve{a_0}.
\eeq 
Multiplying both sides from the left by 
$\ve{U_1} \equiv \ve{S}/\De_1$ gives
\beq{eq:dglapev6}
  (\ve{U_1} - \ve{W_1})\; \ve{a_1} = 
  (\ve{U_1} + \ve{W_0})\; \ve{a_0}.
\eeq
\Eq{eq:dglapev6} is more convenient than \eq{eq:dglapev5} because
matrix multiplication $\ve{S}^{-1}\ve{W}$ is replaced by matrix
addition.\footnote{In fact, adding a matrix with band structure
  \eq{eq:sband} to a lower triangular matrix with structure
  \eq{eq:wmat} takes only {\em two} additions irrespective of the
  dimension of the matrices.}  Note that $\ve{U}$ is a lower diagonal
band matrix so that $\ve{U} \pm \ve{W}$ is still lower triangular
with, in fact, the Toeplitz structure \eq{eq:wmat} preserved.  All
this leads to a very simple and fast evolution algorithm, starting
from $\ve{a_0}$:
\begin{enumerate}
\item At $t_0$, calculate $\ve{a_0}$ from \eq{eq:htoa}, $\ve{W_0}$
  from \eq{eq:dglapev3} and $\ve{U_1}$ as defined above. Then
  construct the  vector $\ve{b_1} \equiv (\ve{U_1} + \ve{W_0})\;
  \ve{a_0}$. 
\item Subsequently, at $t_1$,
      \begin{enumerate}      
      \item Calculate $\ve{W_1}$ and the lower triangular matrix
            $\ve{V_1} = \ve{U_1} - \ve{W_1}$;
      \item Solve the equation $\ve{V_1}\ve{a_1} = \ve{b_1}$ by
            forward substitution, see
            \Ap{app:solveit};
      \item Calculate $\ve{U_2}$ and $\ve{b_2} = (\ve{U_1} +
            \ve{U_2})\ve{a_1} - \ve{b_1}$ for the next
            evolution to $t_2$.\footnote{Using \eq{eq:dglapev6} it is
              a simple exercise to establish 
              this relation between $\ve{b}$, $\ve{U}$ and
              $\ve{a}$. Note that $\ve{b}$ in step
              (2c) is calculated much faster
              than $\ve{b}$ in step (1).} 
      \end{enumerate}
\item Repeat step 2 at $t_2$ and so on.
\end{enumerate}
With this algorithm each evolution step consists of a few vector
manipulations which have an operation count $\order(n)$ and solving
one triangular matrix equation which has an operation count
$\order(n^2/2)$.  The total operation count only very weakly depends
on the order $k$ of the interpolation chosen: quadratic interpolation
is almost for free.

The algorithm can also be used for the coupled evolution of the singlet
quark ($\ve{a}_{\rm s}$) and gluon ($\ve{a}_{\rm g}$) spline
coefficients, provided we make the following replacements in the
formalism:
\[
  \ve{a} \rightarrow        \begin{pmatrix}
                            \ve{a}_{\rm s} \\
                            \ve{a}_{\rm g}
                            \end{pmatrix} \qquad
  \ve{S} \rightarrow        \begin{pmatrix}
                            \ve{S} &        \\
                                   & \ve{S}
                            \end{pmatrix} \qquad
  \ve{W} \rightarrow        \begin{pmatrix}
                            \ve{W}_{\rm qq} & \ve{W}_{\rm qg} \\
                            \ve{W}_{\rm gq} & \ve{W}_{\rm gg}
                            \end{pmatrix}.
\]
In \Ap{app:solveit} is shown how the coupled
triangular equations are solved by extending the forward substitution
algorithm.
The operation count is $4 \times \order(n^2/2)$ so that for $m$ grid
points in $t$ we have in total $\order(2n^2 m)$ operations for the
singlet-gluon evolution and $\order(n^2 m/2)$ operations for each
non-singlet evolution.

Finally, let us express in vector notation the NNLO parton density
discontinuities at the flavour thresholds. The relation between the
singlet and gluon distributions at $\enef$ and $\enef+1$ as given by
\eq{eq:jumpsg} can be written as
\beq{eq:asgjump}
   \begin{pmatrix}
     \ve{S} &       \\
            & \ve{S}
   \end{pmatrix} 
   \begin{pmatrix}
     \ve{a}_{\rm s}\\
     \ve{a}_{\rm g}
   \end{pmatrix}^{(\enef+1)} 
   =
   \begin{pmatrix}
     \ve{S}+\ve{A}_{\rm qq}+\ve{A}_{\rm hq} & \ve{A}_{\rm hg} \\
     \ve{A}_{\rm gq} & \ve{S}+\ve{A}_{\rm gg}
   \end{pmatrix}
   \begin{pmatrix}
     \ve{a}_{\rm s} \\
     \ve{a}_{\rm g}
   \end{pmatrix}^{(\enef)}. 
\eeq
It is easy to solve this linear equation for $\ve{a}^{(\enef+1)}$ when
$\ve{a}^{(\enef)}$ is known (forward evolution) or for
$\ve{a}^{(\enef)}$ when $\ve{a}^{(\enef+1)}$ is known (backward
evolution).
Likewise, we may write for the non-singlet discontinuities
\beq{eq:ansjump}
  \ve{S}\; \ve{a}_{\rm ns}^{(\enef+1)} = 
  \left( \ve{S}+\ve{A}_{\rm qq}\right)\;  
  \ve{a}_{\rm ns}^{(\enef)} + 
  \la \left( \ve{A}_{\rm hq}\; 
  \ve{a}_{\rm s}^{(\enef)} + 
  \ve{A}_{\rm hg}\; \ve{a}_{\rm g}^{(\enef)} \right), 
\eeq
where $\la$ is defined by \eq{eq:epmjump}. 
Also this equation can easily be inverted.

It can be seen from \eq{eq:htoa} and \eq{eq:dglapev4} that $h(y,t)$ is,
by construction, a spline in both the variables $y$ and $t$. However,
it turns out that it is technically more convenient to represent the
pdfs by their
\emph{values} on the grid, instead of by their spline coefficients.
Polynomial interpolation of order $k$ in $y$ and quadratic in $t$
is then done locally on a $k \times 3$ mesh around the interpolation
\siindex{interpolation mesh}%
point. 
The NNLO discontinuities are preserved by storing, at the
flavour thresholds, the pdf values for both $\enef-1$ and~\enef, and
by prohibiting the interpolation mesh to cross a flavour threshold.
Note, however, that the interpolation routine yields a single-valued
function of $t$, so that one has to calculate
$h(y,t_{\cbt}-\eps)$ to view the discontinuity.\footnote{Do not
take $\eps$ too small because \qcdnum\ may snap to the threshold,
see \Se{se:subgrid}.}  

In \qcdnum\ it is possible to evolve on multiple equidistant $y$-grids
which allow for a finer binning at low $y$ (large $x$) where the parton
densities are rapidly varying.
\siindex{multiple evolution grid|)}%
This is illustrated below by a grid $G_0$ which is built-up from three
equidistant sub-grids $G_1$, $G_2$ and~$G_3$ with spacing~$\De/4$,
$\De/2$ and~$\De$, respectively.
\begin{center}
\begin{picture}(360,80)(0,0)

\thicklines

\put( 20,65){\line(1,0){320}}
\put( 20,65){\circle*{2.5}} \put( 30,65){\circle*{2.5}} 
\put( 40,65){\circle*{2.5}} \put( 50,65){\circle*{2.5}} 
\put( 60,65){\circle*{2.5}} \put( 70,65){\circle*{2.5}}
\put( 80,65){\circle*{2.5}} \put( 90,65){\circle*{2.5}} 
\put(100,65){\circle*{2.5}} \put(120,65){\circle*{2.5}} 
\put(140,65){\circle*{2.5}} \put(160,65){\circle*{2.5}}
\put(180,65){\circle*{2.5}} \put(220,65){\circle*{2.5}} 
\put(260,65){\circle*{2.5}} \put(300,65){\circle*{2.5}}
\put(340,65){\circle*{2.5}}

\put( 20,40){\line(1,0){80}}
\put( 20,40){\circle*{2.5}} \put( 30,40){\circle*{2.5}} 
\put( 40,40){\circle*{2.5}} \put( 50,40){\circle*{2.5}}
\put( 60,40){\circle*{2.5}} \put( 70,40){\circle*{2.5}}
\put( 80,40){\circle*{2.5}} \put( 90,40){\circle*{2.5}} 
\put(100,40){\circle*{2.5}}

\put(20, 25){\line(1,0){160}}
\put( 20,25){\circle*{2.5}} \put( 40,25){\circle*{2.5}}
\put( 60,25){\circle*{2.5}} \put( 80,25){\circle*{2.5}}
\put(100,25){\circle*{2.5}} \put(120,25){\circle*{2.5}}
\put(140,25){\circle*{2.5}} \put(160,25){\circle*{2.5}}
\put(180,25){\circle*{2.5}}

\put(20, 10){\line(1,0){320}} 
\put( 20,10){\circle*{2.5}} \put( 60,10){\circle*{2.5}}
\put(100,10){\circle*{2.5}} \put(140,10){\circle*{2.5}}
\put(180,10){\circle*{2.5}} \put(220,10){\circle*{2.5}}
\put(260,10){\circle*{2.5}} \put(300,10){\circle*{2.5}}
\put(340,10){\circle*{2.5}}

\thinlines

\put( 20, 5){\line(0,1){65}}
\put(100,35){\line(0,1){35}}
\put(180,20){\line(0,1){50}}
\put(340, 5){\line(0,1){65}}

\put(260, 5){\line(0,1){15}}
\put(300, 5){\line(0,1){15}}

\put(0,65){\makebox(0,0)[c]{$G_0$}} 
\put(0,40){\makebox(0,0)[c]{$G_1$}} 
\put(0,25){\makebox(0,0)[c]{$G_2$}}
\put(0,10){\makebox(0,0)[c]{$G_3$}}

\put( 20,77){\makebox(0,0)[c]{$y_0$}} 
\put(100,77){\makebox(0,0)[c]{$y_1$}} 
\put(180,77){\makebox(0,0)[c]{$y_2$}}
\put(340,77){\makebox(0,0)[c]{$y_3$}}
\put(280,17){\makebox(0,0)[c]{$\De$}}

\put( 60,53){\makebox(0,0)[c]{(I)}}
\put(140,53){\makebox(0,0)[c]{(II)}}
\put(260,53){\makebox(0,0)[c]{(III)}}

\end{picture}
\end{center}
On such a multiple grid, the parton densities are first evolved on the
grid $G_1$ and the results are copied to the region (I) of~$G_0$. The
evolution is then repeated on the grids~$G_2$ and~$G_3$ followed by a
copy to the regions~(II) and~(III) of~$G_0$, respectively.  We refer
to \Se{se:program4} for spectacular gains in
accuracy that can be achieved by employing these multiple grids.

\siindex{backward evolution|(}%
As remarked above, the evolution algorithm can---at
least in principle---handle both forward and backward evolution in
\ms\ simply by changing the sign of $\De$ in \eq{eq:dglapev4}. This
works very well for linear spline interpolation but it turns out that
backward evolution of quadratic splines can sometimes lead to severe
oscillations. This is illustrated in \Fi{fig:unstable}
%
\begin{figure}[tbh]
\begin{center}
\includegraphics[width=0.99\linewidth]{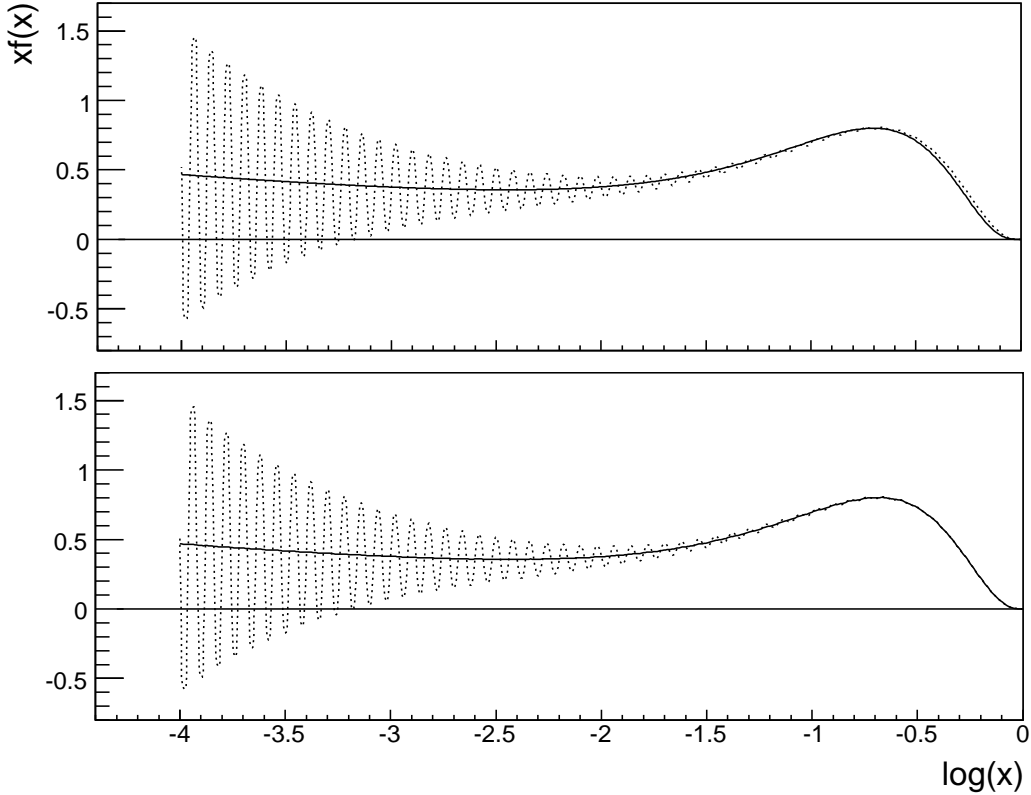}
\end{center}
\caption{\footnotesize A non-singlet parton density $xf(x)$ versus
  $\log(x)$ evolved downward from $\msz = 5$ to $\ms = 2$~\gevs\ in
  the quadratic interpolation scheme showing large oscillations
  (dotted curve).  The full curve in the top plot shows the result of
  downward evolution in the linear interpolation scheme. The full
  curve in the bottom plot shows an improved result obtained by
  iteration, as described in the text.}
\label{fig:unstable} 
\end{figure}
%
where is shown a non-singlet quark density evolved downward from
$\msz = 5$ to~$\ms = 2$~\gevs\ in the quadratic interpolation scheme
(dotted curve). In \qcdnum\ this numerical instability is handled as
follows: (i) evolve downward from \msz\ to~\ms\ in the linear
interpolation
\siindex{spline oscillation}%
scheme (which is stable); (ii) then take \ms\ as the starting scale
and evolve \emph{upward} to \msz\ in the quadratic interpolation
scheme (also stable); (iii) calculate the difference $\De f$ between
the newly evolved pdf and the original one at \msz; (iv) subtract $\De
f$ from the starting value at \msz\ used in (i) and repeat the
procedure.

The full curve in the top plot of \Fi{fig:unstable} shows the result
of downward evolution in the linear interpolation scheme, that is,
without iterations. Oscillations are absent but the evolution is not
very accurate as is evident from the difference between the dotted and
full curve at large $x$. One iteration already much improves the
precision as can be seen from the good match at large $x$ between the
two curves in the bottom plot.  It turns out that one iteration
(\qcdnum\ default), perhaps two, are sufficient while more iterations
tend to spoil the convergence.  Clearly best is to
limit the range of downward evolution by keeping \msz\ low or, 
if possible, to set it at the lowest grid point to avoid
downward evolution altogether.

\Qcdnum\ checks for quadratic spline oscillation as follows. 
We denote 
the values of the quadratic B-spline at 
$(\mbfrac{1}{2} \De, \De, \mbfrac{3}{2} \De)$
by $(b_1,b_2,b_3) = 
(\mbfrac{1}{8},\mbfrac{1}{2},\mbfrac{3}{4})$.  
It is easy to show that quadratic
interpolation mid-between the grid points is given by
$\ve{u} = \ve{D} \ve{a}$, where~$\ve{D}$ is a lower
diagonal Toeplitz band matrix, of bandwidth 3, which is characterised
by the vector~$(b_1,b_3,b_1)$. Likewise,
the linear interpolation of the spline at the mid-points is
calculated from
$\ve{v} = \ve{E} \ve{a}$, where $\ve{E}$ is the lower diagonal
Toeplitz band matrix $(\half b_2, b_2, \half b_2)$.
The maximum deviation
$\eps = \| \ve{u}-\ve{v} \| = \| (\ve{D}-\ve{E})\ve{a} \|$ 
should be small; for pdfs sampled
on a reasonably dense grid, $\eps \approx 0.1$ or less.
For each pdf evolution, $\eps$ is computed at the input scale,
and at the lower and upper end of the \ms\ grid.
An error condition is raised when it exceeds a given limit,
indicating that the spline oscillates, or that the $x$-grid
is not dense enough. 
\siindex{backward evolution|)}%

%% file: sections/program.tex
\section{The QCDNUM Program}\label{se:program}


\subsection{Source Code}\label{se:program1}

\dbindex{\iqcdnum}{web site}%
The \qcdnum\ source code can be downloaded from the web site
\begin{verbatim}
                   http://www.nikhef.nl/user/h24/qcdnum
\end{verbatim}
Unpacking the tar file produces a directory \xtt{qdcnum-xx-yy} with
\xtt{xx-yy} the version number. Sub-directories contain the source
code, example jobs, write-up and a simple script to make a \qcdnum\ 
library, see the \xtt{README} file. The code comes with a utility
package \mbutil\ (including write-up) which is a collection of
\siindex{\imbutil\ package}%
general-purpose routines (some developed privately, some taken from
\cernlib\ and some taken from public source code repositories like
\netlib).  Because \qcdnum\ uses several of these routines, \mbutil\
must also be compiled and linked to your application program. Apart
from this, \qcdnum\ is completely stand-alone. To calculate structure
functions, the \zmstf\ and \hqstf\ add-on packages are provided, see
Sections~\ref{se:zmstf} and~\ref{se:hqstf}.

Before compiling \qcdnum\ you may want to set several parameters which
control the size of internal arrays. These parameters can be found in
the include file \xtt{qcdnum.inc}:
\dbindex{\iqcdnum}{parameters in \texttt{qcdnum.inc}}
\begin{tdeflist}[mxx0\ ]{-1mm}
\item[\xtt{mxg0}]  Maximum number of multiple $x$-grids [5].
\item[\xtt{mxx0}]  Maximum number of points in the $x$-grid [300].
\item[\xtt{mqq0}]  Maximum number of points in the \ms-grid [150].
\item[\xtt{mpt0}]  Maximum number of interpolations
  calculated in a single call [5000].
\item[\xtt{miw0}]  Maximum number of information words in a
  weight store [20].
\item[\xtt{mbf0}]  Maximum number of fast convolution 
  scratch buffers [20].    
\item[\xtt{nwf0}]  Size of the \qcdnum\ dynamic store in words [400000].
\end{tdeflist}
The first 6 parameters are simply dimensions of book-keeping arrays
which you may want to adjust to your needs. More important is the
parameter \xtt{nwf0} that defines the size of an internal
store that contains the weight tables and the tables of parton
densities. How many words are needed
depends on the size of the tables which, in turn,
depends on the size of the current $x$-\ms\ grid. It also depends on how  
many different sets of tables (un-polarised pdfs, polarised pdfs,
fragmentation functions, \mbetc) one wants to store.
In this respect, \qcdnum\ is very user-friendly by always gracefully
grinding to a halt if it runs out of memory, with a message that tells
how large \xtt{nwf0} should be.


\subsection{Application Program}\label{se:program2}

\dbindex{\iqcdnum}{example job|(}%
To illustrate the use of \qcdnum, we present in \Fi{fig:listing}
%
\begin{figure}[p]
\footnotesize
\begin{verbatim}
C      ----------------------------------------------------------------
       program example
C      ----------------------------------------------------------------
       implicit double precision (a-h,o-z)
       data ityp/1/, iord/3/, nfin/0/          !unpolarised, NNLO, VFNS
       data as0/0.364/, r20/2.D0/              !alphas
       external func                           !input parton dists
       dimension def(-6:6,12)                  !flavor decomposition
       data def  /
C--    tb  bb  cb  sb  ub  db   g   d   u   s   c   b   t
C--    -6  -5  -4  -3  -2  -1   0   1   2   3   4   5   6  
     +  0., 0., 0., 0., 0.,-1., 0., 1., 0., 0., 0., 0., 0.,   !dval
     +  0., 0., 0., 0.,-1., 0., 0., 0., 1., 0., 0., 0., 0.,   !uval
     +  0., 0., 0.,-1., 0., 0., 0., 0., 0., 1., 0., 0., 0.,   !sval
     +  0., 0., 0., 0., 0., 1., 0., 0., 0., 0., 0., 0., 0.,   !dbar
     +  0., 0., 0., 0., 1., 0., 0., 0., 0., 0., 0., 0., 0.,   !ubar
     +  0., 0., 0., 1., 0., 0., 0., 0., 0., 0., 0., 0., 0.,   !sbar
     +  78*0.    /
       data xmin/1.D-4/, nxin/100/, iosp/3/            !x grid, splord
       dimension qq(2),wt(2)                           !mu2 grid
       data qq/2.D0,1.D4/, wt/1.D0,1.D0/, nqin/60/     !mu2 grid
       data q2c/3.D0/, q2b/25.D0/, q0/2.0/             !thresh and mu20
       data x/1.D-3/, q/1.D3/, qmz2/8315.25D0/         !output scales
C      ----------------------------------------------------------------
       call qcinit(6,' ')                       !initialise
       call gxmake(xmin,1,1,nxin,nx,iosp)       !x-grid
       call gqmake(qq,wt,2,nqin,nq)             !mu2-grid
       call fillwt(ityp,id1,id2,nw)             !calculate weights
       call setord(iord)                        !LO, NLO, NNLO
       call setalf(as0,r20)                     !input alphas
       iqc  = iqfrmq(q2c)                       !mu2c
       iqb  = iqfrmq(q2b)                       !mu2b
       call setcbt(nfin,iqc,iqb,0)              !thresholds in the VFNS
       iq0  = iqfrmq(q0)                        !starting scale
       call evolfg(ityp,func,def,iq0,eps)       !evolve all pdfs
       csea = 2.D0*fvalxq(ityp,-4,x,q,0)        !charm sea at x,Q2
       asmz = asfunc(qmz2,nfout,ierr)           !alphas(mz2)
       end
C      ----------------------------------------------------------------
       double precision function func(id,x)     !momentum density xf(x) 
C      ----------------------------------------------------------------
       implicit double precision (a-h,o-z)
       if(id.eq.0) func = gluon(x)              !0 = always gluon
       if(id.eq.1) func = dvalence(x)           !1 = defined in def
       ..                                       ..
       if(id.eq.6) func = strangebar(x)         !6 = defined in def
       return
       end 
\end{verbatim}
\caption{\footnotesize Listing of a \qcdnum\ application program
  evolving a complete set of parton densities in the \vfns\ at NNLO.
  The array \xtt{def} defines the light quark valence ($xq-x\bar{q}$)
  and anti-quark ($x\bar{q}$) distributions as an input to the
  evolution. The $x$ dependence of the input densities is coded in the
  function \xtt{func}. After evolution, the pdfs are interpolated to
  some $x$ and \ms\ and \asmz\ is calculated.}
\label{fig:listing} 
\end{figure}
%
the listing of a simple application program. For a detailed
description of the subroutine calls, and for additional routines
not included in the example, we refer to \Se{se:subroutines}.

The first step in a \qcdnum\ based analysis is initialisation
(\xtt{qcinit}), setting up the $x$-\ms\ grid (\xtt{gxmake},
\xtt{gqmake}) and the calculation of the weight tables (\xtt{fillwt}).
The weights depend on the grid definition and the interpolation order
so that \xtt{fillwt} must be called after the grid has been defined.
The weight tables are calculated for LO, NLO and NNLO as
well as for all possible flavour settings in the range $3 \leq \enef
\leq 6$ so that you do not have to call \xtt{fillwt} again when you
set or re-set \qcdnum\ parameters further downstream.  Although the
weight calculation is fast (typically about 10--20~s) it may become a
nuisance in semi-interactive use of \qcdnum\ so that there is a
possibility to dump the weights to disk and read them back in the next
\qcdnum\ run.

In the example code, the weight calculation is followed by setting the
perturbative order (\xtt{setord}) and the input value of \as\ at some
renormalisation scale \Rs\ (\xtt{setalf}).  The call to \xtt{setcbt}
sets the \vfns\ mode and defines the thresholds on the factorisation
scale~\Fs. All the calls that set evolution parameters are destructive
in the sense that they invalidate the parton densities in memory, if
any. In this way all \qcdnum\ results are consistently obtained with
the same value of \as, the same perturbative order, \mbetc

The second step is to evolve the parton densities from input specified
at the scale \msz. It is important to note that \qcdnum\ evolves
parton \emph{momentum} densities $xf(x)$, although all theory in this
write-up is expressed in terms of parton \emph{number} densities
$f(x)$.  The evolution is done by calling the routine \xtt{evolfg}
which evolves $2\enef+1$ input parton densities (quarks plus gluon) in
the \ffns\ or \vfns\ scheme. The routine internally takes care of the
proper decomposition of the input quark densities into singlet and
non-singlets. In the \vfns\ the input scale \msz\ must lie
below the charm threshold \msc\ so that, as a consequence, \msc\ must
lie above the lower boundary of the \ms\ grid.

The flavour composition of each of the input quark densities is given
by a table of weights \xtt{def(-6:6,12)}.  In the example program, six
light quark input densities are defined: three valence densities
$x(q-\bar{q})$ and three anti-quark densities $x\bar{q}$. This is
sufficient input to run evolutions in the \vfns\ scheme.
One is completely free to define the flavour composition of the input
quark densities as long as they form a linearly independent set
(\qcdnum\ checks this).  Note that the flavours are ordered
according to the PDG convention ${\rm d},{\rm u},{\rm s},\ldots$
and not ${\rm u},{\rm d},{\rm s},\ldots$ as often is the case
in other programs.\siindex{PDG convention}%

The $x$ dependence of these momentum densities at \msz\ must be coded
for each identifier in an if-then-else block in the function \xtt{func}.
The sum rules
\siindex{sum rule integrals}
\bea{eq:sumrules}
\int_0^1 xg(x) \der x + \int_0^1 x\qsi(x) \der x & = & 1, \nonumber \\
\int_0^1 [d(x)-\dbar(x)] \der x & = & 1, \nonumber \\
\int_0^1 [u(x)-\ubar(x)] \der x & = & 2 
\eea 
cannot be reliably evaluated by \qcdnum\ since it has no information on
the $x$-dependence of the pdfs below the lowest grid point in $x$. 
These sum rules should therefore be built into the
parametrisation of the input densities. The evolution does, of
course, conserve the sum rules once they are imposed at \msz.  The
easiest way to evolve with a symmetric strange sea is to include
$xs_{\rm v} = x(s-\sbar)$ in the collection of input densities and set
it to zero for all $x$ at the input scale \msz.  In the \vfns\ at LO
or NLO, the generated heavy flavour densities $h = (\cbt)$ are always
symmetric ($xh-x\bar{h} = 0$) but this is not true anymore at NNLO,
which generates a small asymmetry.

After the parton densities are evolved, the results can be accessed by
\xtt{fvalxq}. This routine transforms the parton densities from the
internal singlet/non-singlet basis to the flavour basis and returns the
gluon, a quark, or an anti-quark momentum density, interpolated to $x$
and \ms. Also here the flavours ${\rm d},{\rm u},{\rm s},\ldots$ are
indexed according to the PDG convention.
The last call in the example program evolves the input value of \as\
to the scale $m_Z^2$. This evolution is completely stand-alone and
does not make use of the \ms\ grid. The function \xtt{asfunc} can
thus be called at any point after the call to \xtt{qcinit}.
We refer to \Se{se:subroutines} for more ways to access the \qcdnum\
results, and for ways to change the renormalisation scale
with respect to the factorisation scale.
\dbindex{\iqcdnum}{example job|)}%

\Qcdnum\ has an extensive checking mechanism which maintains internal
consistency and verifies that all subroutine arguments supplied by the
user are within their allowed ranges.
Error messages might pop-up unexpectedly when the renormalisation
scale is changed with respect to the factorisation scale because the
low end of the \ms\ grid may then map onto values of $\Rs < \La^2$.

\dbindex{number of active flavours \enef}{value at threshold}%
Another \qcdnum\ feature is that $\enef = (4,5,6)$ and not $(3,4,5)$
at the heavy flavour thresholds \msh.  This implies, first of all, that
parton evolution in the \vfns\ must start from~$\msz < \msc$ and not from
$\msz \leq \msc$, simply because the number of flavours must be 
\mbox{$\enef =3$} at the starting scale.  There is, however,
no restriction on the
starting (renormalisation) scale of \as\ so that it may very well
coincide with a flavour threshold, either before or after varying the
renormalisation scale with respect to the factorisation scale. If this
happens at NNLO, the input value of \as\ is assumed to include the
discontinuity.
\siindex{discontinuities in \as\ evolution|)}%


\subsection{Validation and Performance}\label{se:program4}

The CPU time that is needed to evolve a pdf on a discrete grid grows
quadratically with the number of grid points in $x$. With linear
(quadratic) interpolation the accuracy increases linearly
(quadratically) with the number of grid points. It follows that an
$r$-fold gain in accuracy will cost a factor of $r^2$ in CPU for
linear interpolation but only a factor of $r$ for quadratic
interpolation. This reduction in cost motivated the inclusion of
quadratic splines in \qcdnum.

To investigate the performance of the two interpolation schemes, we
compare results from \qcdnum\ to those from the $N$-space evolution
program \pegasus~\cite{ref:pegasus}. In this comparison a default set
\siindex{\ipegasus\ program}%
of initial distributions~\cite{ref:salamvogt} is evolved at NNLO from
$\ms = 2$ to $\ms = 10^4$~\gevs\ with $\enef = 4$ flavours.  The dashed
curve in the top plot of \Fi{fig:deltaplot}
%
\begin{figure}[tbh]
\begin{center}
\includegraphics[width=0.9\linewidth]{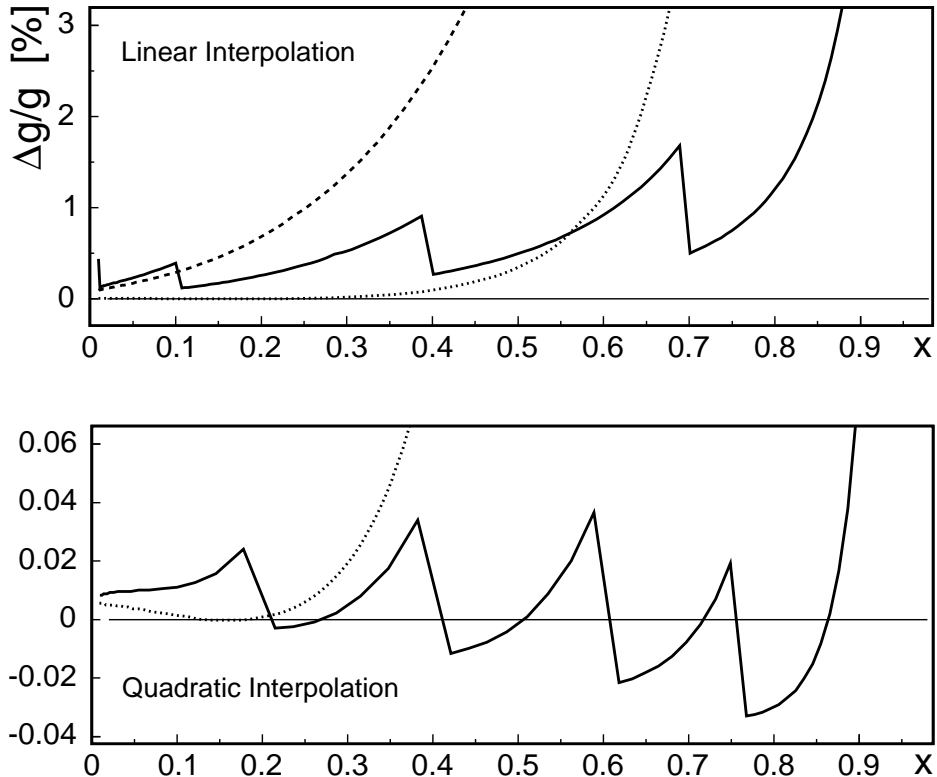}
\end{center}
\caption{\footnotesize
  The relative difference $\De g/g$ (in percent) of gluon densities
  evolved from $\ms = 2$ to $\ms = 10^4$~\gevs\ by \qcdnum\ and
  \pegasus. Top: Evolution with linear splines on a ${200}$ point
  single grid down to $x = 10^{-5}$ (dashed curve) and on multiple
  grids (full curve).  Bottom: Evolution with quadratic splines on a
  100 point single grid (dotted curve, also shown in the top plot) and
  on multiple grids (full curve). Note the different vertical scales
  in the two plots.}
\label{fig:deltaplot} 
\end{figure}
%
shows the relative difference $\De g/g$ versus $x$ for \qcdnum\ 
evolution with linear splines on a single 200 point grid extending
down to $x = 10^{-5}$. The accuracy at low $x$ is satisfactory (few
permille) but deteriorates rapidly to $\De g/g >2$\% for $x > 0.35$.

\siindex{multiple evolution grid|(}%
The precision is much improved by evolving on multiple grids
(\Se{se:dglaplin}) as shown
by the full curve in the top plot of \Fi{fig:deltaplot}. Here the 200
grid points are re-distributed over five sub-grids
with lower limits as given in \Ta{tab:xranges}.
%
\begin{table}[tbh]
\caption{\footnotesize
Lower $x$ limits of multiple grids used in the evolution with linear
and quadratic splines.
}
\begin{center}
\begin{tabular}{lcccccc}
  & $n$ & $x_1$ & $x_2$ & $x_3$ & $x_4$ & $x_5$ \\
\hline
  Linear interpolation &
  200 & $10^{-5}$ &  0.01 &  0.10  &  0.40  &  0.70   \\
  Quadratic interpolation &
  100 & $10^{-5}$ &  0.20 &  0.40  &  0.60  &  0.75  \\
  Relative point density &
      &    1      &   2   &    4   &    8   &   16   \\
\hline
\end{tabular}
\end{center}
\label{tab:xranges}
\end{table}
%
For each successive grid the point density is twice that of the
previous grid. It is seen from \Fi{fig:deltaplot} that the precision
is now better than~2\% for $x < 0.85$.

The dotted curves in \Fi{fig:deltaplot} (top and bottom) correspond to
evolution with quadratic splines on a single 100 point grid. There is
a large improvement in accuracy (more than a factor of 10) compared to
linear splines even though the number of grid points is reduced from
200 to 100. However, also here the precision deteriorates with
increasing~$x$, reaching a level of 2\% at $x = 0.65$. A five-fold
multiple grid with lower limits as listed in \Ta{tab:xranges} yields a
precision $\De g/g < 5 \times 10^{-4}$ over the entire range $x < 0.9$
as can be seen from the full curve in the lower plot of
\Fi{fig:deltaplot}. Note that this is for evolution up to $\ms =
10^4$~\gevs; at lower \ms\ the accuracy is even better since it
increases (roughly linearly) with decreasing $\ln(\ms)$. To fully
validate the \qcdnum\ evolution with \pegasus,\footnote{Similar 
benchmarking between \hoppet~\cite{ref:hoppet} and \pegasus\ is given 
in~\cite{ref:salamvogt} and~\cite{ref:heralhc}, where also pdf
reference tables can be found. We do not provide here benchmark tables
for \qcdnum, but a program that generates such tables and compares
them with \pegasus\ is available upon request from the author.}
we have made additional comparisons in the \ffns\ with $\enef = 3$,
5 or 6 flavours, in the \vfns\ with and without backward evolution,
and with the renormalisation scale set different from the factorisation
scale. This for both un-polarised evolution up to NNLO and polarised
evolution up to NLO.   

As remarked in
\Se{se:dglaplin}, the quadratic spline evolution is not more expensive
in CPU time than linear spline evolution. On the contrary: \qcdnum\ 
runs 4 times \emph{faster} since we need only 100 instead of 200 grid
points. With the multiple grid definition given in \Ta{tab:xranges} for
quadratic splines, the density of the first grid ($x > 10^{-5}$) is 12
points per decade. It follows that for evolution down to $x = 10^{-6}$
($10^{-4}$) a grid with $100+12=112$ ($100-12=88$) points should be
sufficient.
\siindex{multiple evolution grid|)}%

\dbindex{\iqcdnum}{execution speed}%
To investigate the execution speed we did mimic a QCD fit by
performing 1000 NNLO evolutions in the \vfns\ (13 pdfs),
using a 60 point \ms\ grid and the 5-fold 100 point $x$-grid given in \Ta{tab:xranges}. After each evolution, the proton structure
functions $F_2$ and \Fell\ were computed at NNLO
for 1000 interpolation points in the HERA kinematic range.
For this test, \qcdnum, \mbutil, and \zmstf\ were
compiled with the \textsc{gfortran} compiler, using level~2 optimisation
and without array boundary check. The computations took~18.5 CPU seconds
on a 2~GHz Intel Core 2 Duo
processor under Mac~OS-X: 8.5~s for the evolutions and 10~s for the
structure functions.

%% file: sections/callsintro.tex
\section{Subroutine Calls}\label{se:subroutines}

In this section we describe all available \qcdnum\ subroutines and
functions. For convenience a list of these is given in
Table~\ref{tab:subr}.
\begin{table}[p] 
  \caption{ Subroutine and function calls in \qcdnum.}
  \begin{center}
  \begin{tabular*}{0.95\textwidth}{l@{\extracolsep{\fill}}l}
  \\
  Subroutine or function & Description               \\
  \hline
  \xtt{\ \ \ \ QCINIT ( lun, 'filename' )}
  & Initialise                                       \\
  \xtt{\ \ \ \ SETLUN ( lun, 'filename' )}
  & Redirect output                                  \\
  \xtt{SET|GETVAL ( 'opt', val )}
  & Set$|$Get parameters                             \\
  \xtt{SET|GETINT ( 'opt', ival )}
  & Set$|$Get parameters                             \\
  \xtt{\ \ \ \ GXMAKE ( xmi, iwt, n, nxin, *nxout, iord )}
  & Define $x$ grid                                  \\
  \xtt{\ \ \ \ IXFRMX ( x )}
  & Get $i_x$ from $x$                               \\
  \xtt{\ \ \ \ XFRMIX ( ix )}
  & Get $x$ from $i_x$                               \\
  \xtt{\ \ \ \ XXATIX ( x, ix )}
  & Verify grid point                                \\
  \xtt{\ \ \ \ GQMAKE ( qarr, wt, n, nqin, *nqout )}
  & Define $\Fs$ grid                                \\
  \xtt{\ \ \ \ IQFRMQ ( q2 )}
  & Get $i_{\mu}$ from $\Fs$                         \\
  \xtt{\ \ \ \ QFRMIQ ( iq )}
  & Get $\Fs$ from $i_{\mu}$                         \\
  \xtt{\ \ \ \ QQATIQ ( q2, iq )}
  & Verify grid point                                \\
  \xtt{\ \ \ \ GRPARS ( *nx, *x1, *x2, *nq, *q1, *q2, *io )\ } 
  & Get grid definitions                             \\
  \xtt{\ \ \ \ GXCOPY ( *array, n, *nx )}
  & Copy $x$ grid                                    \\
  \xtt{\ \ \ \ GQCOPY ( *array, n, *nq )}
  & Copy $\ms$ grid                                  \\
  \xtt{\ \ \ \ FILLWT ( itype, *idmi, *idma, *nw )}
  & Fill weight tables                               \\
  \xtt{\ \ \ \ FILLWC ( mysub, *idmi, *idma, *nw )}
  & Custom weights                                 \\
  \xtt{\ \ \ \ DMPWGT ( itype, lun, 'filename' )}
  & Dump weight tables                               \\
  \xtt{\ \ \ \ READWT ( lun, 'fn', *idmi, *idma, *nw, *ie )}
  & Read weight tables                               \\
  \xtt{\ \ \ \ NWUSED ( *nwtot, *nwuse, *nwtab )}
  & Memory words used                                \\
   \xtt{SET|GETORD ( iord )}
  & Set$|$Get order                                  \\
  \xtt{SET|GETALF ( alfs, r2 )}
  & Set$|$Get \as\ start value                       \\
  \xtt{\ \ \ \ SETCBT ( nfix, iqc, iqb, iqt )}
  & Set \enef\ or thresholds                         \\
  \xtt{\ \ \ \ GETCBT ( *nfix, *q2c, *q2b, *q2t )}
  & Get \enef\ or thresholds                         \\
  \xtt{SET|GETABR ( ar, br )}
  & Set$|$Get \Rs\ scale                             \\
  \xtt{\ \ \ \ RFROMF ( fscale )}
  & Convert \Fs\ to \Rs\                             \\
  \xtt{\ \ \ \ FFROMR ( rscale )}
  & Convert \Rs\ to \Fs\                             \\
  \xtt{\ \ \ \ ASFUNC ( r2, *nf, *ierr )}
  & Evolve $\as(\Rs)$                                \\
  \xtt{\ \ \ \ EVOLFG ( itype, func, def, iq0, *eps )}
  & Evolve all pdfs                                  \\
  \xtt{\ \ \ \ PDFINP ( subr, iset, offset, *epsi, *nwds )}
  & Pdfs from outside                                \\
  \xtt{\ \ \ \ CHKPDF ( iset )}
  & True if pdf set exists                           \\
    \xtt{\ \ \ \ FVALXQ ( iset, id, x, qmu2, ichk )}
  & Interpolate \ket{g,q,\qbar}                       \\
  \xtt{\ \ \ \ FVALIJ ( iset, id, ix, iq, ichk )}
  & \ket{g,q,\qbar}\ at grid point                    \\
  \xtt{\ \ \ \ FPDFXQ ( iset, x, qmu2, *pdfs, ichk )}
  & All pdfs \ket{g,q,\qbar}                          \\
  \xtt{\ \ \ \ FPDFIJ ( iset, ix, iq, *pdfs, ichk )}
  & All pdfs \ket{g,q,\qbar}                           \\
  \xtt{\ \ \ \ FSUMXQ ( iset, def, x, qmu2, ichk )}
  & Linear combination                               \\
  \xtt{\ \ \ \ FSUMIJ ( iset, def, ix, iq, ichk )}
  & Linear combination                               \\ 
  \xtt{\ \ \ \ FSNSXQ ( iset, id, x, qmu2, ichk )}
  & Interpolate \ket{g,\epm}                           \\
  \xtt{\ \ \ \ FSNSIJ ( iset, id, ix, iq, ichk )}
  & \ket{g,\epm}\ at grid point                      \\
  \xtt{\ \ \ \ FSPLNE ( iset, id, x, iq  )}
  & Spline interpolation                               \\
  \xtt{\ \ \ \ SPLCHK ( iset, id, iq )}
  & Check spline                                       \\   
  \hline
  {\footnotesize Output arguments are pre-fixed with
   an asterisk (\xtt{*}).} & \\
  \end{tabular*} 
  \end{center}
  \label{tab:subr}
  \dbindex{\iqcdnum}{list of subroutines}
\end{table}
In the following we will prefix output variables with an asterisk
(\xtt{*}).  We use the \fortran\ convention that integer variable and
function names start with the letters \xtt{I}--\xtt{N}.  Character
variables are given in quotes as in~\xtt{'opt'}.  Other variables and
functions are in double precision unless otherwise stated.
Note that floating point numbers should be entered in
double precision format:
\begin{verbatim}
         ix = ixfrmx ( x )       ! ok
         ix = ixfrmx ( 0.1D0 )   ! ok
         ix = ixfrmx ( 0.1  )    ! wrong!
\end{verbatim}
Unlike \fortran, \qcdnum\ is case insensitive so that character
arguments like \xtt{'ALIM'} or \xtt{'Alim'} are both valid
inputs.

Most \qcdnum\ functions will, upon error, generate an error message.
The inclusion of function calls in \xtt{print} or \xtt{write}
statements can then cause program hang-up in case the function tries
to issue a message.  Thus:
\dbindex{\iqcdnum}{program hang-up}%
\begin{verbatim}
         write(6,*) 'Glue = ', fvalxq(1,0,x,q,1) ! not recommended
         
         glue = fvalxq(1,0,x,q,1)                ! OK 
         write(6,*) 'Glue = ', glue              ! OK
\end{verbatim}
%

%% file: sections/calls.tex

\subsection{Initialisation} \label{se:subini}


\subrbox{call QCINIT ( lun, 'filename' )}{qcinit}

Initialise \qcdnum\ and define the output stream. Should be called
before anything else.
\begin{tdeflist}[xxfilenamexx\ ]{-1mm}
\item[\xtt{lun}] Output logical unit number. When set to \xtt{6},
  \qcdnum\ messages appear on the standard output. When set to
  \xtt{-6}, the \qcdnum\ banner printout is suppressed on the standard
  output.
\item[\xtt{'filename'}] Output file name. Irrelevant when \xtt{lun} is
  set to \xtt{6} or \xtt{-6}.
\end{tdeflist}


\subrbox{call SETLUN ( lun, 'filename' )}{setlun}

Redirect the \qcdnum\ messages. The parameters are as for \xtt{qcinit}
above. This routine can be called at any time after \xtt{qcinit}.


\subrbox{call SETVAL|GETVAL ( 'opt', val )}{setval}%
\srindex{getval}

Set or get \qcdnum\ floating point parameters. 

\begin{tdeflist}[xepsixx\ ]{-1mm}
\item[\xtt{'null'}] Result of a
  calculation that cannot be performed. Default, 
  \compa{null}{=}{1.D11}.\siindex{\inull\ value}%
\item[\xtt{'epsi'}] The tolerance level in the floating point
  comparison $|x-y| < \eps$, which \qcdnum\ uses to decide if $x$ and
  $y$ are equal. Default, \compa{epsi}{=}{1.D-9}.
\item[\xtt{'epsg'}] Required numerical accuracy of the Gauss integration
  in the calculation of weight tables. Default, \compa{epsg}{=}{1.D-7}.
  \dbindex{Gauss quadrature in \textsc{qcdnum}}{accuracy parameter of}%
\item[\xtt{'elim'}] Allowed difference between a quadratic and
  a linear spline interpolation mid-between the grid points in $x$.
  Default, \compa{elim}{=}{0.5}; larger values may indicate spline
  oscillation. To disable the check, set \compa{elim}{<}{0}. 
\item[\xtt{'alim'}] Maximum allowed value of $\as(\ms)$.  When \as\
  exceeds the limit, a fatal error condition is raised.
  Default, \compa{alim}{=}{10}.\footnote{When you raise 
  \compa{alim}{>}{10} then \as\ will at some point be limited
   by internal cuts in \qcdnum.}  
\item[\xtt{'qmin'}] Smallest possible lower boundary of the \ms\ 
  grid. Default, \compa{qmin}{=}{0.1}~\gevs.
\item[\xtt{'qmax'}] Largest possible upper boundary of the \ms\ 
  grid. Default, \compa{qmax}{=}{1.D11}~\gevs.
\end{tdeflist}

These parameters can be set and re-set at any time after \xtt{qcinit}.


\subrbox{call SETINT|GETINT ( 'opt', ival )}{setint}%
\srindex{getint}

Set or get \qcdnum\ integer parameters. 

\begin{tdeflist}[xepsixx\ ]{-1mm}
\item[\xtt{'iter'}] Set the number of iterations in the backward
  evolution. When set negative, one will evolve backward in the same
  interpolation scheme as the forward evolution (not recommended).
  When set to zero, one will evolve backward in the linear
  interpolation scheme, without iterations (not recommended either).
  A value larger than zero
  gives the number of iterations to perform. Default, \compa{iter}{=}{1}.
  This parameter is only relevant when one works with quadratic
  splines. \siindex{backward evolution}%
\item[\xtt{'lunq'}] Retrieve the \qcdnum\ logical unit number. Useful if 
  one wants to write messages on the same output stream as \qcdnum.
  This option is only available for \xtt{getint} and not for 
  \xtt{setint}.
\item[\xtt{'ntab'}] Number of scratch buffers (maximum 20) generated
  by \xtt{fastini} (fast convolution engine, \Se{se:ustfast}).
  Will have no effect when set after the call to \xtt{fastini}.
  Default, \compa{ntab}{=}{5}. If one wants to generate more than
  20 buffers, the value of \xtt{mbf0} in
  \xtt{qcdnum.inc} should be increased, and \qcdnum\ re-compiled.
\end{tdeflist}


\subsection{Grid Definition} \label{se:subgrid}

A proper definition of the grid in $x$ and $\ms$ is important because
it determines the speed and accuracy of the \qcdnum\ calculations.
The grid definition also governs the partition of the internal store
which contains the weight tables and tables of parton densities.
In addition, the routines set-up the bases of B-splines.

The $x$ grid must be strictly equidistant in the variable $y = -\ln x$
but in \qcdnum\ one can generate multiple equidistant grids
(\Se{se:dglaplin}) to obtain a finer binning at 
\siindex{multiple evolution grid|(}%
low~$y$ (large~$x$). Multiple
grids are generated when the $x$-range is subdivided into regions
with different densities, as is described below.

The \ms\ grid does not need to be equidistant. So one can either enter
a fully user-defined grid or let \qcdnum\ generate one by an
equidistant logarithmic fill-in of a given set of intervals in \ms.


\subrbox{call GXMAKE ( xmin, iwt, n, nxin, *nxout, iord )}{gxmake}

Generate a logarithmic $x$-grid.
\begin{tdeflist}[nxout\ ]{-1mm}
\item[\xtt{xmin}] Input array containing \xtt{n} values of $x$ in
  ascending order: \xtt{xmin(1)} defines the lower end of the grid
  while the other values define the approximate positions where the
  point density will change according to the values set in \xtt{iwt}.
  The list may or may not contain $x = 1$ which is ignored anyway.
\item[\xtt{iwt}] Input integer weights. The point density between
  \xtt{xmin(1)} and \xtt{xmin(2)} will be proportional to
  \xtt{iwt(1)}, that of the next region will be proportional to
  \xtt{iwt(2)} and so on. The weights should be given in ascending
  order and must always be an integer multiple of the previous weight.
  Thus, to give an example, the triplets \{\xtt{1,1,1}\} and
  \{\xtt{1,2,4}\} are allowed but \{\xtt{1,2,3}\} is not.
\item[\xtt{n}] The number of values specified in \xtt{xmin} and
  \xtt{iwt}. This is also the number of sub-grids used internally by
  \qcdnum. 
\item[\xtt{nxin}] Requested number of grid points (not including the
  point $x = 1$). Should of course be considerably larger than \xtt{n}
  for an $x$-grid to make sense.
\item[\xtt{nxout}] Number of generated grid points. This may differ
  slightly from \xtt{nxin} because of the integer arithmetic used to
  generate the grid.
\item[\xtt{iord}] One should set \xtt{iord} = \xtt{2} (\xtt{3}) for
  linear (quadratic) spline interpolation.
\end{tdeflist}
With this routine, one can define a (logarithmic) grid in $x$
with higher point densities at large $x$, where the parton
distributions are strongly varying. Thus
\begin{verbatim}
         xmin = 1.D-4
         iwt  = 1
         call gxmake(xmin,iwt,1,100,nxout,iord)
\end{verbatim}
generates a logarithmic grid with exactly 100 points in the range
$10^{-4} \leq x < 1$, while
\begin{verbatim}
         xmin(1) = 1.D-4
         iwt(1)  = 1
         xmin(2) = 0.7D0
         iwt(2)  = 2
         call gxmake(xmin,iwt,2,100,nxout,iord)
\end{verbatim}
generates a 100-point grid with twice the point density above $x
\approx 0.7$.

A call to \xtt{gxmake} invalidates the weight tables and the pdf
store.


\subrbox{ix =  IXFRMX ( x )\ \ \ \ \ x = XFRMIX ( ix )\ \ \ \ \ 
                                     L = XXATIX ( x, ix )}{ixfrmx}%
\srindex{xfrmix}%
\srindex{xxatix}%

The function \xtt{ixfrmx} returns the index of the closest grid point
at or below $x$.  Returns zero if $x$ is out of range (note that $x =
1$ is outside the range) or if the grid is not defined.
The inverse function is \xtt{x = xfrmix(ix)}.  Also this function
returns zero if
\xtt{ix} is out of range or if the grid is not defined.
To verify that $x$ coincides with a grid point, use the logical
function \xtt{xxatix}, as in
\begin{verbatim}
         logical xxatix
         ix = xfrmix(x)          !x is at or above grid point ix
         if(xxatix(x,ix)) then   !x is at grid point ix 
\end{verbatim}
Note that \qcdnum\ snaps to the grid, that is, $x$ is considered to be
at a grid point $i$ if $|y-y_i| < \eps$ with $y = -\ln x$ and, by
default, $\eps = 10^{-9}$.


\subrbox{call GQMAKE ( qarr, wgt, n, nqin, *nqout )}{gqmake}

Generate a logarithmic \Fs\ grid on which the parton densities are
evolved.\footnote{Note that \as\ is evolved (without using a grid) on
  \Rs\ which may or may not be different from \Fs.}
\begin{tdeflist}[nqout\ ]{-1mm}
\item[\xtt{qarr}] Input array containing \xtt{n} values of $\mu^2$ in
  ascending order: \xtt{qarr(1)} and \xtt{qarr(n)} define the lower
  and upper end of the grid, respectively. The lower end of the grid
  should be above 0.1~\gevs. If $\xtt{n} > \xtt{2}$ then the
  additional points specified in \xtt{qarr} are put into the grid.  In
  this way, one can incorporate a set of starting values
  $\mu^2_0$, or thresholds $\mu^2_{\cbt}$.
\item[\xtt{wgt}] Input array giving the relative grid point density in
  each region defined by \xtt{qarr}. The weights are not restricted by
  integer multiples as in \xtt{gxmake} but can be set to any value in
  the range $0.1 \leq w \leq 10$. With these weights, one can generate a
  grid with higher density at low values of \ms\ where \as\ is
  changing rapidly.
\item[\xtt{n}] The number of values specified in \xtt{qarr} and
  \xtt{wgt} (\xtt{n} $\geq$ \xtt{2}).
\item[\xtt{nqin}] Requested number of grid points. 
  If \xtt{nqin} $\leq$ \xtt{n} then the grid is not generated but taken
  from \xtt{qarr}. This
  feature allows you to read-in your own \ms\ grid.
\item[\xtt{nqout}] Number of generated grid points. This may differ
  slightly from \xtt{nqin} because of the integer arithmetic used to
  generate the grid.
\end{tdeflist}
A call to \xtt{gqmake} invalidates the weight tables and the pdf
store.


\subrbox{iq =  IQFRMQ ( q2 )\ \ \ \ q2 = QFRMIQ ( iq )\ \ \ \  
                                     L = QQATIQ ( q2, iq )}{iqfrmq}%
\srindex{qfrmiq}%
\srindex{qqatiq}%

The function \xtt{iqfrmq} returns the index of the closest grid point
at or below \ms. The inverse function is \xtt{qfrmiq}.  To verify that
\ms\ coincides with a grid point, use the logical function
\xtt{qqatiq}.
As described above for the corresponding $x$ grid routines, a value of
zero is returned if \xtt{q2} and/or \xtt{iq} are not within the range
of the current grid, or if the grid is not defined.
           

\subrbox{call GRPARS ( *nx, *xmi, *xma, *nq, *qmi, *qma, *iord )}{grpars}
Returns the current grid definitions
\begin{tdeflist}[xmin\ ]{-1mm}
\item[\xtt{nx}]  Number of points in the $x$ grid not including
  $x=1$.
\item[\xtt{xmi}] Lower boundary of the $x$ grid.
\item[\xtt{xma}] Upper boundary of the $x$ grid. Is always set to
  \xtt{xma} = \xtt{1}.
\item[\xtt{nq}]  Number of points in the \ms\ grid.
\item[\xtt{qmi}] Lower boundary of the \ms\ grid.
\item[\xtt{qma}] Upper boundary of the \ms\ grid.
\item[\xtt{iord}]  Order of the spline interpolation (\xtt{2} =
  linear, \xtt{3} = quadratic).
\end{tdeflist}


\subrbox{call GXCOPY ( *array, n, *nx )}{gxcopy}
Copy the $x$ grid to a local array
\begin{tdeflist} [array\ ]{-1mm}
\item[\xtt{array}] Local array containing on exit the $x$ grid but not
  the value $x = 1$.
\item[\xtt{n}]  Dimension of \xtt{array} as declared in the calling
  routine.
\item[\xtt{nx}] Number of grid points copied to the local array. A
  fatal error occurs if \xtt{array} is not large enough to contain the
  current grid.
\end{tdeflist}


\subrbox{call GQCOPY ( *array, n, *nq )}{gqcopy}

As above, but now for the \ms\ grid.


\subsection{Weights} \label{se:subweight}

In this section we describe routines to calculate the weight tables,
to dump these to disk and to read them back. The weight tables are
calculated for all orders (LO,NLO,NNLO) and all number of flavours
$\enef = (3,4,5,6)$, irrespective of the current \qcdnum\ settings.
Tables can be created for un-polarised pdfs, polarised pdfs and
fragmentation functions. All these pdf types can exist simultaneously
in memory.  For each type, one gluon table and 12 quark tables are
generated by the routines, in addition to the weight tables.


\subrbox{call FILLWT ( itype, *idmin, *idmax, *nwds )}{fillwt}

Partition the pdf store and fill the weight tables used in the
calculation of the convolution integrals. Both the $x$ and \ms\ grid
must have been defined before the call to \xtt{fillwt}.
\begin{tdeflist}[iselectxx\ ]{-1mm}
\item[\xtt{itype}] Select un-polarised pdfs (\xtt{1}),
  polarised pdfs (\xtt{2}) or fragmentation functions~(\xtt{3}).
  Any other input value will select un-polarised pdfs (default).
  \siindex{pdf type, pdf set}%
\item[\xtt{idmin}] Returns, on exit, the identifier of the first pdf
  table.  Always \compa{idmin}{=}{0}.
\item[\xtt{idmax}] Identifier of the last pdf table in the store.
  Always \compa{idmax}{=}{12}.
\item[\xtt{nwds}]  Total number of words used in memory.
\end{tdeflist}
One can create more than one set of tables tables by calling
\xtt{fillwt} with different values of \xtt{itype}. For instance, the
sequence
\begin{verbatim}
          call fillwt(1,idmin,idmax,nw)   !Unpolarised pdfs
          call fillwt(2,idmin,idmax,nw)   !Polarised pdfs
\end{verbatim}
makes both the un-polarised and the polarised pdfs available.  For each
pdf type, \xtt{fillwt} creates 13 pdf tables. If
there is not enough space in memory to hold all the tables,
\xtt{fillwt} returns with an error message telling how much memory it
needs. One should then increase value of \xtt{nwf0} in the
include file \xtt{qcdnum.inc}, and recompile
\qcdnum.\footnote{This one may have to repeat several times, because
\xtt{fillwt} proceeds in stages and is ignorant of the memory
requirements of the next stage.}
Note that
\xtt{fillwt} acts as a do-nothing when the pdf type already exists in
memory:
\begin{verbatim}
          call fillwt(1,idmin,idmax,nw)   !Unpolarised pdfs
          call fillwt(1,idmin,idmax,nw)   !Do nothing
\end{verbatim}
%


\subrbox{call DMPWGT( itype, lun, 'filename' )}{dmpwgt}

Dump the weight tables (not the pdf tables) of a given pdf type to disk.
When \compa{itype}{=}{0}, all pdf types in memory are dumped.\footnote{
This does not include the weight tables of custom evolution 
(\compa{itype}{=}{4}, see \Se{se:userevol}). Such tables are thus
always dumped on a separate file.}
Fatal error if \xtt{itype}
does not exist.  Additional information about the \qcdnum\ version, grid
definition and partition parameters is also dumped, to protect against
corruption of the dynamic store when the weights are read back in
future \qcdnum\ runs. The dump is unformatted so that the output file
cannot be exchanged across machines.


\subrbox{call READWT( lun, 'fname', *idmin, *idmax, *nwds, *ierr )}%
{readwt}

Read the weight tables from a disk file. Both the $x$ and \ms\ grid
must have been defined before the call to \xtt{readwt}. On exit the
error flag is set as follows:
\begin{tdeflist}[0\ \ ]{-1mm}
  \item[\xtt{0}] Weights are successfully read in.
  \item[\xtt{1}] Read error or input file does not exist.
  \item[\xtt{2}] Input file was written with another \qcdnum\ version.
  \item[\xtt{3}] Key mismatch (should never occur).
  \item[\xtt{4}] Incompatible $x$-\ms\ grid definition.
\end{tdeflist}
When successful (\xtt{ierr} = \xtt{0}), the routine creates
the pdf store(s) and returns on exit the parameters \xtt{idmin},
\xtt{idmax} and \xtt{nwds} as does the subroutine \xtt{fillwt}. 
One will get a fatal error if there is not enough space in memory
to hold all the tables. Like \xtt{fillwt}, \xtt{readwt} acts
as a do-nothing when
a pdf type already resides in memory.

The code below automatically maintains an up-to-date weight file on
disk:
\begin{verbatim}
        call readwt(lun,'polarised.wgt',idmin,idmax,nw,ierr)
        if(ierr.ne.0) then
          call fillwt(2,idmin,idmax,nw)  
          call dmpwgt(2,lun,'polarised.wgt')
        endif
\end{verbatim}


\subrbox{call NWUSED( *nwtot, *nwuse, *nwtab )}{nwused}
Returns the size \xtt{nwtot} of the \qcdnum\ store
(the parameter \xtt{nwf0} in \xtt{qcdnum.inc}), the number
of words used (\xtt{nwuse}) and the size of one pdf table
(\xtt{nwtab}). 
%


\subsection{Parameters} \label{se:subparam}

In this section we describe the \qcdnum\ routines to set evolution
parameters like the perturbative order, flavour thresholds, \as,
\mbetc\ All these parameters have reasonable defaults but one can
change them at any point in the code. Note that a re-definition of
these evolution parameters invalidates the pdf tables of \emph{all}
existing types. The weight tables are not invalidated. 
In this way, all pdfs are always evolved with the same set of
parameters; one cannot, for instance, have both un-polarised and 
polarised pdfs in memory,
and evolve one in NNLO and the other in NLO.


\subrbox{call SETORD|GETORD ( iord )}{setord}%
\srindex{getord}

Set (or get) the order of the \qcdnum\ calculations to \xtt{1},
\xtt{2} or \xtt{3} for LO, NLO and NNLO, respectively. Default,
\xtt{iord} = \xtt{2}.

\vspace{3mm}
\subrbox{call SETALF|GETALF ( alfs, r2 )}{setalf}%
\srindex{getalf}

Set or get for the \as\ evolution the starting value \xtt{alfs} and
the starting renormalisation scale \xtt{r2}. Default $\asmz = 0.118$.


\subrbox{call SETCBT( nfix, iqc, iqb, iqt )}{setthr}

\begin{tdeflist}[\xtt{iqc,b,t}\ ]{-1mm}
\item[\xtt{nfix}] Number of flavours in the \ffns\ mode. If not set to
  \xtt{3}, \xtt{4}, \xtt{5} or \xtt{6}, \qcdnum\ runs in the \vfns\
  mode.
\item[\xtt{iqc,b,t}] Grid indices of the quark mass thresholds
  $\ms_{\cbt}$.  This input is ignored when \qcdnum\ runs in the
  \ffns\ mode, that is, when \xtt{nfix} is set to \xtt{3}, \xtt{4},
  \xtt{5} or~\xtt{6}. There are some restrictions, dictated by
  the evolution and interpolation routines: \compa{iqc}{\geq}{2},
  \compa{iqb}{\geq}{iqc+2} and \compa{iqt}{\geq}{iqb+2}.
\end{tdeflist}
A threshold index value of zero (or larger
than the number of grid points) means `beyond the upper edge of the
grid'.  For instance, (\xtt{iqc,b,t}) = (\xtt{0,0,0}) is like running
in the \ffns\ with $\enef = 3$ while the setting (\xtt{2,4,0}) puts
the top quark threshold beyond the evolution range.  By default, 
\qcdnum\ runs in the \ffns\ with $\enef = 3$.


\subrbox{call GETCBT( *nfix, *q2c, *q2b, *q2t )}{getthr}

Return the current threshold settings. If \xtt{nfix} is non-zero on
return, \qcdnum\ runs in the \ffns\ and the values of
\xtt{q2c},\xtt{b},\xtt{t} are irrelevant. When \xtt{nfix}~=~\xtt{0},
\qcdnum\ runs in the \vfns\ and the routine returns the threshold
values (not the indices) on the \ms\ scale.


\subrbox{call SETABR|GETABR ( ar, br ) }{setabr}%
\srindex{getabr}%

Define the relation between the factorisation scale \Fs\ and the
renormalisation scale \Rs
\[
  \Rs = a_{\rm R}\, \Fs + b_{\rm R}.
\]
Default: \xtt{ar} = \xtt{1} and \xtt{br} = \xtt{0}.


\subrbox{rscale2 = RFROMF( fscale2 )\ \ \ \ \ 
         fscale2 = FFROMR( rscale2 )}{rfromf}%
\srindex{ffromr} 

Convert the factorisation scale \Fs\ to the renormalisation scale \Rs\ 
and \textit{vice versa}.


\subsection{Evolution} \label{se:subevolution}


\subrbox{alphas =  ASFUNC( r2, *nf, *ierr )}{asfunc}

Standalone evolution of \as\ on the renormalisation scale \Rs\ (without
using the \ms\ grid or weight tables). \Qcdnum\ internally keeps track
of \as\ so that there is no need to call this function; it is just
a user interface that gives access to $\as(\Rs)$.
\begin{tdeflist}[ierrxxxxxx\ ]{-1mm}
\item[\xtt{r2}] Renormalisation scale \Rs\ where \as\ is to be
  calculated.
\item[\xtt{nf}] Returns, on exit, the number of flavours at the scale
  \xtt{r2}.
\item[\xtt{ierr} = \xtt{1}] Too low value of \xtt{r2}. Internally, 
  there is a cut \compa{r2}{>}{0.1}~\gevs\ and also a cut on the slope,
  to avoid getting too close to $\La^2$. 
   
\end{tdeflist}
The input scale and input value of \as, the order of the evolution and
the flavour thresholds are those set by default or by the routines
described in \Se{se:subparam}. Note that although \as\ is evolved on
the renormalisation scale the result, in the \vfns, may still depend
on the relation between \Rs\ and \Fs. This is because the position of
the heavy flavour thresholds depends on this relation.


\subrbox{call EVOLFG( itype, func, def, iq0, *epsi )}{evolfg}

Evolve a complete set of parton \emph{momentum} densities from an
input scale \msz. If
\qcdnum\ runs in the \ffns, the gluon and $2\enef$ quark densities
must be given as an input at \msz. In the \vfns, the gluon and $2\enef
= 6$ light quark densities must be given at $\msz < \msc$.

Here and in the following the parton densities are written on the
flavour basis (note the PDG convention) with an indexing defined by
\beq{eq:iqqbar}
  \begin{array}{rrrrrrrrrrrrr}
  -6&-5&-4&-3&-2&-1&\ 0&\ 1&2&3&4&5&6\\
  \hline
  \mbstrut{5mm} \tbar &\ \bbar &\ \cbar &\ \sbar &\ \ubar &\ \dbar &\  g 
     &\ d &\ u &\ s &\ c &\ b &\ t
  \end{array}
\eeq
\siindex{indexing of $q$, $\bar{q}$ basis}
\begin{tdeflist}[nfmax\ ]{-1mm}
\item[\xtt{itype}] Type of evolution: un-polarised~(\xtt{1}), 
  polarised~(\xtt{2}), time-like~(\xtt{3}), or custom~(\xtt{4}).
\item[\xtt{func}] User defined function \xtt{func(j,x)} (see below)
  that returns the input parton momentum density $xf_j(x)$
  at \xtt{iq0}. Must be declared \xtt{external} in the calling
  routine. The index $\xtt{j}$ runs from 0 (gluon) to $2\enef$.
\item[\xtt{def}] Input array dimensioned in the calling routine to
  \xtt{def(-6:6,12)} which contains in \xtt{def(i,j)} the contribution
  of parton species \xtt{i} to the input distribution \xtt{j}, that
  is, \xtt{def(i,j)} specifies the flavour decomposition of all input
  distributions~\xtt{j}. The indexing of \xtt{i} is given in
  \eq{eq:iqqbar}. Internally, \qcdnum\ constructs from \xtt{def} 
  a $2\enef \times 2\enef$ sub-matrix of coefficients and tries to
  invert this matrix. If that fails, the $2\enef$ input densities are
  not linearly independent in flavour space and an error condition
  is raised, see also \eq{eq:efromp}. 
\item[\xtt{iq0}] Grid index of the starting value \msz. When evolving
  in the \ffns\ the input scale can be anywhere inside the range of
  the \ms\ grid. In the \vfns, however, \msz\ should be below the
  charm threshold.
\item[\xtt{epsi}] Maximum deviation of the quadratic spline
  interpolation from linear interpolation mid-between the 
  grid points (see \Se{se:dglaplin}).
  A large value \compa{epsi}{>}{elim} may
  indicate spline oscillation and will cause a fatal error message.
  The value of \xtt{elim} can be set by a call to
  \xtt{setval}. When \compa{elim}{\leq}{0} the error condition
  is disabled so that one can investigate the cause of oscillation.
  Note that, by definition, \compa{epsi}{=}{0} when
  \qcdnum\ is run in
  the linear interpolation scheme. 
\end{tdeflist}
The input function \xtt{func} must be coded as follows
\begin{verbatim}
         double precision function func(ipdf,x)
         implicit double precision (a-h,o-z)
         if(ipdf.eq.0) then
           func = xgluon(x)                     !0	= gluon  xg(x)
         elseif(ipdf.eq.1) then
           func = my_favourite_quark_dstn_1(x)  !1	= quarks xq1(x)
         elseif(ipdf.eq.2) then
           func = my_favourite_quark_dstn_2(x)  !2	= quarks xq2(x)
         elseif(ipdf.eq.3) then
           ..
         endif
         return
         end
\end{verbatim}
Because \xtt{evolfg} will call \xtt{func} only at the grid points
$x_i$, it is possible to feed tabulated values into the evolution
routine as is illustrated by the following code
\begin{verbatim}
         double precision function pdfinput(ipdf,x)
         implicit double precision (a-h,o-z)
         common /input/ table(0:12,nxx)    !table with input values
         ix       = ixfrmx(x)
         pdfinput = table(ipdf,ix)
         return
         end
\end{verbatim}

Here is code that evolves both un-polarised and polarised pdfs.
\begin{verbatim}
         call fillwt(1, idmin, idmax, nw)           !unpolarised 
         call fillwt(2, idmin, idmax, nw)           !polarised
           ..
         call evolfg(1, func1, def1, iq01, epsi1)   !unpolarised
         call evolfg(2, func2, def2, iq02, epsi2)   !polarised
\end{verbatim}


\subsection{External Pdfs} \label{se:pdfinp}

In \qcdnum, one can read up to 5 different pdf sets from some external
source, with type identifiers
running from~\xtt{5} to~\xtt{9}. Before reading an external
\siindex{pdf type, pdf set}%
pdf set, care should be taken that the 
perturbative order, the flavour scheme, the positions of the
thresholds and the input value of~\as\ are set correctly in
\qcdnum. Otherwise one will get the wrong answer when the
pdf set is used later on in structure function or cross-section
calculations. We remind that \emph{all} pdf sets in 
memory---including the external ones---are invalidated when a
\qcdnum\ parameter is re-set by calling one of the routines
in \Se{se:subparam}.  
%
%
\subrbox{call PDFINP ( subr, iset, offset, *epsi, *nwds )}{pdfinp}
\begin{tdeflist}[ix\ iq\ \ \ \ \ \ ]{-1mm}
\item[\xtt{subr}] User supplied subroutine (see below),
  declared \xtt{external} in the calling routine. 
\item[\xtt{iset}] Pdf set identifier in the range \xtt{5}--\xtt{9}.
  If the pdf set already exists, it will be overwritten.
\item[\xtt{offset}] Relative offset at the thresholds \msh.
  This parameter is used to catch discontinuities at the thresholds,
  if any, by sampling the pdfs at $\msh (1 \pm \de)$. 
  A small number like $10^{-3}$ should be sufficient, but
  this depends on how the pdfs are externally represented,
  and how accurate the thresholds are set in \qcdnum. 
\item[\xtt{epsi}] Maximum deviation of the quadratic spline
  interpolation from linear interpolation mid-between the 
  grid points. As for the routine \xtt{evolsg},
  a large value \compa{epsi}{>}{elim} may
  indicate spline oscillation and will cause a fatal error 
  message. Note that, by definition, \compa{epsi}{=}{0} when
  \qcdnum\ is run in
  the linear interpolation scheme.  
\item[\xtt{nwds}] Last word occupied in the store.
  Fatal error if the store is not large enough.  
\end{tdeflist}
The routine \xtt{subr} provides the interface
between \qcdnum\ and the external repository:
\begin{verbatim}
               subroutine SUBR ( x, qmu2, xf )
               implicit double precision (a-h,o-z)
               dimension xf(-6:6)
                 ..  
\end{verbatim}
The output array \xtt{xf(-6:6)} should contain the
values of the gluon and the (anti-)quark momentum densities
at $x$ and \ms, indexed according 
to~\eq{eq:iqqbar}; note the PDG convention.              


\subsection{Pdf Interpolation} \label{se:subpdfstf}

Here we describe routines to access the gluon distribution ($xg$),
the quark and anti-quark distributions ($xq,x\qbar$), or
linear combinations of the quarks and anti-quarks. It is also possible
to directly access the  basis singlet/non-singlet  pdfs
in memory ($x\epm$, defined in \Se{se:decompose}). These routines
perform local polynomial interpolation on a $k \times 3$ mesh
around the interpolation point in $x$ and \ms, where $k$ is
the current interpolation order in $x$.
Fast routines return the value of a pdf at a given grid point
(\xtt{ix},\xtt{iq}). Two routines are provided
to investigate the behaviour of the internal spline representation
in~$x$.

In the routines below, the pdf set identifier
\xtt{iset} selects the pdf set (or type):
un-polarised~(\xtt{1}), polarised~(\xtt{2}), 
fragmentation function~(\xtt{3}), custom~(\xtt{4}), or
external~(\xtt{5}--\xtt{9}).\siindex{pdf type, pdf set}%
%
%
\subrbox{ Lval = CHKPDF( iset )}{chkpdf}
Returns \xtt{.true.} if the pdf set exists in memory. 
Both \xtt{Lval} and \xtt{chkpdf} should be declared 
\xtt{logical} in the calling routine.
%
%
\subrbox{pdf = FVALXQ ( iset, id, x, qmu2, ichk )}{fvalxq}
Returns the gluon density or one of the (anti-)quark densities,
interpolated to $x$ and \ms.
\begin{tdeflist}[ix\ iq\ \ \ \ \ \ ]{-1mm}
\item[\xtt{iset}] Pdf set identifier [\xtt{1}--\xtt{9}].
\item[\xtt{id}]   Gluon, quark or anti-quark identifier,
  indexed as given in \eq{eq:iqqbar}.
\item[\xtt{x}, \xtt{qmu2}] Input value of $x$ and \ms.
\item[\xtt{ichk}] If set to zero, \xtt{fvalxq} will return a
  \xtt{null} value when \xtt{x} or \xtt{qmu2} are outside the grid
  boundaries; if set to a non-zero value a fatal error message will be
  issued. 
\end{tdeflist}
The fast version of this function is:
  \xtt{pdf = }\subr{fvalij}\xtt{( iset, id, ix, iq, ichk )}.
%

%
\subrbox{ call FPDFXQ ( iset, x, qmu2, *pdfs, ichk )}{fpdfxq}
Returns all pdf values in one call. The arguments
are as given above, except
\begin{tdeflist}[ix\ iq\ \ \ \ \ \ ]{-1mm}
\item[\xtt{pdfs}] Output array, dimensioned to \xtt{pdfs(-6:6)} in the
  calling routine. The indexing is given in \eq{eq:iqqbar}.
\end{tdeflist}
The fast version is the subroutine
\subr{fpdfij}\xtt{( iset, ix, iq, *pdfs, ichk )}.
%
%
\subrbox{pdf = FSUMXQ ( iset, def, x, qmu2, ichk )}{fsumxq}
Return a weighted sum of quark densities. The arguments
are as given above, except
\begin{tdeflist}[ix\ iq\ \ \ \ \ \ ]{-1mm}
\item[\xtt{def}] Input array, dimensioned to \xtt{def(-6:6)} in the
  calling routine, containing the coefficients of the linear combination.
  The indexing is as given in \eq{eq:iqqbar} but note that \xtt{def(0)} is
  ignored since it does not correspond to a quark density.
\end{tdeflist}
The fast call is:
\xtt{pdf = }\subr{fsumij}\xtt{( iset, def, ix, iq, ichk )}.
%
%
\subrbox{pdf = FSNSXQ ( iset, id, x, qmu2, ichk )}{fsnsxq}
Return the gluon density or one of the singlet/non-singlet basis
pdfs. The arguments are as given above, except that \xtt{id}
is now indexed as follows:
\beq{eq:isins}
  \begin{array}{lllllllllllll}
  0&\ 1&2&3&4&5&6&\ 7&8&9&10&11&12\\
  \hline
  \mbstrut{5mm} g\ &\ \qsi\ &e^+_2\ &e^+_3\ &e^+_4\ &e^+_5\ &e^+_6
     &\ \qva\ &e^-_2\ &e^-_3\ &e^-_4\ &e^-_5\ &e^-_6
  \end{array}
\eeq
\siindex{indexing of $\epm$ basis}
The fast call is:
\xtt{pdf = }\subr{fsnsij}\xtt{( iset, id, ix, iq, ichk )}.
%
%
\subrbox{ pdf = FSPLNE ( iset, id, x, iq )}{fsplne}
\siindex{spline interpolation}
This routine is identical to \xtt{fsnsxq} above, except that
the local polynomial interpolation in $x$ is replaced by
spline interpolation, as is done in the \qcdnum\ evolution and
convolution routines (note that \xtt{fsplne} does not
interpolate in \ms).
This function is provided as a diagnostic tool to
investigate quadratic spline oscillations, if any, which may not be
visible in the local polynomial interpolation. You do
not need this function to \emph{detect} spline oscillations, since
that is done automatically by \xtt{evolsg} and \xtt{pdfinp}.
%
%
\subrbox{ epsi = SPLCHK ( iset, id, iq )}{splchk}
Returns $\eps = \| \ve{u}-\ve{v} \|$  at
a grid point \xtt{iq}. Here $\ve{u}$ and $\ve{v}$ are the vectors
of quadratic and linear interpolation mid-between the grid
points in $x$, as is described in \Se{se:dglaplin}. 
By definition, $\eps = 0$
for linear interpolation, and should be a small number (like 0.05, say)
for quadratic interpolation. Large values indicate that the spline
oscillates.
\siindex{spline oscillation}

%% file: sections/cengine.tex

\section{Convolution Engine} \label{se:stfuser}

The \qcdnum\ convolution engine provides tools to calculate
structure functions in deep inelastic scattering,
hadron-hadron scattering cross-sections and parton luminosities.
The engine drives the add-on package \zmstf\ that computes the
zero-mass structure functions $F_2$, \Fell\ and $xF_3$ in un-polarised
deep inelastic scattering. It is also used in the \hqstf\ 
package that computes the heavy
flavour contributions to $F_2$ and \Fell\ in the fixed flavour
number scheme~\cite{ref:riemersma}.  Both these packages are included
in the \qcdnum\ distribution and are described in the
Sections~\ref{se:zmstf} and~\ref{se:hqstf} of this write-up.

\siindex{convolution integrals in \textsc{qcdnum}|(}%
From the parton \emph{number} densities $f$ and kernels $K$,
all kind of convolution integrals can be calculated with
the engine, such as
\[ 
  x [ f \otimes K ](x), \ \ \ 
  x [ f \otimes K_a \otimes K_b ](x), \ \ \ 
  x [f_a \otimes f_b ](x), \ \ \  
  x [ f_a \otimes f_b \otimes K ](x), \ \ \mbox{\mbetc}
\]
Here $\otimes$ stands for Mellin convolution as defined
by~\eq{eq:mellinconv}. We refer to \Se{se:numconvol} for
how convolution integrals are computed and where the factor
$x$ in front comes from. We emphasise that the kernel $K$ must be  
defined by convolution with a \emph{number} density. If not, then
it must be transformed as necessary, before it is fed into \qcdnum.
\siindex{convolution integrals in \textsc{qcdnum}|)}%

The steps to be taken in a calculation based on the convolution
engine are the following.
\begin{enumerate}
\item Declare one or more stores and partition these into tables.
      Then fill the tables with weights for all the convolution kernels
      needed in the calculation (\Se{se:ustfwts});
\item Write a function \xtt{myfun(ix,iq)} that returns the structure
      function, cross section or luminosity at a grid point in
      $x$ and $\ms$ (\Se{se:ustfconvol});
\item Pass \xtt{myfun} to a \qcdnum\ routine that will take care of the
      interpolation to any desired $x$ and \ms\ (\Se{se:ustfinterp}).
\end{enumerate}
This procedure is fairly straight-forward and therefore suitable for
prototyping and debugging. However, there is a considerable amount of
overhead so that it is recommended to ultimately move steps~2 and~3 of
the computation to a fast calculation scheme that is described in
\Se{se:ustfast}. By this one will gain at least an order of magnitude
in speed.

Before we present the convolution engine we will first, in the next
section, introduce the rescaling variable $\chi$ to accommodate
generalised mass variable flavour number schemes
\siindex{generalised mass (\textsc{gm}) schemes}%
(\gmvfns, see~\cite{ref:thornetung} for a recent review)
in structure function calculations.
 

\subsection{Rescaling Variable in Convolution Integrals} \label{se:strfun}

The general expression for a structure function can be
written as
\beq{eq:genstf}
  \cF_i(x,\qsq) = \sum_j x \int_{\chi}^1 \frac{\der z}{z}
  f_j(z,\ms)\; C_{ij} \left[ \frac{\chi}{z},\ms,\qsq,\masqh,\as(\ms)
  \right]. 
\eeq
Here the index $i$ labels the structure function (\mbeg\ $F_2$,
$\Fell$, $xF_3$, $F_2^{\rm c}$, $\ldots$) and $j$ labels a parton
number density like the gluon, the singlet and various non-singlets.
The coefficient function $C_{ij}$ depends on $x$, on the scale
variables \ms\ and \qsq, on one or more quark masses \masqh\ and on
the strong coupling constant \as. The variable $\chi = ax$, $a \geq
1$, is a so-called \emph{rescaling} variable which takes into account
the kinematic constraints of heavy quark production, for instance,
\siindex{rescaling variable $\chi$}%
\beq{eq:achi}
  \chi = ax = \left( 1 + \frac{4 \masqh}{\qsq} \right)x.
\eeq
We have $0 \leq \chi \leq 1$ so that the range of $x$ in
\eq{eq:genstf} is restricted to $0 \leq x \leq 1/a$. In the zero-mass
limit $a = 1$, $\chi = x$, and \eq{eq:genstf} reduces to the Mellin
form $x[f \otimes C](x)$.

To calculate the structure function, we first have to evaluate the
convolution integrals (for clarity we drop \as\ and the indices $i,j$)
\beq{eq:genstf2}
  \cF(x,\qsq) = 
  x \int_{\chi}^1 \frac{\der z}{z} f(z,\ms)\; C \left(
  \frac{\chi}{z},\ms,\qsq,\masqh \right).
\eeq
As in \Se{se:numconvol} we denote by $h(y,t)$ a parton momentum
density in the logarithmic scaling variables $y = -\ln x$ and $t = \ln
\mu^2$. In terms of these, and provided that $\chi$ is proportional to
$x$, \eq{eq:genstf2} can be written as a weighted sum of spline
coefficients
\beq{eq:stfsum}
   \cF(y_i,\qsq) = \sum_{j=1}^i W_{ij} A_j
\eeq
with $W_{ij} = w_{i-j+1}$ and
\beq{eq:stfwdef} 
  w_{\ell}  =  e^{-b} \int_{0}^{y_{\ell}-b} \der z \; Y_1(z)
  D(y_{\ell}-b-z,t,\qsq, \masqh) \qquad (1 \leq \ell \leq n).
\eeq
Here $D(y,t,\qsq,\masqh) = e^{-y} C (e^{-y}, e^t, \qsq, \masqh)$ and
$b = \ln(a)$.  It is understood that the integral \eq{eq:stfwdef} is
set to zero in case $y_{\ell} -b \leq 0$. In the massive schemes, $b >
0$ depends on $t$ which implies that the weights must be stored in
2-dimensional $y$-$t$ tables.

We emphasise that convolution integrals found in the literature
must, if necessary, be brought into the general form~\eq{eq:genstf}
by modifying the published Wilson coefficient. An example of such a
modification can be found in~\Ap{se:riemersma}.  


\subsection{Weight Tables}\label{se:ustfwts}

In this section we describe routines that partition a linear store into
tables and fill these tables with weights used in the calculation of
convolution integrals.
It is important to realise that the convolution kernels may contain
singularities, see also \Ap{app:singular}. To deal with such
singularities, we formally decompose a kernel into a regular part
($A$), a singular part ($B$), a product ($RS$) and a delta function
\dbindex{splitting functions}{singularities in}%
\beq{eq:singterms}
  C(x) = A(x) + [B(x)]_+ + R(x)[S(x)]_+ + D(x) \de (1-x).
\eeq
\Qcdnum\ provides routines that can calculate weights for each term
separately (if present) and add these to the weight table of $C$.

For reasons of efficiency and economy of storage, there are four
different types of tables:\siindex{types of weight table}
\begin{tdeflist}[itype = 1 + n\ \ ]{-1mm} 
\item[\xtt{itype} = \xtt{1}] Weights that depend only on $x$. Table
  identifiers run from \xtt{101}--\xtt{199};
\item[\xtt{itype} = \xtt{2}] Weights that depend on $x$ and \enef.
  Identifiers run from \xtt{201}--\xtt{299};
\item[\xtt{itype} = \xtt{3}] Weights that depend on $x$ and \ms.
  Identifiers run from \xtt{301}--\xtt{399};
\item[\xtt{itype} = \xtt{4}] Weights that depend on $x$, \ms\ and
  \enef. Identifiers run from \xtt{401}--\xtt{499}. 
\end{tdeflist}

Although it is a good idea to take out as many
\ms -dependent factors as possible from the convolution kernel,
it is clear from \eq{eq:genstf2} that quark mass parameters
and the relation between \ms\ and \qsq\ may enter via the rescaling
variable $\chi$ and that this dependence can never be factored out of
the convolution integral.  Thus the weight tables of the \textsc{gm}
schemes will, in general, depend on $x$ and \ms\ and must be stored in
type-3 or 4 tables.

\siindex{Gauss quadrature in \textsc{qcdnum}}%
\Qcdnum\ calculates  by Gauss quadrature (\cernlib\ routine D103) the integrals that define the weights. In case the default accuracy
of $\eps = 10^{-7}$ cannot be reached (fatal error message),
this limit can be raised by a call to 
\xtt{setval('epsg',value)}. Note, however,
that problems with the Gauss integration will most likely be caused
by problems with the integrand---such as near-singular behaviour
somewhere in the integration domain---and that this cannot be cured
by relaxing the required accuracy.

In \Ta{tab:ustfwts} we list all available
weight routines.
\begin{table}[bth] 
  \caption{\Qcdnum\ convolution weight table routines.}
  \begin{center}
  \begin{tabular*}{1.0\textwidth}{l@{\extracolsep{\fill}}l}
  \\
  Subroutine or function & Description \\
  \hline
  \xtt{BOOKTAB ( w, nw, itypes, *nwords )}
  & Partition into tables                        \\
  \xtt{MAKEWTA ( w, id, afun, achi )}
  & Regular piece $A(x)$                         \\
  \xtt{MAKEWTB ( w, id, bfun, achi, nodelta )}
  & Singular piece $[B(x)]_+$                    \\
  \xtt{MAKEWRS ( w, id, rfun, sfun, achi, nodelta )\ }
  & Product  $R(x)[S(x)]_+$                      \\
  \xtt{MAKEWTD ( w, id, dfun, achi ) }
  & Delta function $D(x)\de(1-x)$                \\
  \xtt{MAKEWTX ( w, id ) }
  & Weight table for $x[f_a \otimes f_b]$        \\
  \xtt{SCALEWT ( w, c, id ) }
  & Scale weight table                           \\
  \xtt{IDSPFUN ( 'pij', iord, itype ) }                        
  & Splitting function index                     \\
  \xtt{COPYWGT ( w, id1, id2, iadd ) }
  & Copy weight table                            \\
  \xtt{WCROSSW ( w, ida, idb, idc, iadd ) }
  & Double convolution weights                   \\
  \xtt{WTIMESF ( w, fun, id1, id2, iadd ) }
  & Multiply by $f(\ms,\enef)$                   \\
  \xtt{SETWPAR ( w, pars, n ) }
  & Store extra information                      \\
  \xtt{GETWPAR ( w, *pars, n ) }
  & Read extra information                       \\
  \xtt{TABDUMP ( w, lun, 'filename', 'key' ) }
  & Dump to disk                                 \\  
  \xtt{TABREAD ( w, n, lun, 'fn', 'key', *nw, *ierr ) }
  & Read from disk                               \\
  \hline
  {\footnotesize Output arguments are pre-fixed with
   an asterisk (\xtt{*}).} &
  \end{tabular*} 
  \end{center}
  \label{tab:ustfwts}
  \dbindex{\iqcdnum}{list of subroutines}
\end{table}


\subcbox{call BOOKTAB ( w, nw, itypes, *nwords ) }{booktab}

Partition a store \xtt{w} into tables.
\begin{tdeflist}[nwords\ \ \ ]{-1mm} 
\item[\xtt{w}] Double precision array declared in the calling routine.
\item[\xtt{nw}] Dimension of \xtt{w} as declared in the calling routine.
\item[\xtt{itypes}] Integer array dimensioned to
  \xtt{itypes(4)} in the calling routine which contains in
  \xtt{itypes(i)} the number of tables ($\leq$ \xtt{99}) of type \xtt{i}
  to be generated. When \compa{itypes(i)}{=}{0} then no
  tables of type \xtt{i} will be generated.
\item[\xtt{nwords}] Gives, on exit, the number of words used in the
  store. If \xtt{nwords} is negative, then the store is not
  sufficiently large and should be re-dimensioned in the calling
  routine to at least \xtt{-nwords}.
\end{tdeflist}
Note that one can declare and partition as many stores as desired,
one per structure function for instance.


\subcbox{call MAKEWTA ( w, id, afun, achi ) }{makewta}

Calculate the weights for the regular contribution $A(x)$ to a
convolution kernel and add these to table \xtt{id} in the store
\xtt{w}.
\begin{tdeflist}[afun\ \ \ ]{-1mm} 
\item[\xtt{w}] Store declared in the calling routine and previously
  partitioned by \xtt{booktab}.
\item[\xtt{id}] Table identifier. To add results to a type-$n$ table, one
  should use identifiers in the range \xtt{n01}--\xtt{n99}, with
  \compa{n}{=}{1}, \xtt{2}, \xtt{3} or \xtt{4}.
\item[\xtt{afun}] User function (see below) returning the regular
  piece of the convolution kernel. Should be declared \xtt{external}
  in the calling routine.
\item[\xtt{achi}] User function (see below), declared \xtt{external}
  in the calling routine, that returns the value $a$ of the rescaling
  variable $\chi = ax$.
\end{tdeflist}
The function \xtt{afun} provides an interface between \qcdnum\ and the
regular part of the kernel $C(\chi,\ms,\qsq,\masqh)$ and should be
coded as follows.\footnote{We assume here that the kernel conforms to
\eq{eq:genstf}. If not, then \xtt{afun} must take care of this.}
\begin{verbatim}
      double precision function afun(chi,qmu2,nf) !chi = a*x
      implicit double precision (a-h,o-z)                  
      common /fixpar/ par1, par2, .....           !parameters, if any
      Q2   = some_function_of(qmu2,some_params)   !Q2
      afun = cfun(chi,qmu2,Q2,nf,some_params)     !convolution kernel
      return
      end
\end{verbatim}
The function \xtt{achi} should return, as a function of \ms, the
factor $a$ that defines the rescaling variable $\chi = ax$.
\begin{verbatim}
      double precision function achi(qmu2)
      implicit double precision (a-h,o-z)
      common /fixpar/ par1, par2, .....          !parameters, if any
      Q2   = some_function_of(qmu2,some_params)  !Q2
      achi = some_function_of(Q2,some_params)
      return
      end
\end{verbatim}
\Qcdnum\ insists that always \compa{achi}{\geq}{1}, one will get a
fatal error if not. To compute standard Mellin
convolutions $x[f \otimes C](x)$, simply set \compa{achi}{=}{1}
for all \ms.
\begin{verbatim}
      double precision function achi(qmu2)
      implicit double precision (a-h,o-z)
      achi = 1.D0
      return
      end
\end{verbatim}


\subcbox{call MAKEWTB ( w, id, bfun, achi, nodelta ) }{makewtb}

Calculate the weights for the singular contribution $[B(x)]_+$ to a
convolution kernel and add these to a table in the store \xtt{w}.
The arguments and the coding of \xtt{bfun} and \xtt{achi} are as for
\xtt{makewta}. Thus, if a kernel has both a regular and
a singular part, then do
\begin{verbatim}
      call makewa(w,201,afun,achi)   !put weights in id = 201
      call makewb(w,201,bfun,achi,0) !add weights to id = 201
\end{verbatim}
It is seen from \Ap{app:singular}, equation~\eq{eq:plusdef1}, that a
`+' prescription generates a $\de(1-x)$ contribution.  By default,
\xtt{makewtb} includes this contribution, unless you set
\compa{nodelta}{=}{1}. In that case the $\de(1-x)$ contribution is not
calculated and must be entered, perhaps combined with other such
contributions, via a call to \xtt{makewtd}, see below.


\subcbox{call MAKEWRS ( w, id, rfun, sfun, achi, nodelta ) }{makewrs}

Calculate the weights for the product contribution $R(x)[S(x)]_+$ to a
convolution kernel and add these to a table in the store \xtt{w}.
The arguments and the coding of \xtt{rfun}, \xtt{sfun} and \xtt{achi}
are as for \xtt{makewta}.


\subcbox{call MAKEWTD ( w, id, dfun, achi ) }{makewtd}

Calculate the weights for the $\de(1-x)$ contribution to a convolution
kernel and add these to a table in the store \xtt{w}.  The delta
function is multiplied by the function \xtt{dfun}. The arguments and
the coding of \xtt{dfun} and \xtt{achi} is as for \xtt{makewta}.


\subcbox{call MAKEWTX ( w, id ) }{makewtx}

Calculate the weights \eq{eq:sumhacrosshb} for the convolution
$x[f_a \otimes f_b](x)$.
\begin{tdeflist}[afun\ \ \ ]{-1mm} 
\item[\xtt{w}] Store declared in the calling routine and previously
  partitioned by \xtt{booktab}.
\item[\xtt{id}] Table identifier. Because the weight table depends
  only on $x$, it can be stored in a type-1 table, but equally
  well in types-2, 3 or 4, if desired.
\end{tdeflist}


\newpage
\subcbox{call SCALEWT ( w, c, id ) }{scalewt}

Multiply the contents of table \xtt{id} by a constant \xtt{c}.


\subcbox{id =  IDSPFUN ( 'pij', iord, itype ) }{idspfun}

Return the index ($<$ \xtt{0}) of a splitting function weight table
stored internally in \qcdnum. 
\begin{tdeflist}[afun\ \ \ \ ]{-1mm} 
\item[\xtt{'pij'}] Name of the splitting function.
  Valid input strings are
  \begin{verbatim}
      PQQ, PQG, PGQ, PGG, PPL, PMI, PVA.
  \end{verbatim}
\item[\xtt{iord}] Select LO (\xtt{1}), NLO (\xtt{2}) or NNLO
  (\xtt{3}).
\item[\xtt{itype}] Select evolution type: un-polarised (\xtt{1}),
  polarised (\xtt{2}), fragmentation function~(\xtt{3}) or 
  custom (\xtt{4}).  
\end{tdeflist}

The index returned by \xtt{idspfun} is encoded as
\xtt{-(1000*itype+id)}, where \xtt{id} is
the internal table identifier. If the table does not exist, the function returns a value of \xtt{-1}.


\subcbox{call COPYWGT ( w, id1, id2, iadd ) }{copywgt}

Copy the contents of table \xtt{id1} to \xtt{id2}.
\begin{tdeflist}[ida\ \ \ \ \ ]{-1mm} 
\item[\xtt{w}] Store declared in the calling routine.
\item[\xtt{id1}] Input table identifier. One can copy a splitting function 
  weight table from internal \qcdnum\ memory to the store by setting 
  \compa{id1}{<}{0}. Valid identifiers are generated by
  \xtt{idspfun()}, as described above.
\item[\xtt{id2}] Output table identifier with \compa{id2}{\neq}{id1}.
  The output table type may
  be different from the input table type, see below.
\item[\xtt{iadd}] If set to \xtt{0} copy \xtt{id1} to \xtt{id2},
  if set to \xtt{+1} (\xtt{-1}) add (subtract) \xtt{id1} to (from)
  \xtt{id2}. 
\end{tdeflist}
For this routine---and for those described below---the output table
type can be different from the input table type, provided that this
does not lead to a loss of input information. Thus one can copy a
type-1 table to a type-3 table but not the other way around (fatal
error). Note that input splitting function weight tables are
all type-2.


\subcbox{call WCROSSW ( w, ida, idb, idc, iadd )}{wcrossw}

This routine generates a weight table for the convolution
of two kernels $K_a$ and $K_b$.
\siindex{multiple convolution}%
The weight table
is calculated with \eq{eq:wawb} from two input tables~$\ve{W}_{\! a}$
and~$\ve{W}_{\! b}$.
\begin{tdeflist}[ida\ \ \ \ \ ]{-1mm} 
\item[\xtt{w}] Store declared in the calling routine.
\item[\xtt{ida}] Table identifier containing the weights of kernel
  $K_a$.  When \compa{ida}{<}{0} one will access a splitting function
  weight table which is stored internally in \qcdnum. See
  \xtt{idspfun()} above for how to generate a valid splitting
  function identifier.
\item[\xtt{idb}] As above for the weights of kernel $K_b$.
\item[\xtt{idc}] Output table identifier. Cannot be set equal to
  \xtt{ida} or \xtt{idb}.
\item[\xtt{iadd}] If set to \xtt{0} store the result of the
  convolution in \xtt{idc}, if set to \xtt{+1} (\xtt{-1}) add (subtract)
  the result to (from) the contents of \xtt{idc}. 
\end{tdeflist}
The table types of \xtt{ida} and \xtt{idb} may be different, but the
type of \xtt{idc} must be such that it can contain either input table.
Thus if \xtt{ida} is type-2 ($x,\enef$) and
\xtt{idb} is type-3 $(x,\ms)$, then \xtt{idc} must be type-4
($x,\ms,\enef)$. The routine checks this.


\subcbox{call WTIMESF ( w, fun, id1, id2, iadd )}{wtimesf}
 
Multiply a weight table by a function of \ms\ and \enef\ and store the
result in another table.
\begin{tdeflist}[ida\ \ \ \ \ ]{-1mm} 
\item[\xtt{w}] Store declared in the calling routine.
\item[\xtt{fun}] User supplied double precision function \xtt{fun(iq,nf)} 
  declared \xtt{external} in the calling routine.
\item[\xtt{id1}] Input weight table identifier.
  It is possible to access a splitting function weight table by setting
  \compa{id1}{<}{0}. Valid identifiers can be obtained from
  \xtt{idspfun()} described above.
\item[\xtt{id2}] Identifier of the output table. It is allowed to have
  \compa{id1}{=}{id2} (in-place modification of a table), unless
  \xtt{id1} is a splitting function table. The table type of \xtt{id2}
  must be such that no information is lost. The routine checks this.   
\item[\xtt{iadd}] Store the result in \xtt{id2} in case 
  \compa{iadd}{=}{0} or add (subtract) the result to (from) \xtt{id2}
   in case \compa{iadd}{=}{+1} (\xtt{-1}).
\end{tdeflist}
The routine loops over \xtt{iq} and \xtt{nf} and calls
\xtt{fun(iq,nf)} with the following argument ranges,  
depending on the output table type:
\begin{center}
\begin{tabular}{clll}
type & \ \ variables      & \qquad \xtt{iq} range 
                          & \qquad \xtt{nf} range \\
\hline
  1 & \ \ $x$             & \qquad \ \ \xtt{1}--\xtt{1}
                          & \qquad \ \ \xtt{3}--\xtt{3}              \\
  2 & \ \ $x$, \enef      & \qquad \ \ \xtt{1}--\xtt{1} 
                          & \qquad \ \ \xtt{3}--\xtt{6}              \\
  3 & \ \ $x$, \ms        & \qquad \ \ \xtt{1}--\xtt{nq}
                          & \qquad \ \ \xtt{3}--\xtt{3}              \\
  4 & \ \ $x$, \ms, \enef & \qquad \ \ \xtt{1}--\xtt{nq}
                          & \qquad \ \ \xtt{3}--\xtt{6}   
\end{tabular}
\end{center}
With this routine one can, in combination with \xtt{wcrossw},
construct weight tables for combinations of convolution kernels, such
as those given in \eq{eq:cikmdef}. For instance, here is code that
generates a table for
\[
  C^{(2,1)}_{2,+} = C^{(0)}_{2,q} \otimes P^{(1)}_+ + 
                    C^{(1)}_{2,+} \otimes P^{(0)}_{qq} -
                    \be_0\, C^{(1)}_{2,+}.
\]
\begin{verbatim}
    external beta0   !beta function 
      ..
    call WcrossW ( w, idC2Q0, idSpfun('PPL',2,1), idC2P21,  0 )
    call WcrossW ( w, idC2P1, idSpfun('PQQ',1,1), idC2P21, +1 )
    call WtimesF ( w, beta0 , idC2P1            , idC2P21, -1 )
\end{verbatim}


\subcbox{call SETWPAR ( w, par, n ) }{setwpar}

Write extra information to the store, for instance quark masses or
other parameters that you may want to
dump to disk, together with the tables themselves.
\begin{tdeflist}[parxxx \ \ ]{-1mm} 
  \item[\xtt{w}] Store, partitioned by a previous call to \xtt{booktab}.
  \item[\xtt{par}] List of parameters to be written. Should be dimensioned
     to at least \xtt{par(n)} in the calling routine.
  \item[\xtt{n}] Number of items to be written up to a maximum of 
     \compa{miw0}{=}{20}. If necessary, one can change the value of 
     \xtt{miw0} in \xtt{qcdnum.inc} and recompile \qcdnum.
\end{tdeflist}
The parameters can be read back by a call to 
\subc{getwpar}\xtt{(w,par,n)}.


\subcbox{call TABDUMP ( w, lun, 'filename', 'key' ) }{tabdump}

Dump the store \xtt{w} to disk. Apart from the store, information is
written about the \qcdnum\ version, the $x$-\ms\ grid definition and
the current spline interpolation order. The \xtt{key} text string can
be used to stamp the file with a version number or other identifier.
The dump is unformatted so that the file cannot be exchanged
across machines.


\subcbox{call TABREAD ( w, nw, lun, 'filename', 'key', *nwords, *ierr ) }{tabread}

Read a store from disk into the array \xtt{w(nw)}. The size of the
store (in words) is returned in \xtt{nwords}. You will get a fatal
error message if \xtt{w(nw)} is not large enough to contain the store.
Note that the $x$ and \ms\ grids must have been defined before the
call to \xtt{tabread}. On exit, the error flag is set as follows
(non-zero means that nothing has been read in).
\begin{tdeflist}[\ \ \ \ ]{-1mm} 
\item[\xtt{0}] Store successfully read in.
\item[\xtt{1}] Read error or input file does not exist.
\item[\xtt{2}] File written by another \qcdnum\ version.
\item[\xtt{3}] Key mismatch.
\item[\xtt{4}] Incompatible $x$-\ms\ grid definition.
\end{tdeflist}
\Qcdnum\ insists that the key written on the file matches the key
entered as an argument to \xtt{tabread}.\footnote{Note that the key
matching is case insensitive and that leading and trailing blanks are
ignored.} Thus if, for instance, the key is set to a package name and
version number then the user of the package cannot read obsolete files written by
earlier versions, or read files
written by another package.
If you don't want to use keys, just enter an empty string as
a key in the calls to \xtt{tabdump} and \xtt{tabread}.
%


\newpage
\subsection{Convolution}\label{se:ustfconvol}

In \Ta{tab:cengine}
\begin{table}[tbh] 
  \caption{ Calls in the \qcdnum\ convolution engine.}
  \begin{center}
  \begin{tabular*}{1.0\textwidth}{l@{\extracolsep{\fill}}l}
  \\
  Subroutine or function & Description \\
  \hline
  \xtt{FCROSSK ( w, idw, iset, idf, ix, iq ) }
  & Convolution $x[f \otimes K]$                 \\
  \xtt{FCROSSF ( w, idw, iset, ida, idb, ix, iq ) }
  & Convolution $x[f_a \otimes f_b]$             \\
  \xtt{EFROMQQ ( qvec, *evec) }
  & Transform from $q,\bar{q}$ to $e^{\pm}$      \\
  \xtt{QQFROME ( evec, *qvec) }
  & Transform from $e^{\pm}$ to $q,\bar{q}$      \\
  \xtt{NFLAVOR ( iq ) }
  & Returns \enef                                \\
  \xtt{GETALFN ( iq, n, *ierr) }
  & Returns $(\as/2\pi)^n$                       \\
  \hline
  {\footnotesize Output arguments are pre-fixed with
   an asterisk (\xtt{*}).} &
  \end{tabular*} 
  \end{center}
  \label{tab:cengine}
  \dbindex{\iqcdnum}{list of subroutines}
\end{table}
we list the routines that can be used to build a structure function,
cross-section or parton luminosity at a grid point in $x$ and \ms. 

A convolution is always computed with the pdfs in
\qcdnum\ memory, that is, with the gluon density or with one of the
singlet/non-singlet quark densities $\ket{e^{\pm}}$ as defined in
\Se{se:decompose}. 
To translate a linear combination of quarks and anti-quarks to the
$\ket{e^{\pm}}$ basis, and \textit{vice versa}, the routines
\xtt{efromqq} and \xtt{qqfrome} are provided.

For convenience we show here again the indexing \eq{eq:isins} of
the singlet/non-singlet basis
\beq{eq:epmindex}
  \begin{array}{lllllllllllll}
  0&\ 1&2&3&4&5&6&\ 7&8&9&10&11&12\\
  \hline
  \mbstrut{5mm} g\ &\ \qsi\ &e^+_2\ &e^+_3\ &e^+_4\ &e^+_5\ &e^+_6
     &\ \qva\ &e^-_2\ &e^-_3\ &e^-_4\ &e^-_5\ &e^-_6
  \end{array}\raisebox{.7ex}{\ ,}
\eeq
and the indexing~\eq{eq:iqqbar} of the flavour basis
\beq{eq:qqbindex}
  \begin{array}{rrrrrrrrrrrrr}
  -6&-5&-4&-3&-2&-1&\ 0&\ 1&2&3&4&5&6\\
  \hline
  \mbstrut{5mm} \tbar &\ \bbar &\ \cbar &\ \sbar &\ \ubar &\ \dbar &\  g 
     &\ d &\ u &\ s &\ c &\ b &\ t
  \end{array}\raisebox{.7ex}{\ .}
\eeq
%


\subcbox{val = FCROSSK ( w, idw, iset, idf, ix, iq ) }{fcrossk}

Calculate the convolution $x[f \otimes K](x)$ at a grid point
in $x$ and \ms.
\begin{tdeflist}[ix\ iq\ \ \ \ ]{-1mm} 
\item[\xtt{w}] Store declared in the calling routine and previously
  filled with weights.
\item[\xtt{idw}] Identifier of a table in the store \xtt{w}.
\item[\xtt{iset}] Pdf set identifier [\xtt{1}--\xtt{9}].\footnote{
  Un-polarised~(\xtt{1}), polarised~(\xtt{2}), 
  fragmentation function~(\xtt{3}), custom~(\xtt{4}), or
  external~(\xtt{5}--\xtt{9}).} 
\item[\xtt{idf}]  Pdf identifier,
  indexed according to \eq{eq:epmindex}.
\item[\xtt{ix}, \xtt{iq}] Indices of an $x$-\ms\ grid point.
\end{tdeflist}
Splitting function tables cannot be directly accessed by this
routine; they should first be copied to the store by a call
to \xtt{copywgt}.


\subcbox{val = FCROSSF ( w, idw, iset, ida, idb, ix, iq ) }{fcrossf}

Calculate the convolution $x[f_a \otimes f_b](x)$ at a grid point
in $x$ and \ms.
\begin{tdeflist}[ix\ iq\ \ \ \ \ \ ]{-1mm} 
\item[\xtt{w}] Store declared in the calling routine and previously
  partitioned by \xtt{booktab}.
\item[\xtt{idw}] Identifier of a weight table, previously filled by
  a call to \xtt{makewx}.
\item[\xtt{iset}] Pdf set identifier [\xtt{1}--\xtt{9}].   
\item[\xtt{ida}, \xtt{idb}] Pdf identifiers,
 indexed according to \eq{eq:epmindex}.
\item[\xtt{ix}, \xtt{iq}] Indices of an $x$-\ms\ grid point.
\end{tdeflist}
A convolution of a linear combination of pdfs must be calculated
as a sum of pair-wise convolutions, with each term computed by
\xtt{fcrossf}. Note that this is much easier done
with the fast routines described in \Se{se:ustfast}. 


\subcbox{call EFROMQQ ( qvec, *evec, nf ) }{efromqq}

Transform the coefficients of a linear combination of quarks and
anti-quarks from the flavour basis to the singlet/non-singlet basis
as described in \Se{se:decompose}.  
\begin{tdeflist}[qvec\ \ \ ]{-1mm} 
\item[\xtt{qvec}] Input array, dimensioned \xtt{qvec(-6:6)}, filled
  with the coefficients of a linear combination of quarks and
  anti-quarks and indexed according to \eq{eq:qqbindex}.
\item[\xtt{evec}] Output array, dimensioned to \xtt{evec(12)}, filled
  with the coefficients written on the singlet/non-singlet basis,
  indexed according to \eq{eq:epmindex}.
\item[\xtt{nf}] Active number of flavours. This parameter is needed
  to construct the appropriate $2\enef \times 2\enef$ transformation
  matrix that acts on \xtt{qvec(-nf:nf)}.
\end{tdeflist}

Thus if a linear combination of quarks and anti-quarks is written as
\beq{eq:qqfrome}
  \ket{p} = \sum_{i=1}^{\enef} (\al_i \ket{q_i} + \be_i
    \ket{\qbar_i}) = \sum_{i=1}^{\enef} ( d^+_i \ket{e^+_i} + d^-_i
  \ket{e^-_i} )
\eeq
and $\al_i$ and $\be_i$ are stored in the input vector \xtt{qvec},
then the coefficients $d^{\pm}_i$ are returned in the output vector
\xtt{evec}.


\subcbox{call QQFROME ( evec, *qvec, nf ) }{qqfrome}

Transform the coefficients of a linear combination of basis vectors
from the singlet/non-singlet basis to the flavour basis.  The arguments
are as for \xtt{efromqq}. 


\subcbox{nf = NFLAVOR ( iq ) }{nflavor}

Returns the number of active flavours at the grid point \xtt{iq}.
Note that this number is~(\xtt{4},\xtt{5},\xtt{6}) and not
(\xtt{3},\xtt{4},\xtt{5}) at the
thresholds (\xtt{iqc},\xtt{iqb},\xtt{iqt}).


\subcbox{as = GETALFN ( iq, n, *ierr ) }{getalfn}

Returns the value of $(\as/2\pi)^n$ at the factorisation scale \Fs.
Here \as\ is computed from
the Taylor expansion \eq{eq:asexpansion}, appropriately truncated
depending on the current value of the perturbative order (\xtt{iord}).
Because the truncation
\siindex{truncation prescription}%
is different for the pdfs (\Se{se:rscale}) and the structure
functions (\Ap{se:fscale}), the value of
\xtt{n} must be set as follows.
\begin{itemize}
\item If the convolution at $(\rm{LO},\rm{NLO},\rm{NNLO})$ should be
  multiplied by $(\as,\as^2,\as^3)$ then you set \xtt{n} =
  (\xtt{1},\xtt{2},\xtt{3}) in the call to \xtt{getalfn}.
\item If the convolution at $(\rm{LO},\rm{NLO},\rm{NNLO})$ should be
  multiplied by $(1,\as,\as^2)$ then you set \xtt{n} =
  (\xtt{0},\xtt{-1},\xtt{-2}) in the call to \xtt{getalfn}.
\item If \compa{n}{>}{iord},
  the value of  $(\as/2\pi)^n$ is calculated at
  \Rs, instead of at \Fs. (This is already the case for
  \compa{n}{=}{iord}, see \Se{se:rscale}.)
\end{itemize}

To have access to the NNLO discontinuities at the thresholds, one can set
\xtt{iq} positive (includes discontinuity) or negative (does not include
discontinuity), thus:
\begin{verbatim}
    call getalfn (  iqcharm, n, ierr )    !result for nf = 4
    call getalfn ( -iqcharm, n, ierr)     !result for nf = 3
\end{verbatim}
In other words, by preceding \xtt{iq} with a minus sign one effectively
changes the \qcdnum\ default $\enef = (4,5,6)$ at the thresholds to
the alternative $\enef = (3,4,5)$.
\dbindex{number of active flavours \enef}{value at threshold}%
When \xtt{iq} is close to or below the
value of $\La^2$, then \compa{ierr}{=}{1} and \xtt{getalfn} returns the
\xtt{null} value. This also happens when \xtt{iq} is outside the grid
boundaries (\compa{ierr}{=}{2}).

Note that the \qcdnum\ expansion parameter is $\as/2\pi$ but that many
convolution kernels found in the literature are defined for an
expansion in $\as/4\pi$, in which case one must account for the
appropriate factors of $2^n$ somewhere in the calculation.  


\subsection{Interpolation}\label{se:ustfinterp}

With the routines presented above one can write a function
\xtt{stfun(ix,iq)} that returns the value of a structure function or
cross section at a grid point in $x$ and~\ms. The routine
\xtt{stfunxq} then takes care of the interpolation to any value of $x$
and \ms.  This interpolation is done on a $k \times 3$ mesh around the
interpolation point, where $k$ is set to the current spline
interpolation order (2 = linear, 3 = quadratic). Thus $3k$ 
functions have to be computed for each interpolation which becomes
inefficient if there are interpolations with overlapping meshes.  By
processing lists of interpolation points, instead of each point
individually, redundant calculations are avoided
which can lead to considerable gains in computing time. In other
words, \xtt{stfunxq} should not be called in a loop over interpolation
points but be given the \emph{list} of points.


\subcbox{ call STFUNXQ ( stfun, x, qmu2, stf, n, ichk ) }{stfunxq}

Interpolate the function \xtt{stfun(ix,iq)} to a list of $x$ and \ms\
values.
\begin{tdeflist}[xxxqmuu\ \ \ ]{-1mm} 
\item[\xtt{stfun}] Double precision function, declared external in the
  calling routine, that returns the value of a structure function or
  cross-section at (\xtt{ix},\xtt{iq}).
\item[\xtt{x}, \xtt{qmu2}] List of interpolation points, dimensioned
  to at least \xtt{n} in the calling routine.
\item[\xtt{stf}] Contains, on exit, the list of interpolated results.
\item[\xtt{n}] Number of items in \xtt{x}, \xtt{qmu2} and \xtt{stf}.
\item[\xtt{ichk}] If set to \xtt{0} the routine returns a \xtt{null}
  value if $x$ or \ms\ are outside the boundaries of the grid;
  if set non-zero it will insist that all interpolation points are
  inside the grid boundaries.
\end{tdeflist}
Note that the interpolation is done in \ms\ and not in \qsq. You have
to keep track yourself of the relation between these two scales.


\subsection{Fast Computation}\label{se:ustfast}

As already remarked above, the routines provided up to now are fine
for prototyping but are slow because there is quite a lot of
overhead when the calculation is repeated at more than one
interpolation point. Here we describe a set of routines that does
optimised bulk calculations on selected points in the $x$-\ms\ grid.
With these fast routines one can easily gain one or two orders of
magnitude in speed.
To make the calculation flexible it is broken down into small steps
where intermediate results are stored into scratch buffers
(by default, the fast engine generates 5~scratch buffers but one can
have more, if necessary).
An optimised calculation proceeds as follows.
\begin{enumerate}
\item Pass a list of interpolation points in $x$ and \ms\ to a
  \qcdnum\ routine that determines which grid points will be occupied in
  the course of the calculation;
\item Store a pdf or a linear combination of pdfs in a scratch
  buffer;
\item Convolve the pdf with a convolution kernel or a perturbative
  expansion of kernels;
\item Multiply the convolution by a function of $x$ and \ms,
  for instance  by some power of~\as\ or by some kinematic factor;
\item Accumulate the cross section or structure
  function in a final buffer;
\item Pass this buffer to an interpolation routine to get
  a list of interpolated results.
\end{enumerate}
The list of subroutines is given in \Ta{tab:fastcvol}.
\begin{table}[bth] 
  \caption{ Fast convolution engine.}
  \begin{center}
  \begin{tabular*}{0.95\textwidth}{l@{\extracolsep{\fill}}l}
  \\
  Subroutine or function & Description            \\
  \hline 
  \xtt{FASTINI ( x, qmu2, n, ichk )}
  & Pass list of $x$ and \ms\ values              \\
  \xtt{FASTCLR ( id )}
  & Clear buffer                                  \\
  \xtt{FASTEPM ( iset, idf, idout )}
  & Store \ket{g,\epm} in a scratch buffer        \\
  \xtt{FASTSNS ( iset, pdf, isel, idout )}
  & Store singlet/non-singlet component           \\
  \xtt{FASTSUM ( iset, coef, idout )}
  & Store weighted sum of \ket{\epm}              \\
  \xtt{FASTFXK ( w, idw, idf, idout )}
  & Convolution $x[f \otimes K](x)$               \\
  \xtt{FASTFXF ( w, idw, ida, idb, idout )}
  & Convolution $x[f_a \otimes f_b](x)$           \\
  \xtt{FASTKIN ( id, fun )}
  & Scale by a kinematic factor                   \\
  \xtt{FASTCPY ( idin, idout, iadd )}
  & Copy or accumulate result                     \\
  \xtt{FASTFXQ ( id, *f, n )\ }
  & Interpolation                                 \\
  \hline
  {\footnotesize Output arguments are pre-fixed with
   an asterisk (\xtt{*}).} &
  \end{tabular*} 
  \end{center}
  \label{tab:fastcvol}
  \dbindex{\iqcdnum}{list of subroutines}
\end{table}
In principle, the output buffer of any fast routine can serve
as the input buffer of any other fast routine.
There is, however, a little complication related to the amount
of information stored in a buffer. For interpolation purposes,
it is sufficient to store results only at the mesh points; such
a buffer is called~\emph{sparse}. A~convolution routine, on the other
hand, does not only need the values at the mesh points~$x_i$,
but also the values at all points $x_j > x_i$. An input buffer
with such a storage pattern is called \emph{dense}; 
a dense buffer is of course more expensive
to generate than a sparse buffer. Usually one does not have
to worry about sparse and dense buffers, because \qcdnum\ 
has reasonable defaults
on what kind of buffer is accepted as input, and what kind of
buffer is generated on output. One can aways
override the output default and force a routine to generate a
dense or a sparse buffer, as needed.

The pdf set parameter \xtt{iset} in the routines
\xtt{fastepm}, \xtt{fastsns} and \xtt{fastsum} selects
the pdf set, namely,
un-polarised~(\xtt{1}), polarised~(\xtt{2}), 
fragmentation function~(\xtt{3}), custom~(\xtt{4}), or
external~(\xtt{5}--\xtt{9}). 


\subfbox{call FASTINI ( x, qmu2, n, ichk ) }{fastini}

Pass a list of interpolation points and, at the first call,
generate the set of scratch buffers.
\begin{tdeflist}[xxxqmuu\ \ \ ]{-1mm} 
\item[\xtt{x}] Array, dimensioned to at least \xtt{n} in the calling
  routine, filled with $x$ values.
\item[\xtt{qmu2}] As above, but for \ms\ (not \qsq).
\item[\xtt{n}] Number of entries in \xtt{x} and \xtt{qmu2}.
\item[\xtt{ichk}] If non-zero, \xtt{fastini} insists that all
  $x$ and \ms\ are within the grid boundaries.
\end{tdeflist}
By default, 5~scratch buffers (\compa{id}{=}{1}--\xtt{5}) are
generated at the first call (or cleared if they exist). This number
can be changed by calling~\xtt{setint('ntab',ival)}
prior to~\xtt{fastini}. One will get a fatal error if there is not
enough space for the scratch buffers,
in which case one has to increase the value of \xtt{nwf0} in
\xtt{qcdnum.inc}, and recompile \qcdnum.


\subfbox{call FASTCLR ( id ) }{fastclr}

Clear a scratch buffer. Setting \compa{id}{=}{0} will clear all
buffers.


\subfbox{call FASTEPM ( iset, idf, idout ) }{fastepm}

Copy the gluon density or one of the basis pdfs
\ket{\epm} to a scratch table.

\begin{tdeflist}[xxxqmuu\ \ \ ]{-1mm}
  \item[\xtt{iset}] Input pdf set identifier [\xtt{1}--\xtt{9}].
  \item[\xtt{idf}] Pdf identifier [\xtt{0}--\xtt{12}], 
    indexed according to \eq{eq:epmindex}.
  \item[\xtt{idout}] Output scratch table identifier 
    [\xtt{1}--\xtt{5}].    
\end{tdeflist}
By default, \xtt{fastepm} generates a dense buffer; a sparse buffer is generated when you pre-pend the output identifier with a minus sign.
\begin{verbatim}
        call fastEpm(1, 0,  1)      !dense table output
        call fastEpm(1, 0, -1)      !sparse table output
\end{verbatim}


\subfbox{call FASTSNS ( iset, pdf, isel, idout ) }{fastsns}

Decompose a given linear combination of quarks and anti-quarks
into singlet and non-singlet components and copy a specific 
component to a scratch buffer.
\begin{tdeflist}[xxxqmuu\ \ \ ]{-1mm}
  \item[\xtt{iset}] Input pdf set identifier [\xtt{1}--\xtt{9}].
  \item[\xtt{pdf}] Input array, dimensioned \xtt{pdf(-6:6)}, 
  filled with the coefficients of a linear combination of quarks and
  anti-quarks and indexed according to \eq{eq:qqbindex}.
  \item[\xtt{isel}] Selection flag [\xtt{0}--\xtt{7}], see below.
  \item[\xtt{idout}] Output scratch table identifier 
    [\xtt{1}--\xtt{5}].    
\end{tdeflist}
The \xtt{isel} flag selects the gluon density (\xtt{0}),
the singlet component \qsi\ (\xtt{1}), 
the non-singlet component \qnsp\ (\xtt{2}),
the valence component \qva\ (\xtt{3}),
the non-singlet component \qnsm\ (\xtt{4}), the sum
$\qva+\qnsm$ (\xtt{5}),
all non-singlets $\qva + \qnsm + \qnsp$ (\xtt{6}) or
all quarks (\xtt{7}).

Note that the singlet is weighted by an appropriate \enef\
dependent factor, for instance by the average square of the quark
charges in case \xtt{pdf} corresponds to the charge weighted sum
of quarks and anti-quarks. Note also that the gluon density is
multiplied by the same factor, as is required in structure function
calculations, see \Ap{se:zeromasstf}. Setting
\compa{isel}{=}{0} or \xtt{1} is thus not the same as calling
\xtt{fastepm} for the identifiers \xtt{0} or \xtt{1}.

By default, \xtt{fastsns} generates a dense buffer; a sparse buffer is generated when you pre-pend the output identifier with a minus sign.
%
%
\subfbox{call FASTSUM ( iset, coef, idout ) }{fastsum}

Copy a linear combination of basis pdfs 
\ket{\epm} to a scratch table.

\begin{tdeflist}[xxxqmuu\ \ \ ]{-1mm}
\item[\xtt{iset}] Input pdf set identifier [\xtt{1}--\xtt{9}].
\item[\xtt{coef}] Array of coefficients dimensioned
  \xtt{coef(0:12,3:6)} in the calling routine.
\item[\xtt{idout}] Output scratch table identifier 
  [\xtt{1}--\xtt{5}].    
\end{tdeflist}
The array \xtt{coef(i,nf)} is indexed according to \eq{eq:epmindex}.
Here is code that fills \xtt{coef} by transforming a
set of quark coefficients from flavour space to singlet/non-singlet
space:
\begin{verbatim}
        dimension qvec(-6:6), coef(0:12,3:6)
        do nf = 3,6
          coef(0,nf) = 0.D0                   !gluon coefficient
          call efromqq(qvec, coef(1,nf), nf)  !quark coefficients
        enddo
\end{verbatim}
By masking out coefficients, one can copy the singlet component,
or various combinations of non-singlets;
this is exactly what \xtt{fastsns} does.
To copy the gluon distribution, one must set all coefficients to
zero, except \xtt{coef(0,nf)}.

By default, \xtt{fastsum} generates a dense buffer; a sparse buffer is generated when you pre-pend the output identifier with a minus sign.


\subfbox{call FASTFXK ( w, idw, idf, idout ) }{fastfxk}

Calculate the convolution $x[f \otimes K](x)$ at all selected grid points.

\begin{tdeflist}[xxxqmuu\ \ \ ]{-1mm}
\item[\xtt{w}] Store, declared in the calling routine and previously
  filled with weights.
\item[\xtt{idw}] Set of weight identifiers, declared \xtt{idw(4)} in the
  calling routine, see below. 
\item[\xtt{idf}] Input scratch buffer, previously
  filled by \xtt{fastepm}, \xtt{fastsns} or \xtt{fastsum}.
\item[\xtt{idout}] Output scratch table with \compa{idout}{\neq}{idf}.
\end{tdeflist}
One can either convolve with a given weight table or with a
perturbative expansion of weight tables, depending on what one puts in
the array \xtt{idw}:
\begin{enumerate}
\item To convolve with a given weight table, set \xtt{idw(1)} to the
  identifier of that weight table and set
  \xtt{idw(2)}, \xtt{idw(3)} and \xtt{idw(4)} to zero;
\item To convolve with a perturbative expansion, store the (LO,NLO,NNLO)
  weight table identifiers in \xtt{idw(1)},
  \xtt{idw(2)} and \xtt{idw(3)}. Set the identifier to zero if no such
  table exists, as is the case for \Fell\ at LO, for instance.  Declare
  in \xtt{idw(4)} the leading power of \as, that is, multiply
  (LO,NLO,NNLO) by $(1,\as,\as^2)$ if \compa{idw(4)}{=}{0} and by
  $(\as,\as^2,\as^3)$ if \compa{idw(4)}{=}{1}. Note that the perturbative
  expansion is summed up to the current perturbative order, as 
  defined by an upstream call to \xtt{setord}. 
\end{enumerate}
The routine only accepts a dense buffer as input (otherwise fatal error)
and will, by default, generate a sparse buffer as output. If you
pre-pend the output identifier with a minus sign, the output buffer
will be dense. In this way, the output table can serve as an input
to another convolution which allows one to calculate multiple
convolutions in a chain. For example,\footnote{It is more
efficient, however, to first calculate with \xtt{wcrossw} a weight
table for $K_3 = K_1 \otimes K_2$, and use that table
to convolve $K_3$ with $f$.}
\siindex{multiple convolution}%
\begin{verbatim}
        call fastSum( 1, coef,  1 )      ! 1 = f
        call fastFxK( w, idK1,  1, -2 )  ! 2 = f * K1
        call fastFxK( w, idK2,  2,  3 )  ! 3 = f * K1 * K2
\end{verbatim} 


\subfbox{call FASTFXF ( w, idx, ida, idb, idout ) }{fastfxf}

Calculate the convolution $x[f_a \otimes f_b](x)$ at all
selected grid points.

\begin{tdeflist}[xxxqmuu\ \ \ ]{-1mm}
\item[\xtt{w}] Store, declared in the calling routine and previously
  partitioned by \xtt{booktab}.
\item[\xtt{idx}] Identifier of a weight table, previously filled
  by a call to \xtt{makewtx}. 
\item[\xtt{ida}, \xtt{idb}] Identifiers of input scratch tables.
  It is allowed to have \compa{ida}{=}{idb}.
\item[\xtt{idout}] Output scratch table with \compa{idout}{\neq}{ida}
  or \xtt{idb}.
\end{tdeflist}
As above, the routine accepts only dense buffers as input, and generates
a sparse buffer as output, unless the output identifier is pre-pended
by a minus sign, thus,
\siindex{multiple convolution}%
\begin{verbatim}
      call fastSum( 1, coefa, 1 )             ! 1 = fa
      call fastSum( 1, coefb, 2 )             ! 2 = fb
      call fastFxF( w, idwX,  1,  2, -3 )     ! 3 = fa * fb
      call fastFxK( w, idwK,  3,  4 )         ! 4 = fa * fb * K
\end{verbatim}


\subfbox{call FASTKIN ( id, fun ) }{fastkin}

Multiply the contents of a scratch table by a kinematic factor.

\begin{tdeflist}[xxxqmuu\ \ \ ]{-1mm}
\item[\xtt{id}]  Identifier of the input scratch table.
\item[\xtt{fun}] Double precision function,
  declared \xtt{external} in the calling routine.
\end{tdeflist}
The routine loops over the selected grid points and calls the user
supplied function \xtt{fun} that should return the kinematic
factor. The syntax of \xtt{fun} is
\begin{verbatim}
          double precision function fun ( ix, iq, nf, ithresh )
\end{verbatim}
\begin{tdeflist}[xxxqmuu\ \ ]{-1mm}
  \item[\xtt{ix,iq}] Grid point indices.
  \item[\xtt{nf}] Number of flavors at \xtt{iq}. This number is
    bi-valued at the thresholds so that at the charm threshold,
    for instance, \xtt{nf} can be either \xtt{3} or \xtt{4}.\footnote{
    The reader may wonder when
    \qcdnum\ returns the value \xtt{3}, and when the value \xtt{4}.
    This depends on the interpolation point \ms\ to which \xtt{iq}
    is associated: if \ms\ is below (above) $\msc$, then
    \compa{nf}{=}{3} (\xtt{4}).}
    \dbindex{number of active flavours \enef}{value at threshold}%
  \item[\xtt{ithresh}] Set to \xtt{0} if \xtt{iq} is not at a threshold
    and to \xtt{+1} (\xtt{-1}) if \xtt{iq} is at a threshold with the
    upper (lower) number of flavours. This variable can be used to
    take NNLO discontinuities into account, as is shown in the example
    below.
\end{tdeflist}
\begin{verbatim}
      double precision function fkin(ix,iq,nf,ithresh)
        ..
      if(ithresh.ge.0) then
        alfas = getalfn( iq,1,ierr)    !alfas/2pi with discontinuity
      else
        alfas = getalfn(-iq,1,ierr)    !without discontinuity 
      endif
        ..    
\end{verbatim}


\subfbox{call FASTCPY ( idin, idout, iadd ) }{fastcpy}

Copy or accumulate a result in an output buffer.

\begin{tdeflist}[xxxqmuu\ \ \ ]{-1mm}
  \item[\xtt{idin}] Identifier of the input scratch table.
  \item[\xtt{idout}] Identifier of the output scratch table with
    \compa{idout}{\neq}{idin}.
  \item[\xtt{iadd}] Store (\xtt{0}), add (\xtt{1}) or subtract
    (\xtt{-1}) the result to \xtt{idout}.
\end{tdeflist}
The type of output buffer (sparse or dense) is the same as that of
the input buffer, except
that once you have used a sparse input buffer, the output buffer
will be flagged a sparse and will remain so until you set
\compa{iadd}{=}{0} to start a new accumulation in \xtt{idout}.
 

\subfbox{call FASTFXQ ( id, *f, n ) }{fastfxq}

Interpolate the contents of \xtt{id} to the list of $x$ and \ms\
values that was previously passed to \qcdnum\ by the call to
\xtt{fastini}.

\begin{tdeflist}[xxxqmuu\ \ \ ]{-1mm}
\item[\xtt{id}] Identifier of an input scratch buffer.
\item[\xtt{f}] Array dimensioned to at least \xtt{n} in the calling
  routine that will contain, on exit, the interpolated values.
\item[\xtt{n}] Number of interpolations requested. 
\end{tdeflist}
The routine works through the list of interpolation points given in
the call to \xtt{fastini} and exits when it reaches the end of that
list or when the number of interpolations is equal to \xtt{n},
whatever happens first. A dense input buffer is allowed,
but wasteful since it contains a lot of 
information that is not used by \xtt{fastfxq}.


\subsection{Custom Evolution}\label{se:userevol}

\siindex{custom evolution|(}%
In \Se{se:subweight} we have described the \xtt{fillwt} routine to
generate weight tables (and pdf tables) for the evolution of
un-polarised pdfs (\compa{itype}{=}{1}), polarised pdfs (\xtt{2}), and
fragmentation functions (\xtt{3}). But \qcdnum\ can handle yet another
\siindex{pdf type, pdf set}%
type of evolution, with user-defined evolution kernels
(custom evolution, \compa{itype}{=}{4}).
For this one has to provide a
subroutine, described below, that generates the weight tables.  This
routine is passed to \qcdnum\ as an argument of the custom weight
filling routine \subr{fillwc}, after which the custom evolution
becomes available by switching to \compa{itype}{=}{4}.
\begin{verbatim}
       external mysub, func
         ..
       call fillwc( mysub, idmin, idmax, nwords )
         ..
       call evolfg( 4, func, def, iq0, epsi )         
         ..
\end{verbatim}
In \Fi{fig:userevol},
%
\begin{figure}[p]
\footnotesize
\begin{verbatim}
C     -------------------------------------------------
      subroutine myweight(w,nw,nwords,idpij,mxord,idum)
C     -------------------------------------------------

C--   w           (in)   qcdnum store passed by reference
C--   nw          (in)   number of words available
C--   nwords      (out)  number of words used < 0 not enough space
C--   idpij       (out)  list of Pij table identifiers
C--   mxord       (out)  maximum perturbative order LO,NLO,NNLO
C--   idum               not used at present

      implicit double precision (a-h,o-z)
           
      dimension w(*), idpij(7,3), itypes(4)
          
      external AChi
      external PQQR, PQQS, PQQD                         ! PQQ
      external PQGA, PGQA                               ! PQG, PGQ
      external PGGA, PGGR, PGGS, PGGD                   ! PGG
      
      call setUmsg('myweight')   !s/r name for error messages
C-1   Max perturbative order
      mxord = 1
C-2   Partition
      itypes(1) = 2
      itypes(2) = 2 
      itypes(3) = 0
      itypes(4) = 0
      call BookTab(w,nw,itypes,nwords)
C-3   Not enough space       
      if(nwords.le.0) return
C-4   Assign table indices
      idPij(1,1)   =  101                               ! PQQ
      idPij(2,1)   =  201                               ! PQG
      idPij(3,1)   =  102                               ! PGQ
      idPij(4,1)   =  202                               ! PGG
      idPij(5,1)   =  101                               ! PPL
      idPij(6,1)   =  101                               ! PMI
      idPij(7,1)   =  101                               ! PVA
C-5   Fill tables      
      call MakeWRS(w, idPij(1,1), PQQR, PQQS, AChi, 0)
      call MakeWtD(w, idPij(1,1), PQQD, AChi)
      call MakeWtA(w, idPij(2,1), PQGA, AChi)
      call MakeWtA(w, idPij(3,1), PGQA, AChi)
      call MakeWtA(w, idPij(4,1), PGGA, AChi)
      call MakeWRS(w, idPij(4,1), PGGR, PGGS, AChi, 0)
      call MakeWtD(w, idPij(4,1), PGGD, AChi)
C--   Done!            
      call clrUmsg         !clear s/r name for error messages
      return
      end
\end{verbatim}
\caption{\footnotesize Subroutine that
generates weight tables of user-given evolution kernels
(LO only, in this example).
The subroutine is passed to \qcdnum\ via the routine \xtt{fillwc},
as is described in the text.}
\label{fig:userevol} 
\end{figure}
%
we show the listing of a custom weight routine (called \xtt{myweight})
that creates the weight tables of, in fact, un-polarised splitting
functions in LO, see \Ap{app:singular}.  The arguments of such a
subroutine are as follows.
\begin{verbatim}
          subroutine mysub( w, nw, nwords, idpij, mxord, idum )
          implicit double precision (a-h,o-z)
          dimension w(*), idPij(7,3)
          ..      
\end{verbatim}
\begin{tdeflist}[xxxqmuu\ \ ]{-1mm}
\item[\xtt{w}]      The \qcdnum\ store (passed by reference);
\item[\xtt{nw}]     The number of words available in the store
                    (input, passed  by \qcdnum); 
\item[\xtt{nwords}] Number of words used by the tables (output);
\item[\xtt{idpij}]  List of weight table identifiers (output);
\item[\xtt{mxord}]  Maximum perturbative order supported by the
                    tables (output);
\item[\xtt{idum}]   Not used at present.                                                         
\end{tdeflist}
The body of the code in \Fi{fig:userevol} shows the steps to be
taken in a custom weight routine.
\begin{enumerate}
\item Set the maximum order of the custom evolution, 
      here \compa{mxord}{=}{1};
\item Partition the store into tables by a call to \xtt{booktab}.
      Here are booked two type-1 and two type-2 tables;
\item Branch-out if there is not enough space in the store, as
      is signalled by a negative value of \xtt{nwords} returned
      by \xtt{booktab};
\item Return in \xtt{idPij(id,iord)} the identifiers of
      the various splitting function tables. The first index runs
      as follows
      \[
      \begin{array}{ccccccc}
      1 & 2 & 3 & 4 & 5 & 6 & 7\\
      \hline
      \mbstrut{5mm} P_{\rm qq} & P_{\rm qg} & P_{\rm gq}  
      & P_{\rm gg} & P_+  & P_-  & P_{\rm v}  
      \end{array}
      \]
      There are only LO splitting functions in the example,
      with the non-singlet splitting functions all being equal
      to $P_{\rm qq}$ (table identifier \xtt{101});
\item Generate the weight tables by calls to \xtt{makewta},
      \xtt{makewtb}, \xtt{makewrs}, and \xtt{makewd}, as is
      described in \Se{se:ustfwts}. The calls in \Fi{fig:userevol}
      actually accommodate the splitting functions
      given in \eq{eq:losplit}.     
\end{enumerate}
One can have only one set of custom weight tables in memory; a second
call to \xtt{fillwc} will result in a fatal error message.

A custom evolution must obey the DGLAP evolution equations
\eq{eq:dglapsg} and \eq{eq:dglapns} which implies that the evolution
must properly split into singlet/gluon and non-singlet parts. 
We remind the reader that the evolution kernels in \qcdnum\ are defined
by convolution with parton \emph{number} densities, and not parton
momentum densities. These kernels can depend on~$x$, \enef\
and \ms\ and can be stored in tables of types-1, 2, 3 or 4. 
Note, however, that the \ms\ dependence
via \as\ is taken care of in the evolution routine \xtt{evolsg}, through
multiplication of the N$^{\ell}$LO tables by $(\as/2\pi)^{\ell+1}$.
\siindex{custom evolution|)}%


\subsection{Error Messages in Add-On Packages}\label{se:packagerr}

When one writes an add-on package, the problem arises that \qcdnum\
error messages will be labelled with the name of the \qcdnum\ routine
and not with that of the package routine. This can of course become
confusing for the user who should be aware only of the
package routines, and not what is inside.

One solution is that the package catches errors before \qcdnum\ does,
but this would duplicate a good checking mechanism which is already in
place. An easier solution is to pass a string to \qcdnum\ which
contains the name of the package routine so that it will be printed
together with the error message. For this, the routines \xtt{setUmsg} and
\xtt{clrUmsg} are provided. For instance one of the first calls in the
\xtt{zmfillw} routine of the~\zmstf\ package is
\begin{verbatim}
       call setUmsg('ZMFILLW')
\end{verbatim}
so that, upon error, the user gets additional information: 
{\footnotesize
\begin{verbatim}
       ------------------------------------------------------
       Error in BOOKTAB ( W, NW, ITYPES, NT, NWDS ) ---> STOP
       ------------------------------------------------------
       No x-grid available                
       Please call GXMAKE       

       BOOKTAB was called by ZMFILLW
\end{verbatim}}                
The last call in \xtt{zmfillw} is
\begin{verbatim}
       call clrUmsg
\end{verbatim}
that wipes the additional message. This is important because downstream
\qcdnum\ errors would otherwise appear to have always come from
\xtt{zmfillw}.

%% file: sections/ack.tex
\section{Acknowledgements} \label{se:ack}

I am of course indebted to the original authors of \qcdnum, in
particular to M.~Virchaux who introduced me to the program in
1991.\footnote{It was sad to hear that Marc Virchaux passed away in
  November 2004.} I thank M.~Cooper-Sarkar for using preliminary
versions of \qcdnum17\ in her QCD fits and
providing important feedback during the development phase of the
present version. I greatly benefited from the many clarifying
discussions with A.~Vogt and thank him for the code of the NNLO
splitting and coefficient functions. I am grateful to him and
to M.~Cooper-Sarkar, E.~Laenen and R.~Thorne for comments on
the manuscript. This work is part of the research programme of
the Foundation for Fundamental Research on Matter (FOM), which
is financially supported by the Netherlands Organisation for
Scientific Research (NWO).

%% file: sections/appendix.tex

\section{Singularities}\label{app:singular}

In this appendix we denote by $f(x)$ a parton \emph{momentum} density and not a number density. In terms of $f$ the convolution integrals in
the evolution equations read
\beq{eq:genconv}
  I(x) = \int_x^1 \der z\;
  P(z)\; f\left(\frac{x}{z}\right).
\eeq
The LO splitting matrix $P_{ij}^{(0)}$ in \eq{eq:pseries} is written as,
in the notation of \eq{eq:pqqfun},
\beq{eq:splarrays}
 \begin{pmatrix}
  \,P_{\rmq \rmq}^{(0)}   & P_{\rmq \rmg}^{(0)}\, \\
  \,P_{\rmg \rmq}^{(0)}   & P_{\rmg \rmg}^{(0)} \mbstrut{6mm}
  \end{pmatrix}
  =
  \begin{pmatrix}
  \,P_{\rmq \rmq}^{(0)}   & 2\enef P_{\rmq_i \rmg}^{(0)}\, \\
  \,P_{\rmg \rmq_i}^{(0)} & P_{\rmg \rmg}^{(0)} \mbstrut{6mm}
  \end{pmatrix}
  .
\eeq
%
%

The LO un-polarised splitting functions are given by
\dbindex{splitting functions}{at leading order}%
\bea{eq:losplit}
  P_{\rm qq}^{(0)}(x) & = & \frac{4}{3} \left[ \frac{1+x^2}{(1-x)_+}
  + \frac{3}{2} \, \delta(1-x) \right] \nonumber \\
  P_{\rmq_i \rmg}^{(0)}(x) & = & 
  \frac{1}{2} \left[ x^2 + (1-x)^2 \right]
  \nonumber \\
  P_{\rmg \rmq_i}^{(0)}(x) & = & \frac{4}{3} \left[
  \frac{1+(1-x)^2}{x} \right] \nonumber \\
  P_{\rm gg}^{(0)}(x) & = & 
  6 \left[ \frac{x}{(1-x)_+} + \frac{1-x}{x} + x(1-x)
  + \left( \frac{11}{12} - \frac{\enef}{18} \right) \delta(1-x)
  \right].
\eea
For the time-like evolution of fragmentation functions, the
splitting functions $P_{\rmq_i \rmg}^{(0)}$
and~$P_{\rmg \rmq_i}^{(0)}$ are exchanged 
in~\eq{eq:splarrays}~\cite{ref:nasonwebber}.
\siindex{time-like evolution}%
The `+' prescription in \eq{eq:losplit} is defined by
\dbindex{splitting functions}{singularities in|(}%
\beq{eq:plusdef1}
  [f(x)]_+ = f(x) - \delta(1-x) \int_0^1 f(z) \der z
\eeq
so that
\beq{eq:plusdef2}
  \int_x^1 f(z) [g(z)]_+\; \der z = \int_x^1 \left[ f(z) - f(1)
  \right] g(z)\; \der z - f(1) \int_0^x g(z)\; \der z.
\eeq
For reference we give the expressions for $I_{\rm qq}$ and $I_{\rm
  gg}$ obtained from \eq{eq:losplit} and \eq{eq:plusdef1} 
\bea{eq:loconvol}
  I_{\rm qq}^{(0)}(x) & = & \frac{4}{3} \int_x^1\der z\;
  \frac{1}{1-z}\left[ (1+z^2)f\left(\frac{x}{z}\right)-2f(x)\right] +  
  \frac{4}{3}\; f(x) \left[ \frac{3}{2} + 2\ln(1-x)\right]
  \nonumber \\
  I_{\rm gg}^{(0)}(x) & = & 6 \int_x^1 \der z \; \frac{1}{1-z} \left[
  z f\left(\frac{x}{z}\right) - f(x)\right] + 
  6 \int_x^1 \der z \left[\frac{1-z}{z} + z(1-z)\right]
  f\left(\frac{x}{z}\right) + \nonumber \\
  & & 6\; f(x)\left[ \ln(1-x) + \frac{11}{12} - \frac{\enef}{18} 
  \right].  
\eea

To write down a generic expression we decompose a splitting (or
coefficient) function into a regular part ($A$), singular part ($B$),
product of the two ($RS$) and a delta function
\beq{eq:pgeneral}
  P(x) = A(x) + [B(x)]_+ + R(x)[S(x)]_+ + K(x) \de (1-x)
\eeq
where, of course, not all terms have to be present. The following functions are defined in the logarithmic scaling variable $y =
-\ln(x)$:
\beq{eq:funcdef}
  h(y) = f(e^{-y}),\ Q(y) = e^{-y}P(e^{-y}),\ \Abar(y) =
  e^{-y}A(e^{-y})
\eeq
with similar definitions for \Bbar\ and \Sbar; however, $\Rbar(y) =
R(e^{-y})$ and $\bar{K}(y) = K(e^{-y})$ without a factor $e^{-y}$ in
front.  With these definitions \eq{eq:genconv} can be written as
\bea{eq:intuse}
  I(y) & = & \int_0^y \der u\; Q(u)\; h(y-u) = I_1(y) + I_2(y) +I_3(y)
    + I_4(y) \qquad \mbox{with} \nonumber \\ 
  I_1(y) & = & \int_0^y \der u\; \Abar(u)\; h(y-u); \nonumber \\
  I_2(y) & = & \int_0^y \der u\; \Bbar(u)\; \left[ h(y-u) - h(y)
    \right] - h(y) \int_0^x \der z\; B(z); \nonumber \\ 
  I_3(y) & = & \int_0^y \der u\; \Sbar(u)\; \left[ \Rbar(u) h(y-u) -
    \Rbar(0) h(y) \right] - \Rbar(0) h(y) \int_0^x \der z\; S(z);
    \nonumber \\
  I_4(y) & = & \bar{K}(y) h(y).
\eea
where the last integrals of $I_2$ and $I_3$ are still expressed in the
variable $x = \exp(-y)$ to avoid integration extending to infinity in
our expressions.  Note that we are free to swap the arguments $u$ and $y-u$ in \eq{eq:intuse}. 
\dbindex{splitting functions}{singularities in|)}%


\section{Triangular Systems in the DGLAP Evolution}\label{app:solveit}

For the non-singlet evolution we have to solve the equation (see
\Se{se:dglaplin})
\beq{eq:solver1}
   \ve{V} \ve{a} = \ve{b}.
\eeq
The matrix $\ve{V}$ is a lower triangular Toeplitz matrix, that is, a
matrix with the elements~$V_{ij}$ depending only on the difference
$i-j$ as is shown in the $4 \times 4$ example \eq{eq:wmat}.
%
%
This matrix is uniquely determined by storing the first column in a
one-dimensional vector $\ve{v}$ so that  
$V_{ij} = v_{i-j+1}$ for $i \geq j$, and zero otherwise.
\Eq{eq:solver1} is, like any 
other lower triangular system, iteratively solved by forward substitution
\siindex{forward substitution|(}%
\bea{eq:solver3}
   a_1 & = & b_1/v_1 \nonumber \\
   a_i & = & \frac{1}{v_1} \left[ b_i - \sum_{j=1}^{i-1} v_{(i-j+1)}\; 
   a_j \right] \ \mbox{for $i \geq 2$}.
\eea
There is no recursion relation between $a_{i-1}$ and $a_i$
so that in each iteration the sums must be
accumulated, giving an operation
count of \mbox{$n(n+1)/2$} for a system of $n$ equations. This is
as expensive (or cheap) as multiplying
the triangular matrix by a vector.

The substitution algorithm can be extended to solve the coupled
singlet-gluon equation
\beq{eq:solver4}
   \left(
   \begin{array}{cc}
   \ve{V}_{\rm qq} & \ve{V}_{\rm qg} \\ \ve{V}_{\rm gq} & \ve{V}_{\rm gg}
   \end{array}
   \right) \left(
   \begin{array}{c}
   \ve{f} \\ \ve{g}
   \end{array}
   \right) \equiv
   \left(
   \begin{array}{cc}
   \ve{a} & \ve{b} \\ \ve{c} & \ve{d}
   \end{array}
   \right) \left(
   \begin{array}{c}
   \ve{f} \\ \ve{g}
   \end{array}
   \right) = \left(
   \begin{array}{c}
   \ve{r} \\ \ve{s}
   \end{array} \right),
\eeq
where $\ve{a}$ is a short-hand notation for $\ve{V}_{\rm qq}$,
\mbetc\  These matrices are all lower triangular $n \times n$ 
Toeplitz matrices. Writing out this equation in
components it is easy to see that for the first elements
$f_1$ and $g_1$ we have to solve the $2 \times 2$ matrix equation
\beq{eq:solver5}
   \left( \begin{array}{cc} a_1 & b_1 \\ c_1 & d_1 \end{array} \right)
   \left( \begin{array}{c} f_1 \\ g_1 \end{array} \right) =
   \left( \begin{array}{c} r_1 \\ s_1 \end{array} \right) 
   \rightarrow  
   \left( \begin{array}{c} f_1 \\ g_1 \end{array} \right) =
   \frac{1}{a_1 d_1 - b_1 c_1}
   \left( \begin{array}{rr} d_1 & -b_1 \\ -c_1 & a_1 \end{array} \right)
   \left( \begin{array}{c} r_1 \\ s_1 \end{array} \right).
\eeq
For $i \geq 2$ we have to accumulate the sums
\bea{eq:solver6}
    R_i & = & r_i - \sum_{j=1}^{i-1} \left[ a_{(i+1-j)}\;f_j +
    b_{(i+1-j)}\ g_j \right] \nonumber \\
    S_i & = & s_i - \sum_{j=1}^{i-1} \left[ c_{(i+1-j)}\;f_j +
    d_{(i+1-j)}\ g_j \right]
\eea
and solve, for each $i$, the equations
\beq{eq:solver7}
   \left( \begin{array}{cc} a_1 & b_1 \\ c_1 & d_1 \end{array} \right)
   \left( \begin{array}{c} f_i \\ g_i \end{array} \right) =
   \left( \begin{array}{c} R_i \\ S_i \end{array} \right) 
\eeq
The operation count of this algorithm is four times that of
\eq{eq:solver3}, plus some little overhead to solve the $2 \times 2$
matrix equations for each $i$. 
\siindex{forward substitution|)}%

%% file: sections/zmstf.tex

\section{Zero Mass Structure Functions\label{se:zeromasstf}}


\subsection{General Formalism}
\label{se:stfcvol} 

\dbindex{structure functions}{zero-mass structure functions|(}%
The zero-mass structure functions $F_2(x,\qsq)$, $\Fell(x,\qsq)$ and
$xF_3(x,\qsq)$ in un-polarised deep inelastic scattering are calculated
from \eq{eq:genstf} with $\chi = x$. The Wilson coefficients are
\siindex{Wilson coefficient}%
functions of $x$ (and sometimes \enef) only.  We set, for the moment,
the physical scale~\qsq\ equal to the factorisation and
renormalisation
scale $\ms$ and write the singlet/gluon contribution to $F_2$ and
\Fell\ as (there is no contribution to $xF_3$ since this structure
function is a pure non-singlet)
\beq{eq:fsinglet}
  \frac{1}{x} \cF_i^{\rm (s)}(x,\qsq) = [C_{i,{\rm s}} \otimes
  \qsi](x,\ms) + [C_{i, {\rm g}}
  \otimes g](x,\ms) \qquad i = 2,\mbox{L}.
\eeq
Likewise, non-singlet contributions to the structure functions are
given by
\beq{eq:fnsinglet} 
  \frac{1}{x} \cF_i^{({\rm ns})}(x,\qsq) = [C_{i,{\rm ns}} \otimes
  q_{\rm ns}](x,\ms) \qquad i = 2,\mbox{L},3 
\eeq
where the label `ns' stands for the non-singlet indices `+', `$-$' and
`v' as defined by~\eq{eq:defnonsinglet}.  To be precise on notation:
$\cF_2 = F_2$, $\cF_{\rmL} = \Fell$ and $\cF_3 = xF_3$ in
\eq{eq:fsinglet} and \eq{eq:fnsinglet}. A structure function is
calculated by adding the singlet/gluon and non-singlet parts, weighted
by the appropriate combination of electroweak couplings;
we refer to~\cite{ref:epxsec} for how to compute neutral
and charged current cross sections and structure functions
in deep inelastic charged lepton and neutrino scattering.


Like the splitting functions, the coefficient functions are expanded
in powers of \as,
\beq{eq:cseries}
  C_{i,j}^{{\rm N}^{\ell}{\rm LO}} = \sum_{k=0}^{\ell} \asubs^k\;
  C_{i,j}^{(k)} \qquad i = 2, \rmL, 3 \qquad j = \rmg, \rms, +, -, \rmv
\eeq
where $\ell = (0,1,2)$ denotes $({\rm LO},{\rm NLO},{\rm NNLO})$ and
$\asubs = \as/2\pi$. The LO coefficient functions are either zero or
trivial delta functions:
\beq{eq:cfuns0} 
  \begin{array}{lll}
  C_{2,{\rm g}}^{(0)} = 0 \qquad &
  C_{2,{\rm s}}^{(0)} = \de(1-x) \qquad &
  C_{2,{\rm ns}}^{(0)} = \de(1-x) \\
  \mbstrut{5.5mm} 
  C_{\rmL,{\rm g}}^{(0)} = 0 \qquad &
  C_{\rmL,{\rm s}}^{(0)} = 0 \qquad &
  C_{\rmL,{\rm ns}}^{(0)} = 0 \\
  \mbstrut{5.5mm}
  C_{3,{\rm g}}^{(0)} = 0 \qquad &
  C_{3,{\rm s}}^{(0)} = 0 \qquad &
  C_{3,{\rm ns}}^{(0)} = \de(1-x).
  \end{array}
\eeq
The NLO coefficient functions can be found in~\cite{ref:fpzphys}. For
those at NNLO we refer
to~\cite{ref:vnz91,ref:zvn91,ref:zvn92,ref:guillen} and the
parametrisations given in~\cite{ref:f123ns} and~\cite{ref:f12sg}.

The LO coefficient functions for \Fell\ are zero so that the
longitudinal structure function vanishes at LO.  An alternative,
which we call $F_{\rmL}'$, is calculated from the expansion
\siindex{\ifellp\ structure function}%
\beq{eq:clseries}
  C_{\rmL,j}^{{\rm N}^{\ell}{\rm LO}} = \sum_{k=1}^{\ell+1} \asubs^{k}\;
  C_{\rmL,j}^{(k)}. 
\eeq
In this way, $C^{(1)}_{\rmL,j}$ is used already at LO (giving a
non-zero \Fell) and $C^{(2)}_{\rmL,j}$ at NLO. At NNLO the 3-loop
coefficient function $C^{(3)}_{\rmL,j}$ is taken
from~\cite{ref:fln3lo}. As stated in~\cite{ref:fln3lo}, this 3-loop
calculation applies only to electromagnetic current exchange so that
Z$^0$ or W$^{\pm}$ contributions to $F_{\rm L}'$ at NNLO are, at
present, not available.


\subsection{Renormalisation and Factorisation Scale Dependence}
\label{se:fscale}

\dbindex{renormalisation scale dependence}{of structure functions}%
To calculate the \emph{renormalisation} scale dependence
($\Rs \neq \Fs$) we replace, in the expansion of the coefficient
functions, the powers of \asubs\ by the Taylor series given
in~\eq{eq:asexpansion}. If the expansion~\eq{eq:cseries} is used, the
truncation of the right-hand side of~\eq{eq:asexpansion} is to
order~$\asubs$ in NLO and $\asubs^2$ in NNLO. If, for $\Fell'$, the
expansion~\eq{eq:clseries} is used, the truncation is to order $\asubs$
in LO, $\asubs^2$ in NLO and $\asubs^3$ in NNLO, like for the
splitting functions.
\siindex{truncation prescription}%

\siindex{factorisation scale dependence}%
To calculate the \emph{factorisation} scale dependence ($\qsq \neq \Fs$),
the coefficient functions in~\eq{eq:cseries}
and~\eq{eq:clseries} are replaced by~\cite{ref:f123ns,ref:f12sg}
\beq{eq:csgnsexp} 
  \ceeka{0}_{i,j} \ \rar\  \ceeka{0}_{i,j} \qquad \mbox{and} \qquad
  \ceeka{k}_{i,j} \ \rar\  \ceeka{k}_{i,j} + \sum_{m = 1}^k
  \ceeka{k,m}_{i,j} \elF^m \qquad k \geq 1,
\eeq
where $\elF = \ln(\qsq/\Fs)$ and $\Fs = \Rs$. To write compact
expressions for the $\ceeka{k,m}_{i,j}$,
we introduce the following vector notation.
In the non-singlet sector we have a
one-dimensional vector $\ve{C}_i = C_{i,{\rm ns}}$ and a
$1 \times 1$ matrix $\ve{P} = P_{\rm ns}$.
In the singlet/gluon sector we
have a 2-dimensional row-vector and a $2 \times 2$ matrix that 
are given by
\[
\ve{C}_i = \left( C_{i,{\rm s}}\  C_{i,{\rm g}}
           \right) \ \ \mbox{and} \ \ 
\ve{P}   = \begin{pmatrix}
           P_{\rm qq} & P_{\rm qg} \\
           P_{\rm gq} & P_{\rm gg} \end{pmatrix}.
\]

In this vector notation, the functions
$\ceeka{k,m}_{i,j}$ in \eq{eq:csgnsexp} are written as
\bea{eq:cikmdef}                                  
  \ve{C}_i^{(1,1)} & = & \ve{C}_i^{(0)} \otimes
    \ve{P}^{(0)} \nonumber \\ 
  \ve{C}_i^{(2,1)} & = & \ve{C}_i^{(0)} \otimes
    \ve{P}^{(1)} + \ve{C}_i^{(1)} \otimes \left[
    \ve{P}^{(0)} - \be_{0\;} \ve{I} \right] \nonumber \\ 
   \ve{C}_i^{(2,2)} & = & \frac{1}{2}\;\ve{C}_i^{(1,1)} \otimes
    \left[ \ve{P}^{(0)} - \be_{0\;} \ve{I} \right] \nonumber \\
   \ve{C}_i^{(3,1)} & = &  \ve{C}_i^{(0)} \otimes \ve{P}^{(2)} + 
    \ve{C}_i^{(1)} \otimes \left[
    \ve{P}^{(1)} - \be_{1\;} \ve{I} \right] +
    \ve{C}_i^{(2)} \otimes \left[
    \ve{P}^{(0)} - 2\be_{0\;} \ve{I} \right] \nonumber \\
   \ve{C}_i^{(3,2)} & = & \frac{1}{2} \left\{
    \ve{C}_i^{(1,1)} \otimes \left[
    \ve{P}^{(1)} - \be_{1\;} \ve{I} \right] +
    \ve{C}_i^{(2,1)} \otimes \left[
    \ve{P}^{(0)} - 2\be_{0\;} \ve{I} \right] \right\} \nonumber \\
   \ve{C}_i^{(3,3)} & = & \frac{1}{3}
    \ve{C}_i^{(2,2)} \otimes \left[
    \ve{P}^{(0)} - 2\be_{0\;} \ve{I} \right].
\eea 
For $F_2$, $\Fell$ and $xF_3$, 
the coefficients are calculated up to $\ve{C}_i^{(2,2)}$. For $\Fell'$,
on the other hand, all coefficients in \eq{eq:cikmdef} are computed. 
Note, however, that quite some convolutions 
are trivial because the LO coefficient functions are either zero or
$\de$-functions, see~\eq{eq:cfuns0}.

As mentioned above, the expression~\eq{eq:csgnsexp}
applies only when $\Fs = \Rs$. It is therefore not possible to vary
both scales \Rs\ and \qsq\ at the same time. 
\dbindex{structure functions}{zero-mass structure functions|)}%


\subsection{The \zmstf\ Package} \label{se:zmstf}

\siindex{\izmstf|(}%
The \zmstf\ package is a \qcdnum\ add-on with routines that
calculate the structure functions $F_2$, $\Fell$ and $xF_3$ in
un-polarised deep inelastic scattering. The structure functions
are computed as a convolution of the
parton densities with zero-mass coefficient functions,
using the convolution engine described in \Se{se:stfuser}.

The list of subroutines is given in \Ta{tab:zmstf}.
\begin{table}[bth] 
  \caption{ Subroutine and function calls in \zmstf.}
  \begin{center}
  \begin{tabular*}{0.95\textwidth}{l@{\extracolsep{\fill}}l}
  \\
  Subroutine or function & Description \\
  \hline
  \xtt{ZMFILLW ( *nwords )}
  & Fill weight tables                 \\
  \xtt{ZMDUMPW ( lun, 'filename' )}
  & Dump weight tables                 \\
  \xtt{ZMREADW ( lun, 'filename', *nwords, *ierr )}
  & Read weight tables                 \\
  \xtt{ZMDEFQ2 ( a, b )}
  & Define \qsq                        \\
  \xtt{ZMABVAL ( *a, *b )}
  & Retrieve $a$ and $b$ coefficients  \\
  \xtt{ZMQFRMU ( qmu2 )}
  & Convert \Fs\ to \qsq               \\
  \xtt{ZMUFRMQ ( Q2 )}
  & Convert \qsq\ to \Fs               \\
  \xtt{ZSWITCH ( iset )}
  & Switch pdf set                     \\
  \xtt{ZMSTFUN ( istf, def, x, Q2, *f, n, ichk )}
  & Structure functions                \\
  \hline
  {\footnotesize Output arguments are pre-fixed with
   an asterisk (\xtt{*}).} & \\
  \end{tabular*} 
  \end{center}
  \label{tab:zmstf}
\end{table}
Note that error messages are, in most cases, issued by the
underlying \qcdnum\ routines and not by the \zmstf\ routine
itself. However, the calling \zmstf\ routine is mentioned in the error
message so that you know where it came from.

\subnbox{call ZMFILLW ( *nwords )}

Fill the weight tables. The tables are calculated for all flavours $3
\leq \enef \leq 6$ and for all orders LO, NLO, NNLO. On exit, the
number of words occupied by the store is returned in \xtt{nwords}.  If
you get an error message that the internal store is too small to
contain the weight tables, you should increase the value of the
parameter \xtt{nzmstor} in the include file \xtt{zmstf.inc} and
recompile \zmstf.

This routine (or \xtt{zmreadw} below) should be called after an $x$-$\ms$
grid is defined in \qcdnum\ and before the first call to
\xtt{zmstfun}.


\subnbox{call ZMDUMPW ( lun, 'filename' )}

Dump the weights in memory via logical unit number \xtt{lun} to a disk
file. The dump is unformatted so that the weight file cannot be
exchanged across machines.


\subnbox{call ZMREADW ( lun, 'filename', *nwords, *ierr )}

Read weights from a disk file via logical unit number \xtt{lun}. On
exit, \xtt{nwords} contains the number of words read into the store
(fatal error if not enough space, see above) and the flag \xtt{ierr}
is set as follows.
\begin{tdeflist}[istf\ \ ]{-1mm} 
\item[\xtt{0}] Weights are successfully read in.
\item[\xtt{1}] Read error or input file does not exist.
\item[\xtt{2}] Incompatible \qcdnum\ version.
\item[\xtt{3}] Incompatible \zmstf\ version.
\item[\xtt{4}] Incompatible $x$-\ms\ grid definition.
\end{tdeflist}
These errors will not generate a program abort so that one should check
the value of~\xtt{ierr}, and take the appropriate action if it is
non-zero.


\subnbox{call ZMDEFQ2 ( a, b )}

Define the relation between the factorisation scale \Fs\ and \qsq
\[
  \qsq = a \Fs + b.
\]
The \qsq\ scale can only be varied when the renormalisation and factorisation scales are set equal in \qcdnum. The default setting is \compa{a}{=}{1} and \compa{b}{=}{0}. The ranges are limited to \range{0.1}{\leq}{a}{\leq}{10} and \range{-100}{\leq}{b}{\leq}{100}.

A call to \xtt{zmabval(a,b)} reads the coefficients back from
memory. To convert between the scales use:
\begin{verbatim}
             Q2   = zmqfrmu(qmu2)
             qmu2 = zmufrmq(Q2)
\end{verbatim} 


\subnbox{call ZSWITCH ( iset )}

By default, the structure functions are calculated from the
un-polarised parton densities, evolved with \qcdnum\ 
(\compa{iset}{=}{1}). With this routine one can switch to
the custom evolution (\xtt{4}), or to one of the external
pdf sets (\xtt{5}--\xtt{9}). Switching to polarised pdfs
(\xtt{2}) or to fragmentation functions (\xtt{3}) does not
make sense and will produce an error message.


\subnbox{call ZMSTFUN ( istf, def, x, Q2, *f, n, ichk )}

Calculate a structure function for a linear combination of parton
densities. 
\begin{tdeflist}[defhmzczh\ \ \ ]{-1mm} 
\item[\xtt{istf}] Structure function index
  (\xtt{1},\xtt{2},\xtt{3},\xtt{4}) = $(\Fell,F_2,xF_3,\Fell')$.
\item[\xtt{def(-6:6)}] Coefficients of the quark linear combination
  for which the structure function is to be calculated. The indexing
  of \xtt{def} is given in \eq{eq:iqqbar}.
\item[\xtt{x}, \xtt{Q2}] Input arrays containing a list of $x$ and
  \qsq\ (not \ms) values.
\item[\xtt{f}] Output array containing the list of structure
  functions.
\item[\xtt{n}] Number of items in \xtt{x}, \xtt{Q2} and \xtt{f}.
\item[\xtt{ichk}] If set to zero, \xtt{zmstfun} will return a
  \xtt{null} value when $x$ or \ms\ are outside the grid boundaries;
  otherwise you will get a fatal error message. A \ms\ point that is
  close or below the QCD scale $\La^2$ is considered to be outside the
  grid boundary.
\end{tdeflist}
To calculate a structure function for more than one interpolation
point, it is recommended to not execute \xtt{zmstfun} in a loop but to pass the entire list of interpolation points in a single
call. The loop is then internally optimised for greater speed.
\siindex{\izmstf|)}%

%% file: sections/hqstf.tex

\section{Heavy Quark Structure Functions\label{se:riemersma}}

\dbindex{structure functions}{heavy flavour contributions|(}%
A NLO calculation of the heavy quark contributions to the $F_2$
and \Fell\ structure functions in deep inelastic charged lepton-proton
scattering is given in~\cite{ref:riemersma}. 
Only electromagnetic exchange contributions are taken into account.
In this calculation, a heavy flavour $h$ is not taken to be a
constituent of the incoming proton but is, instead, assumed to be
exclusively produced in the hard scattering process. 
Quarks with pole mass $m < m_h$ are taken to be mass-less so that
the input light quark densities should have
been evolved in the~\ffns\ with $\enef = (3,4,5)$ for 
$h = (\cbt)$~\cite{ref:eric}.

A heavy flavour contribution to $F_2$ or $F_L$ is calculated from
\beq{eq:riemersma}
  F_k^h(x,Q^2)   =   
  \frac{\alpha_s}{2\pi} \left\{\, e^2_h\, g \otimes {\cal C}^{(0)}_{k,g} + 
  \frac{\alpha_s}{2\pi} \left(\,  e^2_h\, g \otimes {\cal C}^{(1)}_{k,g} + 
                           e^2_h\, \qsi \otimes {\cal C}^{(1)}_{k,q} +
              q_{\rm p}   \otimes {\cal D}^{(1)}_{k,q}\, \right)\, \right\},
\eeq
where $e_h$ is the charge of the heavy quark (in units of the positron
charge), $g$ is the gluon density, $\qsi$ is the singlet density and 
\[
  q_{\rm p} = \sum_{i=1}^{\enef} e^2_i \; (q_i + \bar{q}_i)
\]
is the charge-weighted proton quark distribution for \enef\ light
flavours.  The first term in~\eq{eq:riemersma} is the LO contribution
from the photon-gluon fusion process $\gamma^*g \rightarrow h\bar{h}.$
The last three terms correspond to the NLO sub-process $\gamma^*g
\rightarrow h\bar{h}g$ and $\gamma^*q \rightarrow
h\bar{h}q$.\footnote{In the LO and the first two NLO terms the virtual
photon couples to the heavy quark, hence the factor $e^2_h$ in
\eq{eq:riemersma}. The last NLO term describes the process where the
virtual photon couples to a light quark which subsequently branches
into a $h\bar{h}$ pair via an intermediate gluon: hence the appearance
of the charge weighted sum, $q_{\rm p}$, of light quark
distributions.} For the heavy quark coefficient functions ${\cal C}$
and $\cal D$ in \eq{eq:riemersma} we refer to
\cite{ref:riemersma}.\footnote{ Some of these coefficient functions
are given as interpolation tables (taken from code provided by
S.~Riemersma) since they are too complex to be cast into analytical
form.  Note that in~\cite{ref:riemersma} the coefficient functions are
convolved with parton momentum densities and not with number
densities.}

In terms of a number density $f(x,\ms)$, the convolution integrals in
\eq{eq:riemersma} are defined by
\beq{eq:fotimesc}
  f \otimes {\cal C} =
      \int_{ax}^1 \frac{dz}{z}\; zf(z,\ms)
       \; {\cal C} ( x/z, Q^2,\ms, m_h^2 ) 
\eeq
where $a = 1 + 4m_h^2/Q^2$ and $\ms$ is the factorisation (equals
renormalisation) scale which is usually set to $\ms = Q^2$ or $\ms =
Q^2 + 4m_h^2$. The kinematic domain where the heavy quarks contribute
is restricted by the requirement that the square of the $\gamma^*$p
centre of mass energy must be sufficient to produce the $h\bar{h}$
pair: $W^2 = M^2 + Q^2 (1-x)/x \geq M^2 + 4m_h^2$ so that the lower
integration limit $ax \leq 1$ in \eq{eq:fotimesc}.  It turns out that
the dependence of the coefficient functions on the relation between
\qsq\ and \ms\ cannot be factorised so that each setting of the scale
parameters needs its own set of weight tables. To calculate the
renormalisation scale dependence, the powers of $\asubs = \as/2\pi$
\dbindex{renormalisation scale dependence}{of structure functions}%
in~\eq{eq:riemersma} are replaced by the Fourier 
expansion~\eq{eq:asexpansion}, truncated to~$\asubs$ in LO, 
and to~$\asubs^2$ in NLO. Note that one can vary either
\Rs\ or \qsq\ with respect to \Fs, but not both at the same time. 

The convolution integral \eq{eq:fotimesc} is not of the general
form~\eq{eq:genstf}: (i) the factor $x$ in front is missing; (ii) the
pdf is $xf(x)$ and not $f(x)$ and (iii) the argument of $C$ is $x/z$ and
not~$\chi/z$.  This mismatch is cured by presenting
to \qcdnum\ the modified kernel
\[
  \tilde{C}(\chi, \ms, \qsq, \masqh) \equiv \frac{a}{\chi}\; C  
  \left( \frac{\chi}{a}, \ms, \qsq, \masqh \right),\mbox{\ with\ \ }
  \chi \equiv a x.
\]  
To make the heavy quark calculation available in \qcdnumnew\ (as it
was in \qcdnumold) we provide the add-on package \hqstf\ described
below. 
\dbindex{structure functions}{heavy flavour contributions|)}%


\subsection{The \hqstf\ Package} \label{se:hqstf}

\siindex{\ihqstf|(}%
The \hqstf\ package calculates up to NLO the heavy flavour
contributions to the $F_2$ or \Fell\ structure functions from pdfs
evolved in the \ffns\ scheme with \enef\ light flavours. The list of
subroutines is given in \Ta{tab:hqstf}.
\begin{table}[bth] 
  \caption{ Subroutine and function calls in \hqstf.}
  \begin{center}
  \begin{tabular*}{0.95\textwidth}{l@{\extracolsep{\fill}}l}
  \\
  Subroutine or function & Description \\
  \hline
  \xtt{HQFILLW ( istf, qmass, aq, bq, *nwords )}
  & Fill weight tables                 \\
  \xtt{HQDUMPW ( lun, 'filename' )}
  & Dump weight tables                 \\
  \xtt{HQREADW ( lun, 'filename', *nw, *ierr )}
  & Read weight tables                 \\
  \xtt{HQPARMS ( *qmass, *aq, *bq )}
  & Retrieve parameters                \\
  \xtt{HQQFRMU ( qmu2 )}
  & Convert \Fs\ to \qsq               \\
  \xtt{HQMUFRQ ( Q2 )}
  & Convert \qsq\ to \Fs               \\
  \xtt{HSWITCH ( iset )}
  & Switch pdf set                     \\
  \xtt{HQSTFUN ( istf, icbt, def, x, Q2, *f, n, ichk )}
  & Structure functions                \\
  \hline
  {\footnotesize Output arguments are pre-fixed with
   an asterisk (\xtt{*}).} & \\
  \end{tabular*} 
  \end{center}
  \label{tab:hqstf}
\end{table}
We will only describe here the routines \xtt{hqfillw} and
\xtt{hqstfun}, the other ones being similar to those in the \zmstf\
package.
%
%
\subnbox{call HQFILLW ( istf, qmass, aq, bq, *nwords )}
Fill the weight tables. To be called before anything else.
\begin{tdeflist}[nwordsxxx\ \ ]{-1mm} 
  \item[\xtt{istf}] Select structure function: \xtt{1} = \Fell,  \xtt{2} = 
    $F_2$ and \xtt{3} = both.
  \item[\xtt{qmass(3)}] Input array with the \cbt\ quark masses in GeV.
    If a quark mass is set to
    $m_h < 1$~GeV, no tables will be generated for that quark.
  \item[\xtt{aq}, \xtt{bq}] Defines the relation $\qsq = a \Fs + b$.  
  \item[\xtt{nwords}] Gives, on exit, the number of words used in the store.
\end{tdeflist}
One will get a fatal error if the store is not large enough to hold
all tables. In that case you can increase the value of \xtt{nhqstor}
in the include file \xtt{hqstf.inc} and recompile \hqstf. The values
of the mass and scale parameters can be retrieved at any time after
the call to \xtt{hqfillw} (or \xtt{hqreadw}) by a call to
\xtt{hqparms(qmass,aq,bq)}.  
%

\subnbox{call HQSTFUN ( istf, icbt, def, x, Q2, *f, n, ichk )}
Calculate the heavy quark contribution to a structure function.
\begin{tdeflist}[defxxxxxxx\ \ ]{-1mm} 
  \item[\xtt{istf}] Calculate \Fell\ (\xtt{1}) or $F_2$ (\xtt{2}).
  \item[\xtt{icbt}] Select contribution from charm (\xtt{1}), bottom
    (\xtt{2}) or top (\xtt{3}).
  \item[\xtt{def(-6:6)}] Coefficients of the quark linear combination
    for which the structure function is to be calculated. The indexing
    of \xtt{def} is given in \eq{eq:qqbindex}.
  \item[\xtt{x}, \xtt{Q2}] Input arrays containing a list of $x$ and
    \qsq\ (not \ms) values. 
  \item[\xtt{f}] Output array containing the list of structure
    functions.
  \item[\xtt{n}] Number of items in \xtt{x}, \xtt{Q2} and \xtt{f}.
  \item[\xtt{ichk}] If set to zero, \xtt{hqstfun} will return a
    \xtt{null} value when $x$ or \ms\ are outside the grid boundaries;
    otherwise one will get a fatal error message.
\end{tdeflist}
The routine checks that for \xtt{icbt} = (1,2,3) = (c,b,t) the pdfs
were evolved in the \ffns\ with $\enef = (3,4,5)$ flavours. 
When \xtt{icbt} is pre-pended by a minus sign, the check
on the \ffns\ remains active but that on the number of flavours
is switched off.
    
Here is a snippet of code that, in combination with \zmstf, calculates
the d,u,s contribution, the charm contribution and the
total $F_2$ (neglecting bottom and top) in charged lepton-proton
scattering (the pdfs should have
been evolved with $\enef = 3$ flavours).
\begin{verbatim}
      dimension x(100),Q2(100),F2dus(100),F2c(100),F2p(100)
      dimension proton(-6:6)
      data proton /4.,1.,4.,1.,4,.1.,0.,1.,4.,1.,4.,1.,4./ !divide by 9
        ..
      call zmstfun(2,    proton, x, Q2, F2dus, 100, ichk)
      call hqstfun(2, 1, proton, x, Q2, F2c  , 100, ichk)
      do i = 1,100
        F2p(i) = F2dus(i) + F2c(i)
      enddo    
\end{verbatim}
\siindex{\ihqstf|)}%

%% file: sections/references.tex
%
%